\documentclass[fleqn,10pt]{wlscirep}
\usepackage[utf8]{inputenc}
\usepackage[T1]{fontenc}
\usepackage{float} 
\usepackage{subcaption} 
\usepackage{placeins} 
\usepackage{threeparttable} 
\usepackage{amsmath, bm} 
\usepackage[style=nature]{biblatex} 
\title{Systems-level health of patients living with end-stage kidney disease using standard lab values}

\author[1,*]{Glen Pridham}
\author[2]{Karthik K.\ Tennankore}
\author[3]{Kenneth Rockwood}
\author[2]{George Worthen}
\author[1,$\dagger$]{Andrew~D.~Rutenberg}

\affil[1]{{Department of Physics and Atmospheric Science}, {Dalhousie University}, {Halifax}, {B3H 4R2}, {Nova Scotia}, {Canada}}

\affil[2]{{Dalhousie University and Nova Scotia Health}, {5820 University Avenue}, {Halifax}, {B3H 1V8}, {Nova Scotia}, {Canada}}

\affil[3] {{Division of Geriatric Medicine}, {Dalhousie University}, {Halifax}, {B3H 2E1}, {Nova Scotia}, {Canada}}

\affil[*]{glen.pridham@dal.ca}
\affil[$\dagger$]{adr@dal.ca}

\begin{abstract} 
We present a systems-level analysis of end-stage kidney disease (ESKD) with a dynamical network analysis of 14 commonly measured blood-based biomarkers in patients undergoing regular haemodialysis. Utilizing a validated pipeline for declining homeostatic systems, our approach learns a dynamical model together with an invertible transformation that simplifies the behaviour of observed biomarkers into natural variables. Within the natural variables, we identified two distinct dynamical behaviours: (i) stochastic accumulation, the random accumulation of abnormal values, and (ii) mallostasis, a deterministic drift towards worse health. These behaviours are identified by persistent fluctuations indicating weak stability, or a gradual shift in homeostatic set point, respectively. Both lead to worsening natural variable values, making the natural variables salient survival predictors with preferred directions of increasing risk. When this worsening is transformed back into observable biomarkers, it generates a coherent spectrum of worsening medical signs characteristic of a medical syndrome. Specifically, we found that small modules of natural variables corresponded to two existing syndromes commonly afflicting ESKD patients: protein-energy wasting and sepsis. We also identified new prospective syndromes. Our findings suggest that natural variables are robust, systems-level biomarkers, capturing the complex, holistic changes in health associated with ESKD.
\end{abstract}
\begin{document}

\flushbottom
\maketitle

\thispagestyle{empty}


\section{Introduction}
End-stage kidney disease (ESKD), like chronic kidney disease (CKD), is a systemic illness resulting from multiple causes and affecting multiple sub-systems \cite{Zoccali2017-vp, Lousa2020-js, Pandey2023-ou} including metabolic, immune, neuroendocrine and cardiopulmonary \cite{Zoccali2017-vp}. This results in complex syndromes, such as protein energy wasting (PEW) \cite{Zoccali2017-vp, Fouque2008-ax}, cardio-renal syndrome \cite{Zoccali2011-ax}, and mineral and bone disorders \cite{Zoccali2011-ax}. For example, PEW includes interactions between malnutrition, uremic toxins, hypercatabolism and chronic inflammation, and emerges from multiple underlying factors that result in a persistent wasting state leading to a reduced quality of life and increased risk of hospitalization or death \cite{Fouque2008-ax}. Sub-system dysfunction due to CKD --- e.g.\ renal function, inflammation and metabolism --- each have a substantial and growing collection of inter-dependent biomarkers but there is no clear rule for how to combine or prioritize this information \cite{Lousa2020-js}. Needed are systems-level biomarkers that can capture the widespread changes to health that occur in CKD, together with interpretable quantitative models able to describe how these biomarkers evolve in time.

To achieve this goal, we characterize patient health by their biological system's ability to preserve stability, i.e. to recover from perturbations that challenge homeostasis. For this we use the Stochastic Finite-difference (SF) model \cite{mallostasis, twins}. The model parameterizes the dynamical behaviour of a system in terms of equilibrium (fixed-point) values, $\vec{\mu}$, and an interaction network within and between biomarkers, $\boldsymbol{W}$. The eigen-decomposition of $\boldsymbol{W}$ determines the stability of the system \cite{Ledder2013-em} in terms of its canonical coordinates: ``natural'' variables \cite{mallostasis}. Each natural variable has a characteristic recovery rate which defines its stability. The slowest of these determines the overall system stability.

Stability is a mathematical expression of resilience. Whereas stability of a network can be obscured by compensatory interactions, these interactions are absent between natural variables --- leaving bare stability, and thus resilience. Our prior work suggests that stability is preserved in the short-term, but that there is a long-term drift in the homeostatic fixed points towards worse health --- a phenomenon we named ``mallostasis'' \cite{mallostasis}. Mallostasis is consistent with allostatic load theory, which posits that the demands of short-term stability leads to long-term failure, through e.g.\ chronic over-activation of the flight-or-flight stress response \cite{Juster2010-kw}. The degree of resilience (stability) can also be important for determining declining health, either weak stability \cite{twins, mallostasis} or instability \cite{Avchaciov2022-ws}. Identifying and characterizing specific mechanisms of homeostatic erosion in ESKD would help to clarify the roles of stability in declining health and would also identify the variables that are useful for monitoring health and planning treatment.

Our study population is comprised of Canadian ESKD patients receiving haemodialysis. We include a small longitudinal cohort from Nova Scotia (main dataset; $N=713$), and a large cross-sectional cohort from the rest of Canada (validation dataset; $N=61036$). Using the main dataset, we model the dynamical behaviour of a system of 14 standard blood tests measured approximately every 6 weeks. We use this model to identify the natural variables, then we determine their relevance to patient health, their modes of action, and their underlying biological meaning. We find that the natural variables efficiently capture changes to health that occur during ESKD, and are specific to biological syndromes. Each natural variable can be estimated using common blood tests and hence is a prospective systems-level biomarker with clinical applicability. 
\section{Model}
We model a system near a stable point as,
\begin{align}
    \vec{y}_{i n+1} &= \vec{y}_{i n} + \boldsymbol{W}\Delta t_{i n+1}(\vec{y}_{i n} - \hat{\vec{\mu}}_{i n}) +\hat{\vec{\epsilon}}_{i n+1}, \nonumber\\
    \hat{\vec{\epsilon}}_{i n+1} &\sim \mathcal{N}(0,\boldsymbol{\hat{\Sigma}}|\Delta t|_{i n+1} ) \nonumber \\
    \hat{\vec{\mu}}_{i n} &\equiv \hat{\vec{\mu}}_{0}+\boldsymbol{\hat{\Lambda}}\vec{x}_{i n}+ \hat{\vec{\mu}}_{t} t_{in}
    \label{eq:sf}
\end{align}
where $\vec{y}_{in}$ represents the $i$th individual's set of biomarkers measured at time $t_{in}$. The model estimates: a parameterized dynamical equilibrium, $\hat{\vec{\mu}}$, where the system reaches a steady-state (`set point'); a causal, network parameter, $\boldsymbol{W}$; and a noise term, $\boldsymbol{\hat{\Sigma}}$ which controls the strength and correlation of fluctuations, making $\boldsymbol{\hat{\Sigma}}$ sensitive to the response to external stressors (robustness) as well as additional effects not in the model (e.g.\ individual variability). The dynamical equilibrium, $\hat{\vec{\mu}}$, is allowed to vary according to a set of covariates for each individual, $\vec{x}_{in}$. The model estimates linear interactions between the covariates and $\hat{\vec{\mu}}$ via $\boldsymbol{\hat{\Lambda}}$, and a linear drift rate with age via $\hat{\vec{\mu}}_{t}$.

The key analysis step is to use an invertible transformation $\boldsymbol{P}$ to diagonalize $\boldsymbol{W}$ such that the equations decouple into `natural variables':
\begin{align}
    z_{ijn+1} &= z_{ijn} + \lambda_j\Delta t_{in+1}(z_{ijn} - {\mu}_{ijn}) +{\epsilon}_{ijn+1}, \label{eq:z}
\end{align}
where $\vec{z}_n\equiv \boldsymbol{P}^{-1}\vec{y}_n$, $\lambda_j \equiv P_{j\cdot}^{-1} \boldsymbol{W} P_{\cdot j}$, ${\vec{\mu}}_{n} \equiv \boldsymbol{P}^{-1}\hat{\vec{\mu}}_{n}$ and ${\vec{\epsilon}}_{in} \equiv \boldsymbol{P}^{-1}\hat{\vec{\epsilon}}_{in}$ (meaning $\vec{\epsilon}_{i n+1} \sim \mathcal{N}(0,\boldsymbol{\Sigma}|\Delta t|_{i n+1})$).  The dynamical behaviour is simple: each stable $z$ moves independently towards the steady-state, $\mu(t)$, with a speed proportional to $\lambda$. (Unstable $z$ would be repulsed from $\mu(t)$.) Stability is determined by the sign of $\lambda$: positive is unstable, and negative is stable. (Although $\lambda$ can be complex, the convergence rate depends only on the real part and hence we use the shorthand $|\lambda|$ to mean $|Re(\lambda)|$.) For stable systems, $|\lambda|^{-1}$ sets the recovery timescale: small $|\lambda|$ means slow recovery, long auto-correlation time, and long memory. This permits variables with small $|\lambda|$ to build up stochastic fluctuations, which are incorporated into the variance (Eqs.~\ref{eq:ss}). The noise pushes each $z$ randomly up or down, but can be correlated across the different $z$ via $\boldsymbol{\Sigma}$. 

In the limit $\Delta t\to 0$ Eq.~\ref{eq:z} becomes a stochastic differential equation which once solved for the mean \cite{mallostasis} yields
\begin{align}
    \langle z \rangle(t) &= \big(\langle z_0 \rangle  - \frac{\mu_{t}}{\lambda} - \mu_0\big)e^{\lambda t} + \big( \frac{\mu_{t}}{\lambda} + \mu_0 + \mu_{t} t \big), \label{eq:meant}
\end{align}
the first term is a decaying memory of the initial mean and the second term is the (lagged) homeostatic set point, which is permitted to drift linearly with time via $\mu_t$. 
For example, $z_1$, had a long transient period followed by an equilibrium near $\mu\approx \mu_0$, Figure~\ref{fig:samplefits}. 

So long as the system is stable, with $\lambda < 0$, a steady-state is reached after time $t \gg |\lambda|^{-1}$. The steady-state statistics are 
\begin{subequations} \label{eq:ss}
\begin{align}
    \lim_{\text{steady-state}} \frac{d}{dt} \langle z \rangle &= \mu_t, \label{eq:meantss}
\end{align}  
\begin{align}
    \lim_{\text{steady-state}}\text{Var}(z)(t) &= \frac{\sigma^2}{2|\lambda|},~\text{and} \label{eq:var}
\end{align}  
\begin{align}
    \lim_{\text{steady-state}}\text{ACF}(z)(t,t+\Delta t) &= e^{\lambda|\Delta t|}, \label{eq:acf}
\end{align}
\end{subequations}
where $\sigma^2\equiv \langle \epsilon^2\rangle$ ($\sigma_j \equiv {\Sigma}_{jj}$), $\Delta t$ is the time lag between observations, $\text{Var}$ is the variance and $\text{ACF}$ is the auto-correlation function. The recovery timescale $|\lambda|^{-1}$ is the auto-correlation time.


\section{Results}
\subsection{System health} \label{sec:system}
We modelled 14 longitudinal `raw' serum (blood-based) biomarkers across the domains of kidney function, dialysis clearance, electrolytes, immune function, anemia and metabolic function. These were regularly measured every  6~weeks (approx.) over an observation window of 3~months to 5~years (3~months excludes individuals with acute kidney injury; 5~years was the study half-life). Additional biomarkers were included when testing for associations. Where ambiguous, biomarkers include the prefix ``pre'' if they were measured before dialysis sessions (default) or ``post'' if measured after. See the supplemental for details.

\begin{figure} 
     \centering
        \includegraphics[width=\textwidth]{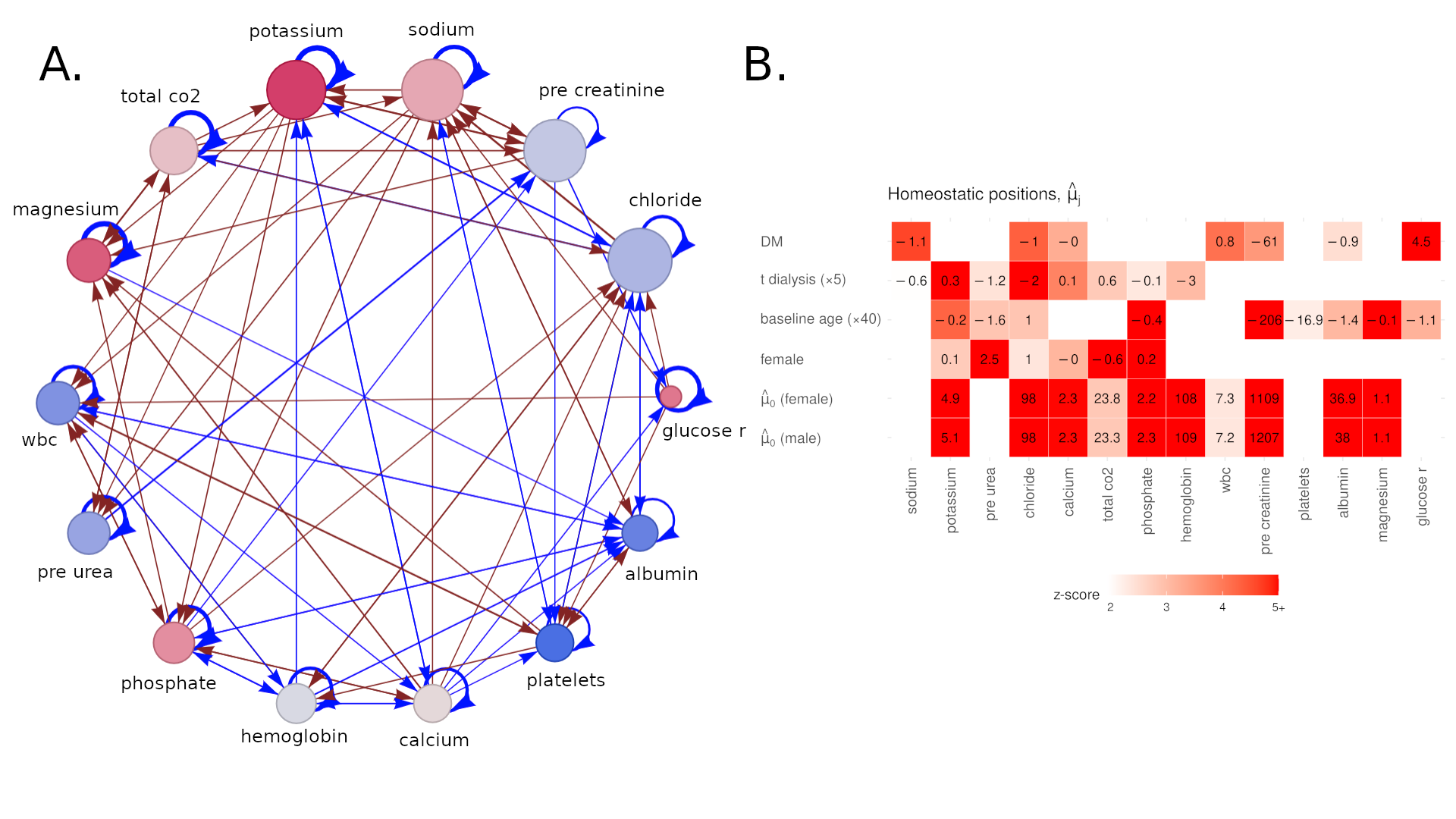}  
    \caption{Key parameter estimates for measured biomarkers.  \textbf{A.} Interaction network $\boldsymbol{W}$. Link width and node size are proportional to link strength; red has strongest noise, blue has weakest. The network links encode conditional dependencies between the variables: $X\to Y$ means that the next observation of $Y$ depends on the current observed value of $X$ (blue links are negative, red are positive). This permits abnormal biomarker values to push other biomarker values --- capturing both compensatory behaviour and propagation of dysfunction.  Potassium and sodium are key nodes since they have high connectivity (large) and strong noise (red). Creatinine and chloride are also well connected. Node size is $n_k = \sqrt{\sum_{j\neq k} W_{jk}^2 + \sum_{j\neq k} W_{kj}^2 }$; node colour is rank of noise from red to blue, $c_k \sim \sqrt{\text{diag}(\hat{\boldsymbol{\Sigma}})}$. \textbf{B.} Homeostatic positions, $\hat{\vec{\mu}}(t)$.  Baseline positions, $\mu_0$, appear to be reasonably close to medical target values. However several are changing with age or with time on dialysis, for example potassium (t\_dialysis row; drift rate per 5 years). Sex=0 for male and 1 for female. DM=1 for diabetes mellitus positive and 0 for negative (DM positive are instructed to maintain higher glucose to avoid hypoglycemia during dialysis \cite{Kalantar-Zadeh2015-ld}). Non-significant parameters have been excluded ($p > 0.05$). }
%
    \label{fig:network}
\end{figure}

The estimated causal interaction network between the biomarkers is shown in Figure~\ref{fig:network}. Links indicate conditional dependencies between measurements e.g.\ high chloride today predicts high sodium in 6 weeks time. Potassium and sodium are key network nodes since they have both strong noise and many connections. Strong noise (red nodes) indicates that a variable is introducing a lot of information into the system, while high-connectivity (large) nodes mediate by pushing and pulling information to and from other biomarkers. Given the extensive connectivity, an event that suddenly increases any one of these variables can cause delayed complex changes to many other variables.

\begin{figure} 
     \centering
        \includegraphics[width=\textwidth]{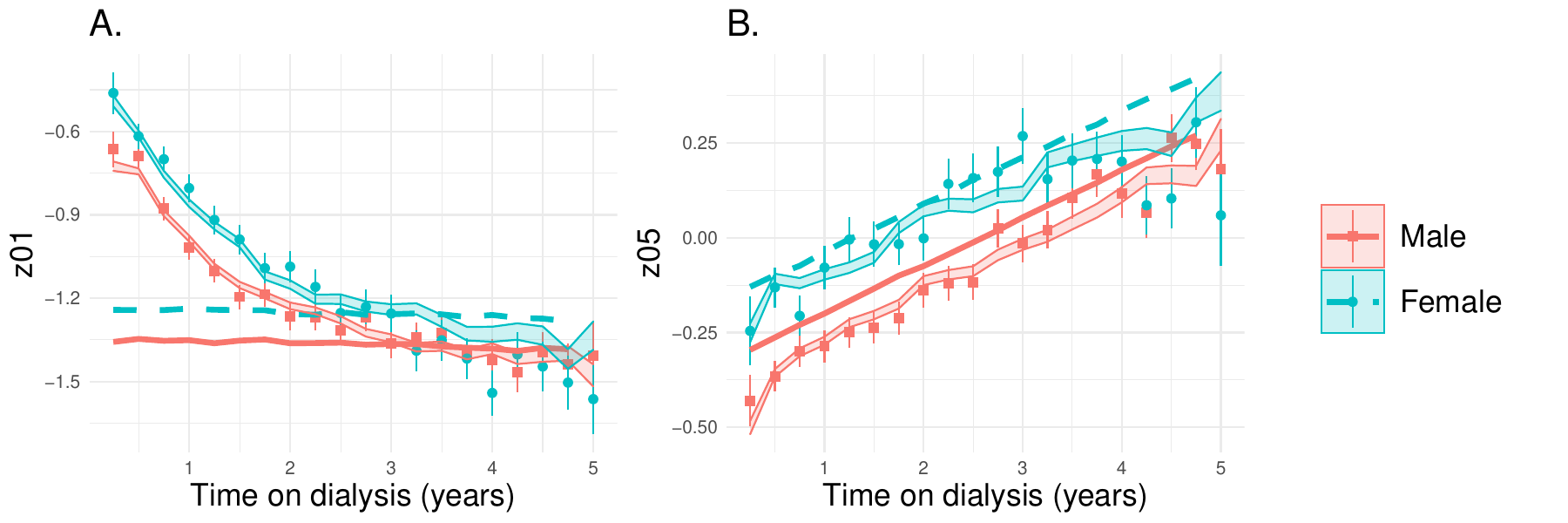}  
    \caption{Key natural variable data together with simulated data from fit parameters (mean $\pm$ standard error). Points: data grouped by 3 month bins. Bands: simulated data. Lines: dynamical equilibrium ($\mu(t)$).  The simulated data uses known initial values and parameter estimates from our dynamical model (Eq.~\ref{eq:sf}) and survival model (Eq.~\ref{eq:h}), with no additional tuning. We find good agreement between the simulated and real data for the natural variables (\textbf{A.} $z_1$ and \textbf{B.} $z_5$). This indicates that we correctly capture the population-level behaviour \textit{in silico}. The natural variables with faster recovery times quickly reached steady state, e.g. $z_5$, whereas the slowest natural variables, e.g. $z_1$, did not. Note that $\langle z \rangle = \mu(t)-\mu/|\lambda|$ in the steady-state.}
    \label{fig:samplefits}
\end{figure}

Eigen-decomposition permits us to greatly simplify the network by decomposing it into a linear combination of sub-networks (e.g.\ see \cite{twins}). Associated with this decomposition is a set of natural variables, $z$, which are the raw biomarkers transformed into the canonical basis wherein they recover independently. The population-level behaviour is illustrated by $z_1$ and $z_5$ in Figure~\ref{fig:samplefits}. The $z$ are sorted by their recovery speed, $|\lambda|$, from slowest ($z_1$) to fastest ($z_{14}$): small $|\lambda|$ indicates slow equilibration time and long auto-correlation time. Complex conjugate pair eigenvectors did occur for $z_2$/$z_3$ and $z_6$/$z_7$, which we represent as $z_2\equiv Re(z_2)$, $z_3\equiv Im(z_2)$, $z_6\equiv Re(z_6)$, and $z_7\equiv Im(z_6)$. (The imaginary components of the associated eigenvalues were small and are ignored.)

End stage kidney disease has poor prognosis \cite{Hashmi2023-aq}; in the present study we found that the half-life of individuals was 5~years. Survival is a good proxy for disease severity, as well as overall health. We found that survival was well fit by a time-dependent Weibull distribution with a proportional hazard factor,
\begin{align}
    h(t|\vec{x}(t)) &= \nu h_0 t^{\nu-1}e^{\vec{\beta}^T\vec{x}(t)} \label{eq:h}
\end{align}
where $h$ is the hazard, $t$ is the time-to-death, $\nu$ is the Weibull shape parameter, $h_0\equiv \text{scale}^{-\nu}$ is the baseline hazard, $\vec{\beta}$ is the vector of proportional hazard coefficients and $\vec{x}$ is a vector of predictors including baseline static covariates (age, sex, diabetes status) as well as the set of longitudinal biomarkers being modelled (raw biomarkers, natural variables or principal components). The $z$ variables fit a linear proportional hazard assumption whereas the raw biomarkers frequently did not. (See supplemental.)

\begin{figure}[h]
     \centering
        \includegraphics[width=\textwidth]{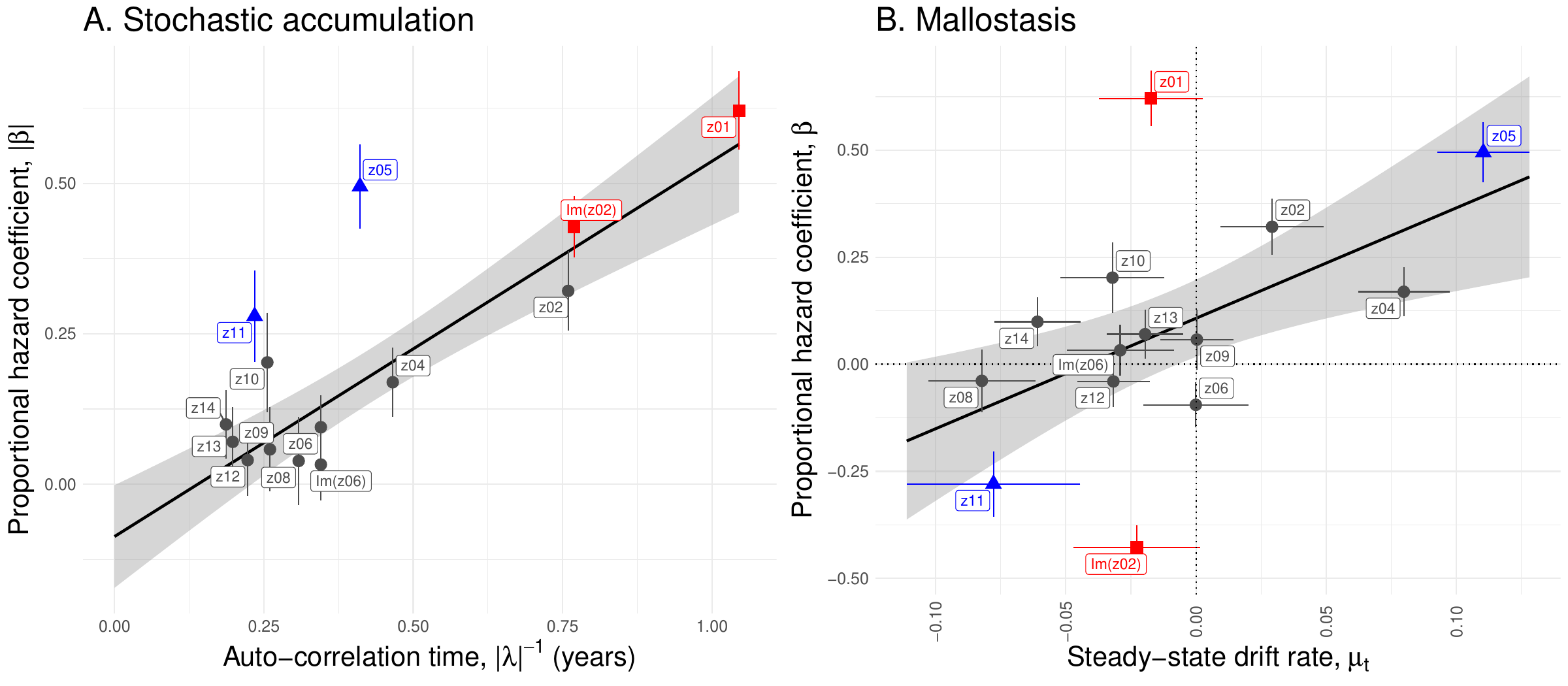} 
    \caption{Two modes of death: associations between dynamical parameters and mortality risk. Survival association strength is quantified by the univariate proportional hazard coefficient ($\beta$ in Eq.~\ref{eq:h}). \textbf{A.} For most variables, the survival hazard was correlated with the auto-correlation time, $|\beta|\propto|\lambda|^{-1}$. This suggests a mode of death characterized by persistent fluctuations at the individual level: which we name `stochastic accumulation'. Note the outliers: $z_5$ and $z_{11}$ (blue triangles), and the exemplars: $z_1$ and $Im(z_2)$ (red squares). \textbf{B.} In contrast, mallostasis occurs when a population experiences a predictable, steady-state decline in homeostatic set point towards worse health ($\mu_t\propto\beta$) \cite{mallostasis}. The outliers in \textbf{A} exhibit the strongest effect in \textbf{B} (blue triangles) and vice versa (red squares). This suggests that each $z$ is a combination of the two modes of death, with a tendency for slow variables (small $|\lambda|$) to prefer stochastic accumulation and faster variables (larger $|\lambda|$) to prefer mallostasis. Best fit black lines exclude respective outliers; each $z$ has been scaled to unit variance; sign of each $z$ is quasi-arbitrary via eigen-decomposition. $\beta$ was also strongly correlated with non-parameteric survival risk (C-index, Pearson $\rho = 0.94$, $p=7\cdot 10^{-7}$).}
    \label{fig:mallo}
\end{figure}

We looked for associations between health (survival) and dynamical behaviour for the model parameters of each of the $z_j$ natural variables, including: $\lambda_j$ and $\mu_{tj}$ versus $\beta_j$. We identify two strong associations: (i) \textit{stochastic accumulation} of individual poor health ($|\beta|\propto|\lambda|^{-1}$), and (ii) ``mallostasis'': a deterministic drift in homeostatic set point due to evolution of the disease ($\beta\propto\mu_t$) \cite{mallostasis}; Figure~\ref{fig:mallo} illustrates. Remarkably, the exemplars of stochastic accumulation were outliers of mallostasis and vice versa. Evidently, small $\lambda$ is incompatible with large $\mu_t$, and each $z$ has a preferred death mode depending on its position in $(\lambda,\mu_t)$--parameter space. 
Assuming our dynamical model is correct, each individual evolves stochastically over time. The expected hazard then differs from Eq.~\ref{eq:h} since it must be averaged over all possible (stochastic) paths. $z$ is normally-distributed at any given time, hence the expected hazard is
\begin{align}
    \langle h(t|z_0,z(t))\rangle_{z(t)}&= \nu h_0 t^{\nu-1}\exp{(\beta^2 \text{Var}(z)(z_0,t)/2 + \beta \langle z \rangle(z_0,t))}, \label{eq:ht}
\end{align}
for each particular natural variable, $z$, averaging over all possible paths that started at $z_0$ (details in supplemental). We see that death proceeds via either the variance or the mean of $z$. The former is due to individual differences which average to zero at the population level, whereas the latter are population-level effects that apply to everybody. Within our dynamical model, individual differences can only occur through the noise term, $\sigma$, but will persist for a time determined by the resilience parameter, $\lambda$, which dictates the auto-correlation time ($|\lambda|^{-1}$). To be lethal, fluctuations need to either be extremely strong (large $\sigma$) or to last a long time (small $|\lambda|$).  In stochastic accumulation, the slowest dimensions have the longest auto-correlation times, permitting them to accumulate fluctuations and thus individual differences (Eq.~\ref{eq:acf}). This suggests that stochastic accumulation can be an effective failure mode that maximizes the hazard via $\beta^2 \text{Var}(z)$. Conversely, population-level changes occur through $\langle z \rangle$ whose behaviour is dictated by the position of the equilibrium and the steady-state drift rate, $\mu_t$. This indicates that mallostasis is an effective failure mode that maximizes the hazard via $\beta \langle z \rangle$.

\begin{figure} 
     \centering
                \includegraphics[width=\textwidth]{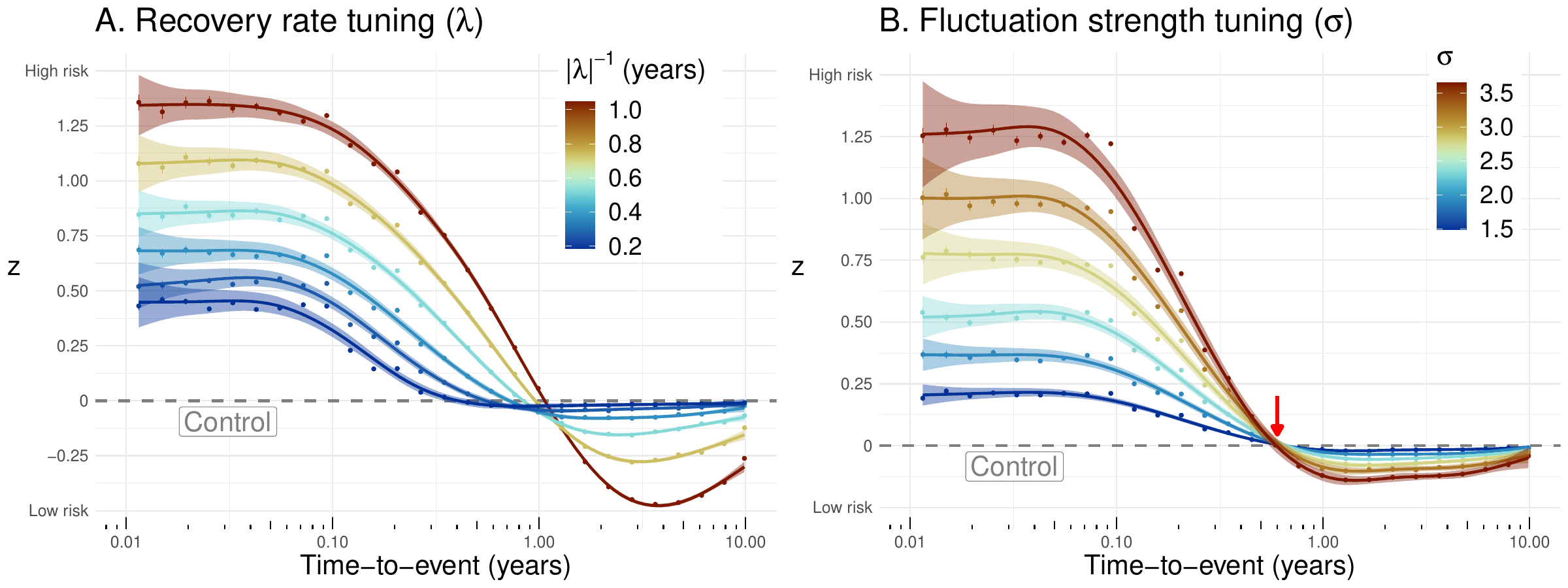}  
    \caption{Simplified simulation study of stochastic accumulation that retrospectively interprets survival trajectories after each individual has died. \textbf{A.} $|\lambda|^{-1}$ sets the timescale over which health deficits can accumulate via Eq.~\ref{eq:acf}. When $|\lambda|^{-1}$ is large, deficits can accumulate for a long period of time permitting individuals to develop very abnormal values (e.g.\ red line, $|\lambda|^{-1}=1~\text{year}$, similar to $z_1$). In contrast when $|\lambda|^{-1}$ is small, values become abnormal just before death (e.g.\ blue line, $|\lambda|^{-1}=0.2~\text{years}$, similar to $z_{14}$). This gradual accumulation of dysfunction leading to death characterizes stochastic accumulation, with $|\beta|\propto|\lambda|^{-1}$. \textbf{B.} The fluctuation strength $\sigma$ controls how abnormal an individual can be within the timescale set by $|\lambda|^{-1}$, without affecting the timescale (arrow). The fluctuations also increase the noise, making it difficult to discriminate individual health trajectories. The dashed grey line indicates the mean of a hypothetical control which doesn't feel the effects of $z$ (i.e.\ $\beta=0$). Coloured lines are natural splines \cite{Wickham2016-kw}.  Parameters: $\beta = 0.53$, $h_0 =0.044$, $\nu = 1.67$, $\mu_t=0$, $\sigma = 2.5$ (mean of the $z$), and $\lambda = -3.1$ (also mean). The ranges of $\sigma$ and $\lambda$ indicated by the legends  correspond to the observed ranges across the various $z$. All simulated values started in steady-state. Complete analysis in supplemental.}
    \label{fig:sim}
\end{figure}

By varying parameters in a simplified simulation, we confirm our interpretation of Eq.~\ref{eq:ht}. In Figure~\ref{fig:sim}A, we see that $|\lambda|^{-1}$ controls the horizon time of the terminal decline trajectory. We interpret that $|\lambda|^{-1}$ sets the timescale over which an individual accumulates signs of dysfunction, i.e.\ health deficits, prior to death --- with large $|\lambda|^{-1}$ permitting strong, persistent changes and small $|\lambda|^{-1}$ leading to a quick death (or quick recovery). Abnormal values indicate health deficits and will drive the observed biomarkers via the mapping $\boldsymbol{P}$. How abnormal the biomarkers are able to get within the time interval set by  $|\lambda|^{-1}$ depends on the fluctuation strength, Figure~\ref{fig:sim}B. Stronger fluctuations lead to values which are more abnormal at death, but they are non-specific and include noise and healthy variability. This is likely why there was no significant correlation between fluctuation strength and survival ($\beta$, $|\beta|$, and the C-index all had $p>0.3$ non-significant Spearman correlation with $\sigma$). Rather, stochastic accumulation and its corresponding mortality effect is characterized by \textit{persistent} fluctuations with long auto-correlation times, $|\lambda|^{-1}$ and persistent biomarker abnormality.



\begin{figure}[H] 
     \centering
     \includegraphics[width=\textwidth]{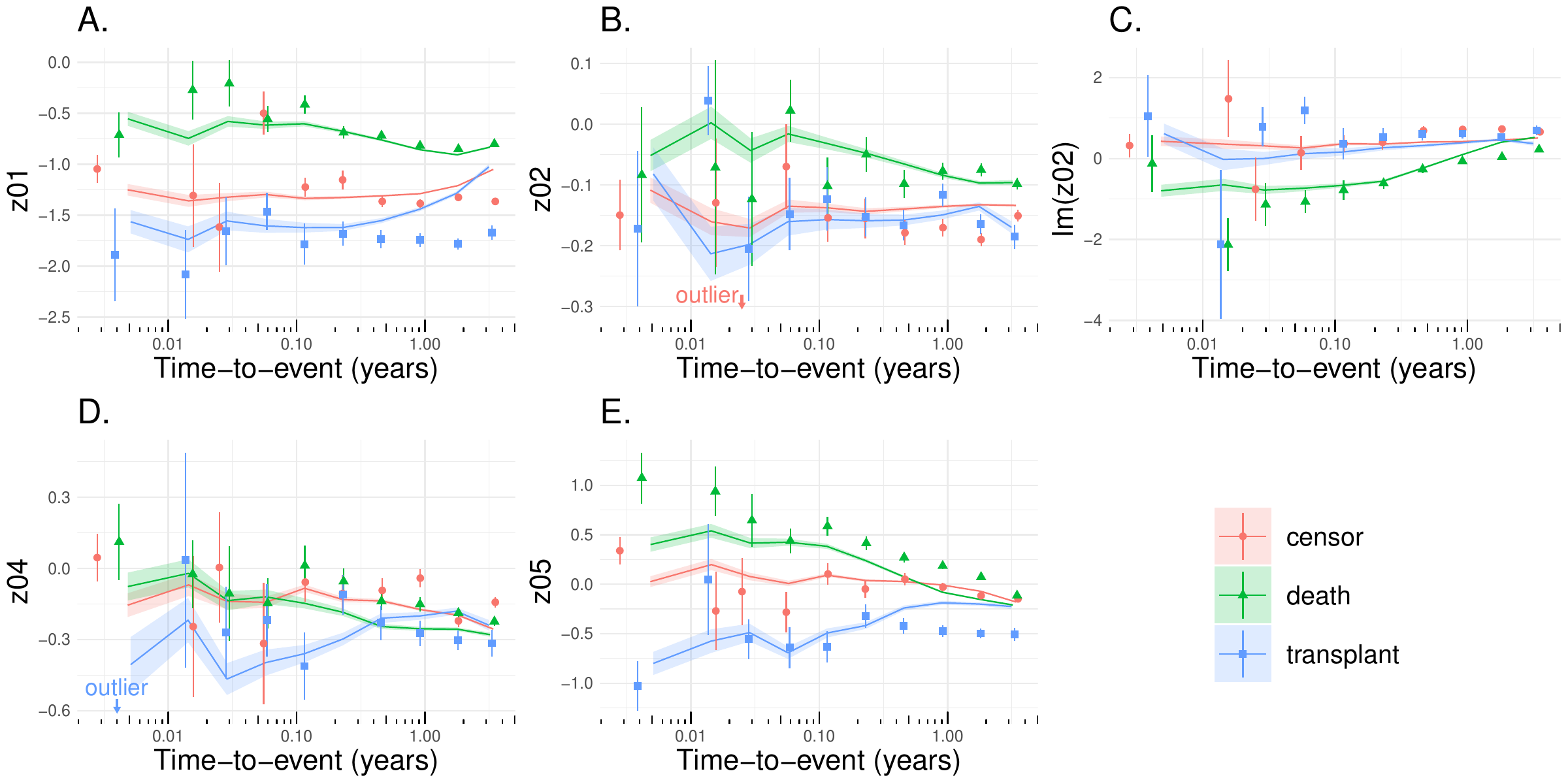}  
    \caption{Terminal decline exhibits log-linear drift in each natural variable with saturation shortly prior to death (mean $\pm$ standard error). \textbf{A.} - \textbf{E.} are for $z_1$ to $z_5$, respectively. Points are data binned by time-to-death with cuts (0,2,4,...,$2^8=256$)~weeks.  The simulation captures the correct qualitative behaviour (bands, from full model in supplemental), and fits reasonably well.  We performed multivariate simulation on 10000 individuals using parameter estimates together with initial conditions sampled from the population. Parameters were estimated by maximum likelihood; survival predictors were selected via the likelihood-ratio test. Two outliers with big error bars have been excluded where indicated (colour-coded; see Supplemental Figure~\ref{fig:si:ttd} for all $z$.)}
    \label{fig:ttd}
\end{figure}

In Figure~\ref{fig:ttd} we see that the characteristic terminal decline phenomenon exhibited by the full simulated model is also be observed in the real data. For the $z$ that are strong survival predictors, we see a distinct log-linear behaviour with saturation (flattening) shortly before death. On a linear scale this would lead to a divergence right before death. The simulated values (bands) agree well considering they they aren't directly fitted.



\subsection{Risk Dynamics} \label{sec:q}
\begin{figure}[H] 
     \centering
        \includegraphics[width=\textwidth]{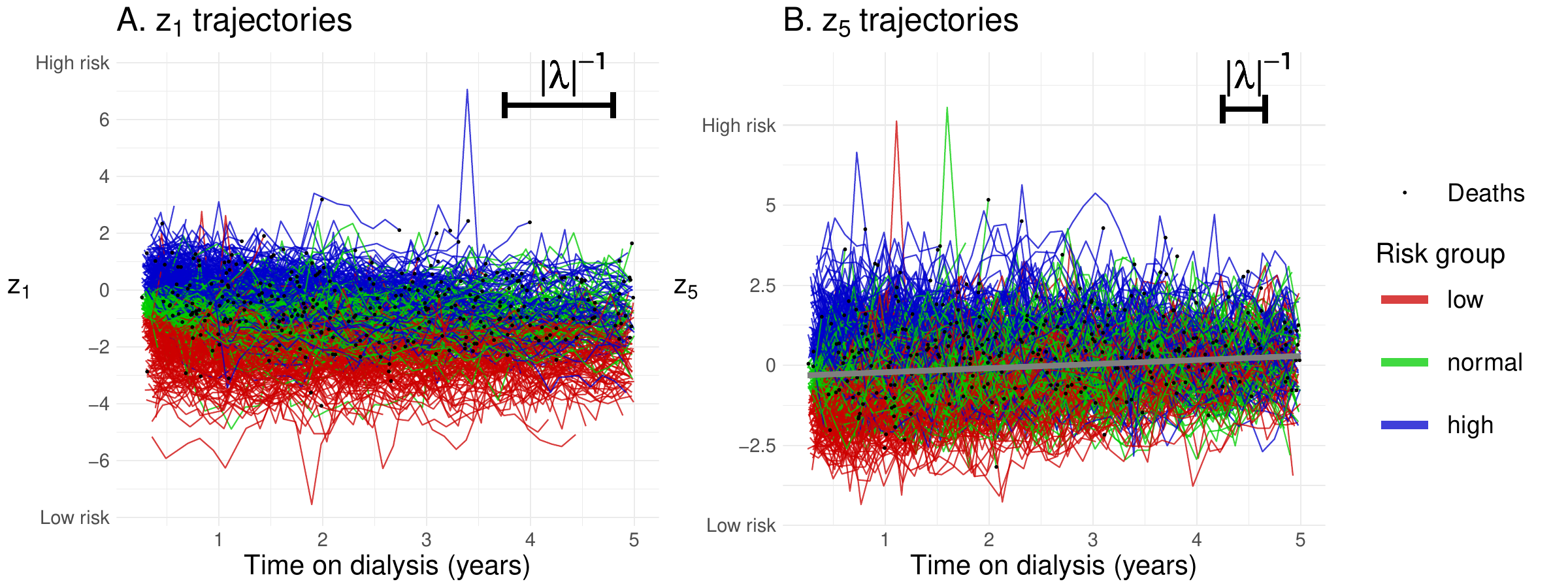} 
    \caption{Individual trajectories for $z_1$ and $z_5$. The respective auto-correlation time for each $z$ is indicated in the top right ($|\lambda|^{-1}$). \textbf{A.} $z_1$ has a small eigenvalue, giving it a long auto-correlation time, and a small drift rate making individuals change slowly and stochastically. Blue is thinning out from attrition (deaths, black points), and green is gradually drifting up. Some individuals visually worsen gradually prior to death, consistent with stochastic accumulation. \textbf{B.} $z_5$ has a larger eigenvalue and a large drift rate giving it a shorter auto-correlation time and causing a deterministic drift shared across the population (grey line), respectively. Colours are tertiles based on baseline value at 3~months.}
    \label{fig:traj}
\end{figure}

Individual trajectories were consistent with population-level trends. In Figure~\ref{fig:traj} we present individual trajectories for the exemplar variables: $z_1$ for stochastic accumulation and  $z_5$ for mallostasis. We label individuals by tertiles, representing low-normal-high risk groups. The tertiles remain stratified in $z_1$ but quickly mix together in $z_5$  --- a consequence of $|\lambda_5| \gg |\lambda_1|$. In $z_1$ we see risk groups gradually moving up and then thinning out due to deaths, consistent with stochastic accumulation of health deficits. For $z_5$, we see frequent transitions between risk strata. By year~4, the $z_5$ strata appear to be randomly distributed around the equilibrium, $\mu(t)$ (grey line), which is gradually pulling the entire population towards higher risk. This is consistent with $z_5$ being dominated by mallostasis. 

\begin{figure}
     \centering
        \includegraphics[width=\textwidth]{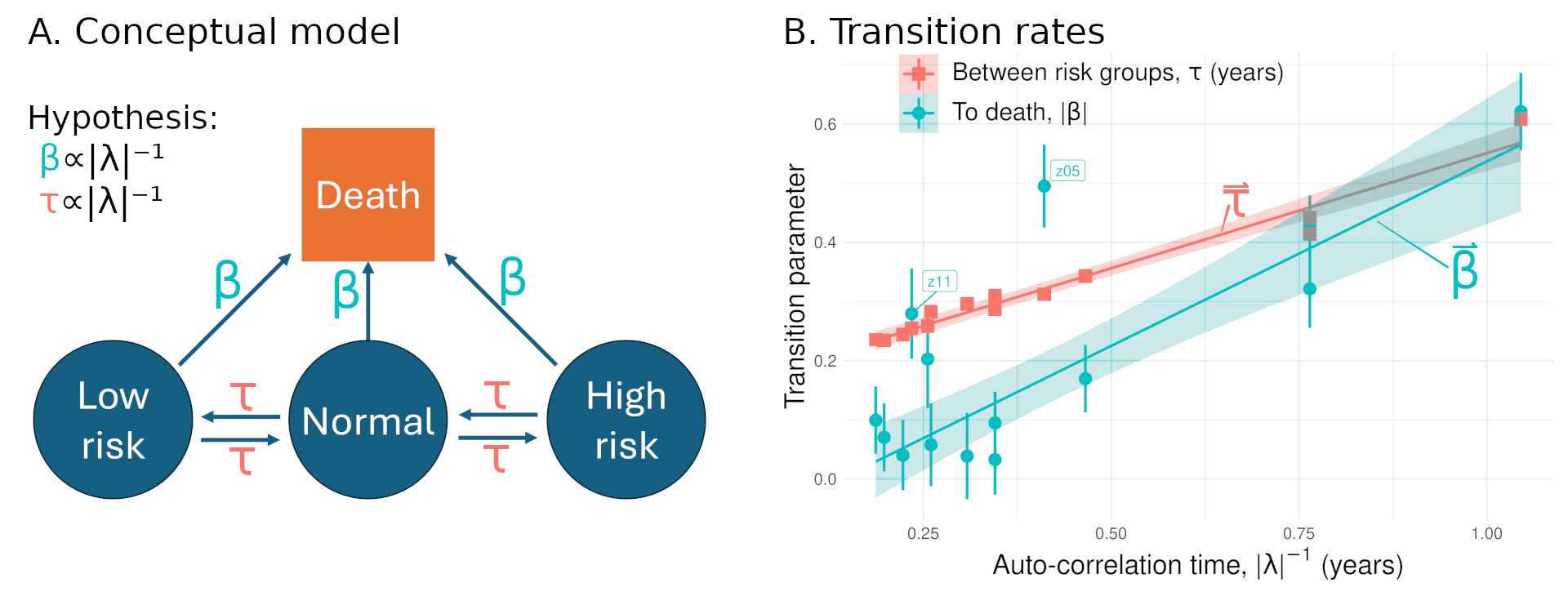}  
    \caption{Dynamical risk model.  \textbf{A.} Individuals transition with exponential probability (fixed-rate) between risk categories according to a single parameter, $\tau$. Individuals die with Weibull probability proportionate to proportional hazard coefficient $\beta$ as estimated earlier. We hypothesize that the auto-correlation time, $|\lambda|^{-1}$, is the central mediator of risk --- at least for the stochastically-accumulating $z$ --- since  persistent fluctuations are the most hazardous. \textbf{B.} Across the different $z$, both $\tau$ (red squares) and $\beta$ (blue points) are strongly, linearly associated with $|\lambda|^{-1}$, supporting our hypothesis. Each $\beta_j$ has been scaled by the standard deviation of $z_j$; $z_5$ and $z_{11}$ were excluded from fits (lines) since they have strong mallostatic effects.}
    \label{fig:rates}
\end{figure}

If stochastic accumulation is governed by accumulating fluctuations leading to death, as we hypothesized in Section~\ref{sec:system}, then we expect that $|\lambda|^{-1}$ should also set the timescale for risk transitions. Indeed, in Figure~\ref{fig:rates} a simple exponential model with constant-rate transitions between risk tertiles confirms these expectations. Excluding the mallostatic $z_5$ and $z_{11}$, transition time and $|\beta|$ were strongly linearly correlated (Pearson $\rho = 0.91$, $p=4\cdot 10^{-5}$). This correlation was mediated by the auto-correlation time, $|\lambda|^{-1}$, which strongly  correlates with both $|\beta|$ ($\rho = 0.93$) and transition time ($\rho = 0.98$). 

\subsection{The natural variables form dynamical modules} \label{sec:interp}

\begin{figure}[h] 
     \centering
        \includegraphics[width=\textwidth]{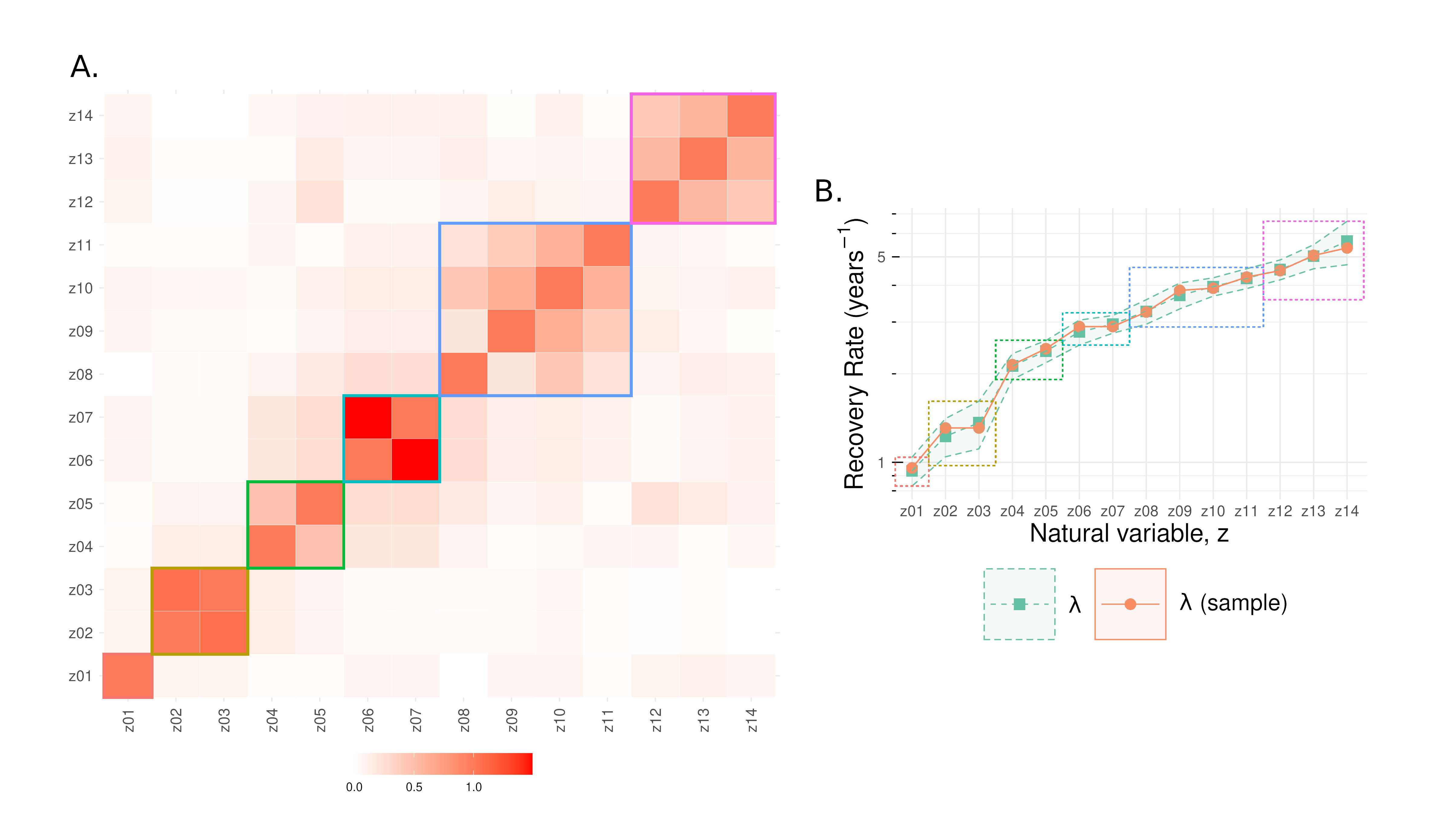}  
    \caption{Dynamical modules in the noise \textbf{A} and eigenvalues \textbf{B} (coloured outlines). \textbf{A.} Normalized noise covariance modulus, $|\Sigma_{ij}|/\sqrt{|\Sigma_{ii}||\Sigma_{jj}|}$, suggests that the $z$ live in multi-dimensional sub-spaces. The outlined modules are (approximately) independent of each other, but within each module the associated $z$ are strongly coupled by the noise.  \textbf{B.} The modules are associated with similar recovery rates ($-Re(\lambda)$). Orange points are the eigenvalues of the final network, whereas the green squares are the bootstrap estimates (with 95\% confidence intervals).}
    \label{fig:modules}
\end{figure}

We observed that the natural variables ($z$) formed distinct modules through mutual stochastic events, as indicated by the block-diagonal structure of the noise, Figure~\ref{fig:modules}A. These modules have strongly correlated noise as well as very similar recovery rates (Figure~\ref{fig:modules}B). Each block represents an independent dynamical module since Eq.~\ref{eq:z} ensures independent recovery of each $z$, and Figure~\ref{fig:modules}A demonstrates (approximately) independent stochastic events between modules. We hypothesize that each module is capturing an underlying latent multi-dimensional biological process characterized by a distinct timescale. 

\begin{figure}[h]
     \centering
        \includegraphics[width=\textwidth]{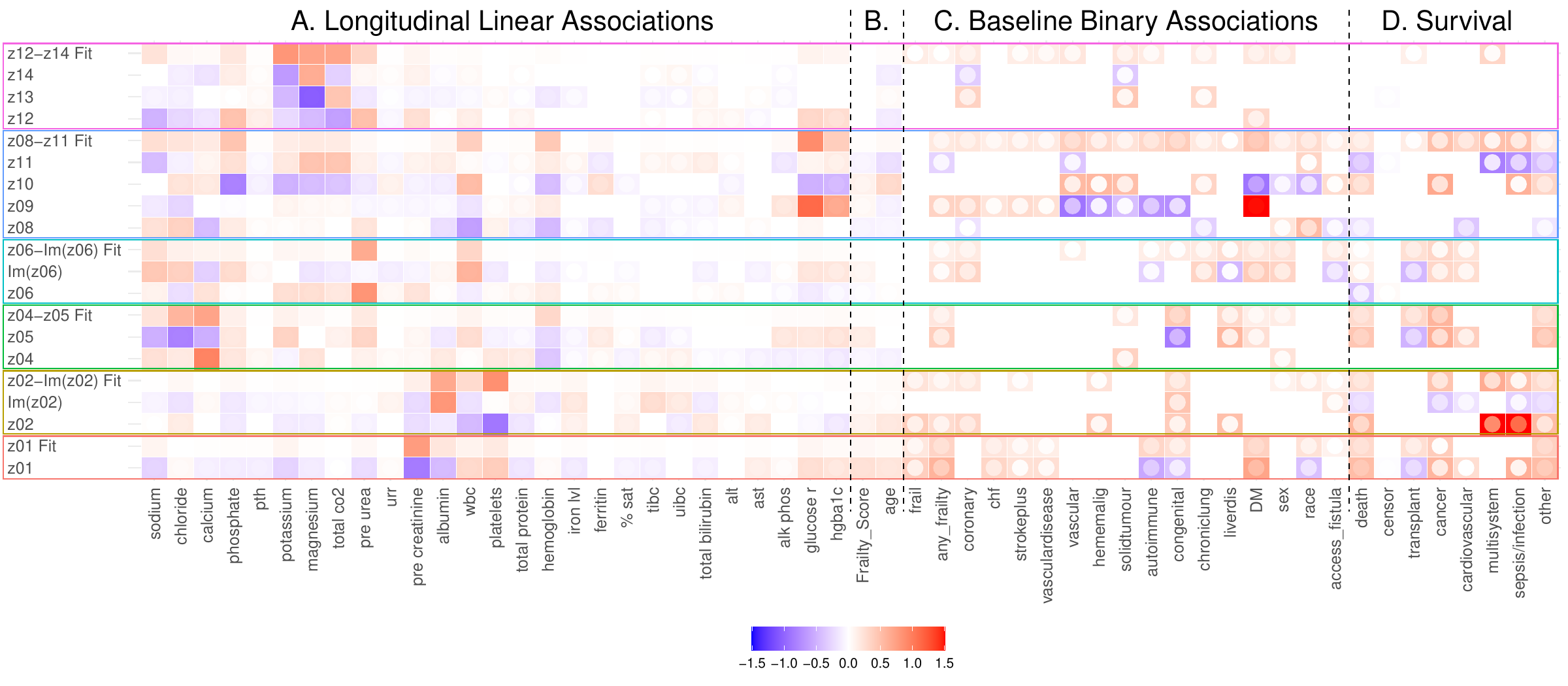}
    \caption{Associations for the natural variables, $z$, by module. \textbf{A.} linear regression against (continuous) longitudinally-measured biomarkers ($z$ within module are predictors). \textbf{B.} linear regression against ordinal baseline variables. \textbf{C.} logistic regression against binary baseline variables. \textbf{D.} Competing time-to-event regression \cite{De_Wreede2011-tj}. The variables are grouped by modules (coloured outlines that correspond to those in Figure~\ref{fig:modules}.). Each module has a row of regression coefficients for each $z$ within the module and an overall fit quality row (e.g.\ z04-z05 Fit). To read, pick a module and read first the coefficients from left to right (row 1), e.g.\ $z_1$ is strongly negatively associated with both albumin and creatinine, and moderately positively associated with indicators of inflammation (WBC and platelets), suggesting it is sensitive to wasting. Then read the next row in that module, until you reach the overall fit quality, this tells us how well the module can describe each variable e.g.\ $z_1$ is a strong predictor of survival. Inner point is 95\% confidence interval closest to 0; non significant points have been whited out (no multiple-comparison corrections). ``Fit'' score depends on variable type: $R^2$ for linear, $2\times\text{AUC}-1$ for binary, and $2\times\text{C}-1$ for survival (all range from 0: useless, to 1: perfect). C-index neglects competing risks. Colour scale is truncated at 1.5 for visualization (true values: $z_9$-DM: $1.9\pm0.2$, $z_2$-sepsis: $2.8\pm0.9$, and $z_2$-multisystem: $3\pm1$). All continuous variables (\textbf{A}) were scaled to zero mean, unit variance.}
    \label{fig:zcor}
\end{figure}

Using regression, we sought to interpret the biological meaning of each module based on its ability to predict a panel of known biomarkers, baseline conditions and time-to-event outcomes. The resulting regression coefficients are presented in Figure~\ref{fig:zcor}, together with overall fit scores. The coefficients describe what information is present and the fit scores summarize how essential that information is to the outcome. We see that $z_1$ is associated with biomarkers indicating poor nutrition and chronic inflammation, the hallmarks of a wasting syndrome \cite{Fouque2008-ax}. $z_1$ is also associated with baseline age, frailty and diabetes, and is strongly associated with death (positive) and transplant (negative). From Sections~\ref{sec:system} and \ref{sec:q} we can infer that $|\lambda_1|^{-1} \approx 1~\text{year}$ sets the timescale over which these signs are evolving. $z_1$ is a prospective biomarker for this slowly-evolving syndrome. 

We next see that $z_2$ forms a module with its complex conjugate, $z_3$. This module, represented in Figure~\ref{fig:zcor} as $(z_2,Im(z_2))\equiv (Re(z_2),Im(z_2))$, has a very strong signal related to albumin and platelets, and is associated with overall risk of death --- with very strong associations with sepsis and multisystem collapse as the cause of death (in practice, multisystem collapse often includes sepsis). Next is the $z_4$/$z_5$ module which seems to be primarily related to calcium ($z_4$) and electrolytes: sodium, potassium and chloride ($z_5$) (note that $z_4$ and $z_5$ were positively correlated). The module is a strong predictor of overall death and specifically death via cancer; it is also one of the strongest predictors of death via cardiovascular event.

These three modules ostensibly capture the dominant survival effects, as seen in the cumulative survival prediction plot (Supplemental Figure~\ref{fig:si:cumc}; this can also be seen in Figure~\ref{fig:zcor} by the strength of the `death' column of each `Fit' row). 

\subsection{Validation using an external cohort}
In Section~\ref{sec:interp} we presented evidence that the natural variables are biomarkers of specific underlying biological processes. These processes are latent in the sense that they are not directly measured, but their effects can be inferred through the natural variables. To validate this hypothesis, we demonstrate that a restricted set of raw biomarkers from a different dataset can still be used to estimate natural variables with consistent properties from what we have observed. Full details in supplemental.

We used a large, cross-sectional validation cohort of Canadian dialysis patients (non--Nova Scotian). The dataset includes only 6 biomarkers of the 14 which we used in the primary analysis. We used linear regression to estimate the transformation from this set of 6 biomarkers into the $z$ natural variables using the main dataset, applied this transformation to the validation dataset, and analyzed these ``emulated'' natural variables, $\hat{z}$, to see if they recapitulate our key results. The key natural variable is $z_1$ since: (i) it has the longest auto-correlation time and therefore should be most similar between longitudinal and cross-sectional measurements (most discriminating), (ii) it depends heavily on creatinine and albumin, both of which are in the validation dataset (best emulated), and (iii) it was the strongest survival predictor in the main dataset (most relevant).


We found that $z_1$ was an equally-strong survival predictor in both the main and validation datasets (Harrell's C-indices \cite{Harrell1982-qt}: $0.65\pm0.02$ and $0.65\pm0.01$, respectively). $z_1$ depends heavily on both creatinine and albumin, but is a better survival predictor than either (Supplemental Figure~\ref{fig:si:cumcvalidation}). Survival hazard saturated for both creatinine and albumin, whereas the saturation effects cancelled in $z_1$ which instead increases linearly with survival hazard (Supplemental Figures~\ref{fig:phz} and \ref{fig:phy}). This suggests that $z_1$ is the primary underlying mortality process that drives the observed survival associations of (serum) albumin and creatinine both here and possibly by prior researchers as well \cite{Lowrie1990-ru, Kalantar-Zadeh2001-mq, Karaboyas2020-nx}. This evidence supports the interpretation that $z_1$ is repeatable and biologically meaningful. 

\begin{figure}[!ht] 
     \centering
        \includegraphics[width=\textwidth]{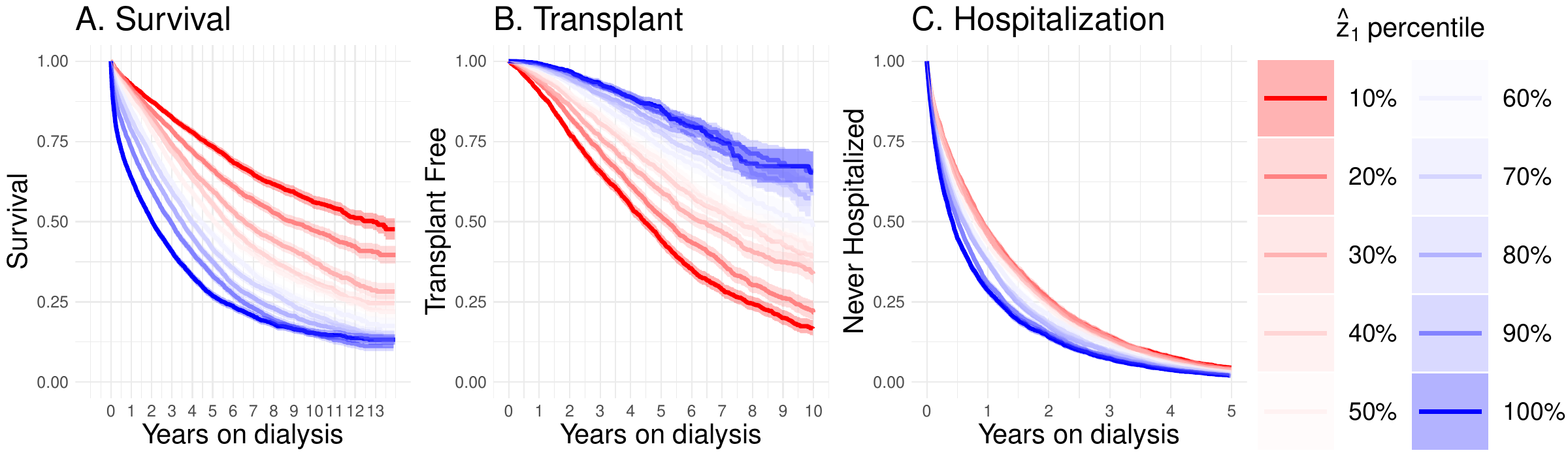}  
    \caption{Validation of $\hat{z}_1$ using the validation population. Time-to-event survival curves for \textbf{A.} death, \textbf{B.} transplant and \textbf{C.} first hospitalization (since starting dialysis). Stratified by decile. We see that $\hat{z}_1$ is a strong predictor of all three outcomes. $\hat{z}_1$ approximates $z_1$ ($R^2=0.86$), using available biomarkers.}
    \label{fig:svalid}
\end{figure}

The predictive power of $\hat{z}_1$ is illustrated using Kaplan-Meier curves for common outcomes in Figure~\ref{fig:svalid}. The emulated $z_1$ is highly predictive of survival, hospitalization and transplant (only the healthiest patients are eligible for transplant).

\FloatBarrier

\section{Discussion}
CKD (chronic kidney disease) is a systemic disease \cite{Zoccali2017-vp,Pandey2023-ou} that engenders causes, signs and symptoms spanning multiple biological sub-systems \cite{Hashmi2023-aq}. We have embraced this complexity \cite{Cohen2022-gt} and applied a system-level dynamical model \cite{mallostasis} to capture the time evolution of CKD patients undergoing haemodialysis. We treat a multivariate collection of blood-based biomarkers as a biological system maintaining homeostasis against stochastic stressor events. By analyzing the natural variables associated with this system (network) we greatly simplify its dynamical behaviour, and we can infer fundamental connections between the time evolution of the natural variables, $z$, and health. Within the $z$ variables, we observed that two dynamical modes, each consistent with a loss of homeostasis, drive an increase in risk of death. The first mode, `stochastic accumulation', was characterized by weak stability (poor resilience) and long auto-correlation times during which individual differences were able to build and persist. The second mode is `mallostasis': the gradual erosion of homeostasis that leads to a steady-state decline in health over time \cite{mallostasis}. The associated signals are moderately-fast $z$ that are well regulated and follow a declining homeostatic set point ($\mu(t)$). We further observe that the natural variables ($z$) are collections of biomarker signs that overlap with the medical concept of a syndrome \cite{Calvo2003-vr}. Indeed, several can be associated with existing CKD syndromes, suggesting the existence of prospectively new syndromes as well. 

The first death mode, stochastic accumulation, arises from the buildup of abnormal individual differences originating from the stochastic stress term, $\boldsymbol{\Sigma}$. These are systemic vulnerabilities that can be identified through our analysis pipeline by small $|\lambda|$, indicating poor resilience. These vulnerabilities are strongly associated with health, where $|\beta|\propto|\lambda|^{-1}$. Remarkably, the population showed specific preferred directions of failure for each natural variable --- as evidenced by the linear Cox model --- leading to specific lab signs which gradually worsen over time. For example, $z_1$ is associated with increased risk of death and decreased levels of creatinine and albumin. This contrasts with previous models that have assumed loss of resilience to be symmetrical \cite{Yashin2007-py,Liu2021-zu} --- whereby both abnormally high and low natural variable values would be equally dangerous. We show that this is not the case for ESKD (see also \cite{mallostasis}). This indicates that human biology doesn't dysregulate randomly, but tends to dysregulate in specific directions leading to specific, worsening medical signs. 

The second death mode is mallostasis \cite{mallostasis}, wherein the homeostatic set point drifts towards worsening health, causing the entire population to deterministically increase in risk over the course of the study period. The exemplars for this effect were $z_5$ and $z_{11}$. Existing theories suggest mechanisms for mallostasis, including wear-and-tear of the adaptive stress response (``allostatic load'') \cite{Juster2010-kw}, and the saturation of repair processes \cite{Alon2023-lu,Karin2019-gf}. These theories suggest that mallostatic $z$ are pushed towards tolerance thresholds beyond which catastrophic failure can occur. The $z_5$ trajectories in Figure~\ref{fig:traj}B are consistent with this: most deaths occur near the mean, ostensibly due to other causes, but occasionally with extreme values, ostensibly due to death via $z_5$. This effect was seen primarily in the intermediate $\lambda$, consistent with the competing effects of homeostatic stability versus the effects of occasional failures. The drift likely originates from either a deterministic progression of the disease or treatment, or by dysfunction entering from unobserved network nodes whose effects are captured by $\mu_t$ \cite{mallostasis}. The drift is deterministic, for example $z_5$ should worsen at a rate of $\mu_t = 0.13\pm0.02~\text{years}^{-1}$: representing a $\exp{(\beta \mu_t)}\sim5$\% increase in risk of death for each year of dialysis. Since $z_5$ is strongly associated with electrolyte concentrations, this increasing hazard translates into a loss of robustness against stressors due to worsening electrolyte balance. 

Stochastic accumulation and mallostasis are key phenomena that describe the expected proportional hazard. We anticipate they will generalize to all normal, stochastic variables with a proportional hazard survival effect, which has a vast range of applications. These phenomena should help to understand declining health at all timescales, including aging through Gompertz' law \cite{Kirkwood2015-zm}, and terminal decline in the final years \cite{Stolz2021-wd} and days \cite{Bruera2014-oz} of life.

Whereas the recovery dynamics for each $z$ are independent according to Eq.~\ref{eq:z}, the perturbations or stressors that push $z$ away from normal form modules that were captured by the noise (covariance) parameters, $\boldsymbol{\Sigma}$. We manually identified six modules with highly correlated noise, $\boldsymbol{\Sigma}$, which had similar recovery rates, $Re(\lambda)$ (Figure~\ref{fig:modules}). Multi-dimensional processes should have the same or similar recovery rates to maintain coherence, hence the modules are suggestive of latent underlying biological processes that the natural variables are capturing. We hypothesize that each $z$-module associated with survival is a biomarker for a distinct syndrome. (Modules not associated with survival could capture benign biological functioning, but survival effects could also be supplanted by $h_0(t)$ since large $\lambda$ would indicate tight regulation at 6~week timescales.) Any worsening $z$-module will translate into coherently increasingly abnormal serum biomarker values, with the effects typically spread out across several biomarkers (via $\boldsymbol{P}$). This makes each $z$-module a ``syndrome'' from the perspective of the observed biomarkers, which by definition is a collection of physical findings without a clear cause \cite{Calvo2003-vr}. Our contribution is in being able to automatically identify, quantify and prognosticate these new and existing syndromes. 

The first natural variable, $z_1$, appears to be a biomarker of an underlying protein wasting syndrome characterized by low albumin and creatinine, together with elevated white blood cell count and platelets. These biomarkers are consistent with PEW (protein energy wasting), a common clinical syndrome characterized by chronic inflammation and malnutrition, leading to cachexia \cite{Fouque2008-ax}. $z_1$ appears to evolve via stochastic accumulation of dysfunction, caused by external stressors, with a long horizon time on the order of years between when an individual first starts looking abnormal and when they die. The changes appear to be random and cannot be readily predicted based on how long an individual has been on dialysis. This makes $z_1$ an important syndrome for patient management, but its long auto-correlation indicates that it has a long and likely refractory decline trajectory. If so, $z_1$-syndrome should be easy to identify but hard to treat. This could make $z_1$ an important biomarker of clinical decline for shared care decisions, such as preparation for either conservative or palliative care.

The next natural variable module, $(z_2,Im(z_2))$, appear to be biomarkers of sepsis syndrome. Sepsis is a 2-dimensional process since it relies on both the pathogen threat and the internal response of the immune system \cite{Singer2016-ax}. It appears that the real part is capturing response to external threat via inflammation, particularly low platelets, and the imaginary part is capturing the body's subsequent response --- shifting liver protein production away from albumin. A systemic review showed that low platelets have a common, strong association with mortality, particularly via sepsis \cite{Hui2011-pw} due to platelets being consumed through multiple mechanisms \cite{Vardon-Bounes2019-os}. $z_2$ shares many behaviours consistent with sepsis syndrome and is able to predict death via sepsis with a high hazard rate. 

Among the new prospective syndromes, $z_5$ had the strongest association with mortality --- second only to $z_1$ overall. This would indicate that the $z_4$/$z_5$ module is an important prospective syndrome for clinical care. Electrolyte changes, elevated urea, elevated WBC and low hemoglobin were the key biomarker signs of $z_4$/$z_5$. Electrolyte disturbances are commonly associated with cardiovascular events \cite{Samanta2019-dm}, cancer \cite{Rosner2014-jo} and liver disease \cite{Jimenez2017-ko}, all of which were strongly associated with $z_5$. The $z_4$/$z_5$ module may therefore represent a disruption of electrolyte homeostasis, possibly originating from a variety of causes. The mallostatic nature of $z_5$ would support a connection to dialysis treatment, such as a loss of potassium to the dialysate \cite{Samanta2019-dm}.

The natural variables also provide a natural choice for monitoring and communicating ESKD dialysis patient health. For this purpose we have shared both the exact transformation using the 14 biomarkers in the main dataset as well as the approximations for smaller sets of biomarkers. Other problems related to systemic disease are worth interrogating using our approach, in particular early detection of CKD during its silent period \cite{Lopez-Giacoman2015-ae}. Prior network analysis of CKD has focused on `omics data, which could benefit from our dynamical understanding \cite{Pandey2023-ou} in the future. Here we have focused on clinical applicability by monitoring ESKD health using clinically-available data.

We note two potential sources of error. Our population is undergoing active treatment and hence are subject to the idiosyncrasies of their healthcare provider and environment, although our results appear to generalize across Canadians. Since our data are observational, interventional data would be useful to refine the interaction network and survival effects, e.g.\ albumin is modifiable but doesn't necessarily improve survival outcomes \cite{Lodebo2018-od}. Our perspective is that interventions should target the biology underlying each $z$-module, which may differ from the biomarkers used to estimate and score the $z$ natural variables.

To understand systemic diseases such as CKD (chronic kidney disease), which have diffuse signals spread across multiple biological sub-systems, we have generated networks that capture the holistic, systemic character of human health and eventual failure. We have quantified homeostasis. We assembled biomarkers, $z$, that can identify and monitor specific syndromes --- prospective targets of interventions that improve health outcomes. The $z$-modules are independent both in recovery dynamics and (approximately) in the noise, suggesting that interventions that affect any one module will not affect others. Our approach automatically identified protein energy wasting (PEW) and sepsis as two major syndromes which lead to increased risk of death, in addition to a collection of additional prospective syndromes of varying risk. While not all of these targets will be amenable to treatment, the ease with which our model identifies them from routine blood tests reassures us that this is only the beginning of a more complete systems level understanding that will allow us to identify, characterize and develop treatments for the syndromes that emerge from CKD.
\section{Methods}

\subsection{Data}
The main dataset was gathered from patients receiving dialysis in Nova Scotia over the period of Jan 13, 2009 to Dec 26, 2020 ($N=713$). Validation data are incident dialysis patients in Canada, including related treatments and acute hospitalizations, over the period Jan 1, 2005 to Dec 31, 2018 ($N=61036$). All patients signed informed consent.

\subsection{Statistics and Models}
All analysis and statistics used \texttt{R} version 4.1.1  \cite{R_Core_Team2021-uq}. We used the event history analysis (\texttt{eha}) package for parametric survival \cite{eha}. For competing risks we used the \texttt{mstate} package \cite{De_Wreede2011-tj}. Our primary model was fit using linear regression via the SF model, available on GitHub at \url{https://github.com/GlenPr/stochastic_finite-difference_model}. Included in the GitHub page are CSV files containing the exact parameters for the emulator as well as the simulation parameters for the primary simulation. The SF model is described in detail elsewhere \cite{mallostasis,twins}.

\section*{Acknowledgements}
A.R.\ thanks the Natural Sciences and Engineering Research Council of Canada (NSERC) for operating Grant RGPIN-2019-05888.

\section*{Author contributions statement}
GP, KKT, KR and ADR conceived the project. KKT extracted the main dataset. GW extracted and assisted with the validation dataset. GP performed the analysis and drafted the manuscript.  All authors reviewed the manuscript. 

\printbibliography

@BOOK{Alon2023-lu,
  title     = "Systems Medicine: Physiological Circuits and the Dynamics of
               Disease",
  author    = "Alon, Uri",
  publisher = "CRC Press",
  year      =  2023,
  isbn      = "9781000960679"
}

@ARTICLE{Jimenez2017-ko,
  title     = "Electrolyte and {Acid--Base} Disturbances in {End-Stage} Liver
               Disease: A Physiopathological Approach",
  author    = "Jim{\'e}nez, Jos{\'e} V{\'\i}ctor and Carrillo-P{\'e}rez, Diego
               Luis and Rosado-Canto, Rodrigo and Garc{\'\i}a-Ju{\'a}rez,
               Ignacio and Torre, Aldo and Kershenobich, David and
               Carrillo-Maravilla, Eduardo",
  journal   = "Dig. Dis. Sci.",
  publisher = "Springer",
  volume    =  62,
  number    =  8,
  pages     = "1855--1871",
  month     =  aug,
  year      =  2017,
  url       = "https://doi.org/10.1007/s10620-017-4597-8",
  issn      = "0163-2116, 1573-2568",
  doi       = "10.1007/s10620-017-4597-8"
}

@ARTICLE{Karin2019-gf,
  title     = "Senescent cell turnover slows with age providing an explanation
               for the Gompertz law",
  author    = "Karin, Omer and Agrawal, Amit and Porat, Ziv and Krizhanovsky,
               Valery and Alon, Uri",
  journal   = "Nat. Commun.",
  publisher = "nature.com",
  volume    =  10,
  number    =  1,
  pages     = "5495",
  month     =  dec,
  year      =  2019,
  url       = "http://dx.doi.org/10.1038/s41467-019-13192-4",
  issn      = "2041-1723",
  pmid      = "31792199",
  doi       = "10.1038/s41467-019-13192-4",
  pmc       = "PMC6889273"
}

@ARTICLE{Singer2016-ax,
  title     = "The Third International Consensus Definitions for Sepsis and
               Septic Shock (Sepsis-3)",
  author    = "Singer, Mervyn and Deutschman, Clifford S and Seymour,
               Christopher Warren and Shankar-Hari, Manu and Annane, Djillali
               and Bauer, Michael and Bellomo, Rinaldo and Bernard, Gordon R
               and Chiche, Jean-Daniel and Coopersmith, Craig M and Hotchkiss,
               Richard S and Levy, Mitchell M and Marshall, John C and Martin,
               Greg S and Opal, Steven M and Rubenfeld, Gordon D and van der
               Poll, Tom and Vincent, Jean-Louis and Angus, Derek C",
  journal   = "JAMA",
  publisher = "jamanetwork.com",
  volume    =  315,
  number    =  8,
  pages     = "801--810",
  month     =  feb,
  year      =  2016,
  url       = "http://dx.doi.org/10.1001/jama.2016.0287",
  issn      = "0098-7484, 1538-3598",
  pmid      = "26903338",
  doi       = "10.1001/jama.2016.0287",
  pmc       = "PMC4968574"
}

@ARTICLE{Wilhelm-Leen2009-xe,
  title     = "Frailty and chronic kidney disease: the Third National Health
               and Nutrition Evaluation Survey",
  author    = "Wilhelm-Leen, Emilee R and Hall, Yoshio N and K Tamura, Manjula
               and Chertow, Glenn M",
  journal   = "Am. J. Med.",
  publisher = "Elsevier",
  volume    =  122,
  number    =  7,
  pages     = "664--71.e2",
  month     =  jul,
  year      =  2009,
  url       = "http://dx.doi.org/10.1016/j.amjmed.2009.01.026",
  issn      = "0002-9343, 1555-7162",
  pmid      = "19559169",
  doi       = "10.1016/j.amjmed.2009.01.026",
  pmc       = "PMC4117255"
}

@ARTICLE{Samanta2019-dm,
  title     = "Arrhythmias and Sudden Cardiac Death in End Stage Renal Disease:
               Epidemiology, Risk Factors, and Management",
  author    = "Samanta, Rahul and Chan, Christopher and Chauhan, Vijay S",
  journal   = "Can. J. Cardiol.",
  publisher = "Elsevier",
  volume    =  35,
  number    =  9,
  pages     = "1228--1240",
  month     =  sep,
  year      =  2019,
  url       = "http://dx.doi.org/10.1016/j.cjca.2019.05.005",
  issn      = "0828-282X, 1916-7075",
  pmid      = "31472819",
  doi       = "10.1016/j.cjca.2019.05.005"
}

@ARTICLE{Vardon-Bounes2019-os,
  title     = "Platelets Are Critical Key Players in Sepsis",
  author    = "Vardon-Bounes, Fanny and Ruiz, St{\'e}phanie and Gratacap,
               Marie-Pierre and Garcia, C{\'e}dric and Payrastre, Bernard and
               Minville, Vincent",
  journal   = "Int. J. Mol. Sci.",
  publisher = "mdpi.com",
  volume    =  20,
  number    =  14,
  month     =  jul,
  year      =  2019,
  url       = "http://dx.doi.org/10.3390/ijms20143494",
  issn      = "1422-0067",
  pmid      = "31315248",
  doi       = "10.3390/ijms20143494",
  pmc       = "PMC6679237"
}

@ARTICLE{Hui2011-pw,
  title     = "The frequency and clinical significance of thrombocytopenia
               complicating critical illness: a systematic review",
  author    = "Hui, Phil and Cook, Deborah J and Lim, Wendy and Fraser, Graeme
               A and Arnold, Donald M",
  journal   = "Chest",
  publisher = "Elsevier",
  volume    =  139,
  number    =  2,
  pages     = "271--278",
  month     =  feb,
  year      =  2011,
  url       = "http://dx.doi.org/10.1378/chest.10-2243",
  issn      = "0012-3692, 1931-3543",
  pmid      = "21071526",
  doi       = "10.1378/chest.10-2243"
}

@Manual{splines2,
    title = {{splines2}: {R}egression Spline Functions and Classes},
    author = {Wenjie Wang and Jun Yan},
    year = {2021},
    url = {https://CRAN.R-project.org/package=splines2},
    note = {{R} package version 0.4.3},
  }

@ARTICLE{Yashin2007-py,
  title     = "Stochastic model for analysis of longitudinal data on aging and
               mortality",
  author    = "Yashin, Anatoli I and Arbeev, Konstantin G and Akushevich, Igor
               and Kulminski, Aliaksandr and Akushevich, Lucy and Ukraintseva,
               Svetlana V",
  journal   = "Math. Biosci.",
  publisher = "Elsevier",
  volume    =  208,
  number    =  2,
  pages     = "538--551",
  year      =  2007,
  url       = "http://dx.doi.org/10.1016/j.mbs.2006.11.006",
  issn      = "0025-5564",
  pmid      = "17300818",
  doi       = "10.1016/j.mbs.2006.11.006",
  pmc       = "PMC2084381"
}

@MISC{Wickham2016-kw,
  title     = "ggplot2: Elegant Graphics for Data Analysis",
  author    = "Wickham, Hadley",
  publisher = "Springer-Verlag New York",
  year      =  2016,
  url       = "https://ggplot2.tidyverse.org",
  isbn      = "9783319242774"
}

@MISC{Hashmi2023-aq,
  title     = "{End-Stage} Renal Disease",
  booktitle = "{StatPearls}",
  author    = "Hashmi, Muhammad F and Benjamin, Onecia and Lappin, Sarah L",
  publisher = "StatPearls Publishing",
  month     =  feb,
  year      =  2023,
  url       = "https://www.ncbi.nlm.nih.gov/pubmed/29763036",
  address   = "Treasure Island (FL)",
  pmid      = "29763036"
}

@article{twins,
    author = {Pridham, Glen and Rutenberg, Andrew D},
    title = "{Dynamical network stability analysis of multiple biological ages provides a framework for understanding the aging process}",
    journal = {The Journals of Gerontology: Series A},
    pages = {glae021},
    year = {2024},
    month = {01},
    issn = {1758-535X},
    doi = {10.1093/gerona/glae021},
    url = {https://doi.org/10.1093/gerona/glae021},
    eprint = {https://academic.oup.com/biomedgerontology/advance-article-pdf/doi/10.1093/gerona/glae021/55452242/glae021.pdf},
}

@ARTICLE{Lousa2020-js,
  title     = "New Potential Biomarkers for Chronic Kidney Disease
               {Management---A} Review of the Literature",
  author    = "Lousa, Irina and Reis, Fl{\'a}vio and Beir{\~a}o, Idalina and
               Alves, Rui and Belo, Lu{\'\i}s and Santos-Silva, Alice",
  journal   = "Int. J. Mol. Sci.",
  publisher = "Multidisciplinary Digital Publishing Institute",
  volume    =  22,
  number    =  1,
  pages     = "43",
  month     =  dec,
  year      =  2020,
  url       = "https://www.mdpi.com/1422-0067/22/1/43",
  issn      = "1422-0067, 1422-0067",
  doi       = "10.3390/ijms22010043"
}

@ARTICLE{Karaboyas2020-nx,
  title    = "Estimating the Fraction of {First-Year} Hemodialysis Deaths
              Attributable to Potentially Modifiable Risk Factors: Results from
              the {DOPPS}",
  author   = "Karaboyas, Angelo and Morgenstern, Hal and Li, Yun and Bieber,
              Brian A and Hakim, Raymond and Hasegawa, Takeshi and Jadoul,
              Michel and Schaeffner, Elke and Vanholder, Raymond and Pisoni,
              Ronald L and Port, Friedrich K and Robinson, Bruce M",
  journal  = "Clin. Epidemiol.",
  volume   =  12,
  pages    = "51--60",
  month    =  jan,
  year     =  2020,
  url      = "http://dx.doi.org/10.2147/CLEP.S233197",
  issn     = "1179-1349",
  pmid     = "32021471",
  doi      = "10.2147/CLEP.S233197",
  pmc      = "PMC6974411"
}

@ARTICLE{Lopez-Giacoman2015-ae,
  title    = "Biomarkers in chronic kidney disease, from kidney function to
              kidney damage",
  author   = "Lopez-Giacoman, Salvador and Madero, Magdalena",
  journal  = "World J Nephrol",
  volume   =  4,
  number   =  1,
  pages    = "57--73",
  month    =  feb,
  year     =  2015,
  url      = "http://dx.doi.org/10.5527/wjn.v4.i1.57",
  issn     = "2220-6124",
  pmid     = "25664247",
  doi      = "10.5527/wjn.v4.i1.57",
  pmc      = "PMC4317628"
}

@ARTICLE{Zoccali2017-vp,
  title     = "The systemic nature of {CKD}",
  author    = "Zoccali, Carmine and Vanholder, Raymond and Massy, Ziad A and
               Ortiz, Alberto and Sarafidis, Pantelis and Dekker, Friedo W and
               Fliser, Danilo and Fouque, Denis and Heine, Gunnar H and Jager,
               Kitty J and Kanbay, Mehmet and Mallamaci, Francesca and Parati,
               Gianfranco and Rossignol, Patrick and Wiecek, Andrzej and
               London, Gerard and {European Renal and Cardiovascular Medicine
               (EURECA-m) Working Group of the European Renal Association --
               European Dialysis Transplantation Association (ERA-EDTA)}",
  journal   = "Nat. Rev. Nephrol.",
  publisher = "nature.com",
  volume    =  13,
  number    =  6,
  pages     = "344--358",
  month     =  jun,
  year      =  2017,
  url       = "http://dx.doi.org/10.1038/nrneph.2017.52",
  issn      = "1759-5061, 1759-507X",
  pmid      = "28435157",
  doi       = "10.1038/nrneph.2017.52"
}

@ARTICLE{Fouque2008-ax,
  title     = "A proposed nomenclature and diagnostic criteria for
               protein--energy wasting in acute and chronic kidney disease",
  author    = "Fouque, D and Kalantar-Zadeh, K and Kopple, J and Cano, N and
               Chauveau, P and Cuppari, L and Franch, H and Guarnieri, G and
               Ikizler, T A and Kaysen, G and Lindholm, B and Massy, Z and
               Mitch, W and Pineda, E and Stenvinkel, P and Trevinho-Becerra, A
               and Wanner, C",
  journal   = "Kidney Int.",
  publisher = "Elsevier",
  volume    =  73,
  number    =  4,
  pages     = "391--398",
  month     =  feb,
  year      =  2008,
  url       = "https://www.sciencedirect.com/science/article/pii/S0085253815529992",
  issn      = "0085-2538",
  doi       = "10.1038/sj.ki.5002585"
}

@ARTICLE{De_Wreede2011-tj,
  title     = "mstate: An {R} Package for the Analysis of Competing Risks and
               {Multi-State} Models",
  author    = "de Wreede, Liesbeth C and Fiocco, Marta and Putter, Hein",
  journal   = "J. Stat. Softw.",
  publisher = "jstatsoft.org",
  volume    =  38,
  pages     = "1--30",
  month     =  jan,
  year      =  2011,
  url       = "https://www.jstatsoft.org/article/view/v038i07",
  issn      = "1548-7660, 1548-7660",
  doi       = "10.18637/jss.v038.i07"
}

@ARTICLE{Fulop2010-ho,
  title    = "Aging, frailty and age-related diseases",
  author   = "Fulop, T and Larbi, A and Witkowski, J M and McElhaney, J and
              Loeb, M and Mitnitski, A and Pawelec, G",
  journal  = "Biogerontology",
  volume   =  11,
  number   =  5,
  pages    = "547--563",
  month    =  oct,
  year     =  2010,
  url      = "http://dx.doi.org/10.1007/s10522-010-9287-2",
  issn     = "1389-5729, 1573-6768",
  pmid     = "20559726",
  doi      = "10.1007/s10522-010-9287-2"
}

@ARTICLE{Rockwood2005-gl,
  title    = "A global clinical measure of fitness and frailty in elderly
              people",
  author   = "Rockwood, Kenneth and Song, Xiaowei and MacKnight, Chris and
              Bergman, Howard and Hogan, David B and McDowell, Ian and
              Mitnitski, Arnold",
  journal  = "CMAJ",
  volume   =  173,
  number   =  5,
  pages    = "489--495",
  month    =  aug,
  year     =  2005,
  url      = "http://dx.doi.org/10.1503/cmaj.050051",
  issn     = "0820-3946, 1488-2329",
  pmid     = "16129869",
  doi      = "10.1503/cmaj.050051",
  pmc      = "PMC1188185"
}

@ARTICLE{Rockwood2020-oa,
  title    = "Using the Clinical Frailty Scale in Allocating Scarce Health Care
              Resources",
  author   = "Rockwood, Kenneth and Theou, Olga",
  journal  = "Can. Geriatr. J.",
  volume   =  23,
  number   =  3,
  pages    = "210--215",
  month    =  sep,
  year     =  2020,
  url      = "http://dx.doi.org/10.5770/cgj.23.463",
  issn     = "1925-8348",
  pmid     = "32904824",
  doi      = "10.5770/cgj.23.463",
  pmc      = "PMC7458601"
}

@BOOK{Hastie2017-wl,
  title     = "The elements of statistical learning: data mining, inference,
               and prediction",
  author    = "Hastie, T and Tibshirani, R and Friedman, J",
  publisher = "Springer",
  volume    = "2nd",
  year      =  2017
}

@Manual{eha,
    title = {eha: Event History Analysis},
    author = {Göran Broström},
    year = {2023},
    note = {R package version 2.11.1},
    url = {https://cran.r-project.org/package=eha},
  }

@MISC{R_Core_Team2021-uq,
  title       = "R: A Language and Environment for Statistical Computing",
  author      = "{R Core Team}",
  institution = "R Foundation for Statistical Computing",
  year        =  2021,
  url         = "https://www.R-project.org/",
  address     = "Vienna, Austria"
}

@ARTICLE{Juster2010-kw,
  title     = "Allostatic load biomarkers of chronic stress and impact on
               health and cognition",
  author    = "Juster, Robert-Paul and McEwen, Bruce S and Lupien, Sonia J",
  journal   = "Neurosci. Biobehav. Rev.",
  publisher = "Elsevier",
  volume    =  35,
  number    =  1,
  pages     = "2--16",
  month     =  sep,
  year      =  2010,
  url       = "http://dx.doi.org/10.1016/j.neubiorev.2009.10.002",
  issn      = "0149-7634, 1873-7528",
  pmid      = "19822172",
  doi       = "10.1016/j.neubiorev.2009.10.002"
}

@ARTICLE{mallostasis,
  title     = "Network dynamical stability analysis reveals key ``mallostatic''
               natural variables that erode homeostasis and drive age-related
               decline of health",
  author    = "Pridham, Glen and Rutenberg, Andrew D",
  journal   = "Sci. Rep.",
  publisher = "Nature Publishing Group",
  volume    =  13,
  number    =  1,
  pages     = "1--12",
  month     =  dec,
  year      =  2023,
  url       = "https://www.nature.com/articles/s41598-023-49129-7",
  issn      = "2045-2322, 2045-2322",
  doi       = "10.1038/s41598-023-49129-7"
}

@ARTICLE{Calvo2003-vr,
  title    = "Diagnoses, syndromes, and diseases: a knowledge representation
              problem",
  author   = "Calvo, Franz and Karras, Bryant T and Phillips, Richard and
              Kimball, Ann Marie and Wolf, Fred",
  journal  = "AMIA Annu. Symp. Proc.",
  volume   =  2003,
  pages    = "802",
  year     =  2003,
  url      = "https://www.ncbi.nlm.nih.gov/pubmed/14728307",
  issn     = "1942-597X, 1559-4076",
  pmid     = "14728307",
  pmc      = "PMC1480257"
}

@ARTICLE{Lowrie1990-ru,
  title     = "Death risk in hemodialysis patients: the predictive value of
               commonly measured variables and an evaluation of death rate
               differences between facilities",
  author    = "Lowrie, E G and Lew, N L",
  journal   = "Am. J. Kidney Dis.",
  publisher = "Elsevier",
  volume    =  15,
  number    =  5,
  pages     = "458--482",
  month     =  may,
  year      =  1990,
  url       = "http://dx.doi.org/10.1016/s0272-6386(12)70364-5",
  issn      = "0272-6386",
  pmid      = "2333868",
  doi       = "10.1016/s0272-6386(12)70364-5"
}

@BOOK{Ledder2013-em,
  title     = "Mathematics for the Life Sciences",
  author    = "Ledder, Glenn",
  publisher = "Springer New York",
  year      =  2013,
  url       = "https://link.springer.com/book/10.1007/978-1-4614-7276-6",
  doi       = "10.1007/978-1-4614-7276-6"
}

@ARTICLE{Pandey2023-ou,
  title    = "Network medicine: an approach to complex kidney disease
              phenotypes",
  author   = "Pandey, Arvind K and Loscalzo, Joseph",
  journal  = "Nat. Rev. Nephrol.",
  volume   =  19,
  number   =  7,
  pages    = "463--475",
  month    =  jul,
  year     =  2023,
  url      = "http://dx.doi.org/10.1038/s41581-023-00705-0",
  issn     = "1759-5061, 1759-507X",
  pmid     = "37041415",
  doi      = "10.1038/s41581-023-00705-0",
  pmc      = "4904833"
}

@ARTICLE{Cohen2022-gt,
  title     = "A complex systems approach to aging biology",
  author    = "Cohen, Alan A and Ferrucci, Luigi and F{\"u}l{\"o}p, Tam{\`a}s
               and Gravel, Dominique and Hao, Nan and Kriete, Andres and
               Levine, Morgan E and Lipsitz, Lewis A and Olde Rikkert, Marcel G
               M and Rutenberg, Andrew and Stroustrup, Nicholas and Varadhan,
               Ravi",
  journal   = "Nature Aging",
  publisher = "Nature Publishing Group",
  volume    =  2,
  number    =  7,
  pages     = "580--591",
  month     =  jul,
  year      =  2022,
  url       = "https://www.nature.com/articles/s43587-022-00252-6",
  issn      = "2662-8465, 2662-8465",
  doi       = "10.1038/s43587-022-00252-6"
}

@ARTICLE{Avchaciov2022-ws,
  title     = "Unsupervised learning of aging principles from longitudinal data",
  author    = "Avchaciov, Konstantin and Antoch, Marina P and Andrianova,
               Ekaterina L and Tarkhov, Andrei E and Menshikov, Leonid I and
               Burmistrova, Olga and Gudkov, Andrei V and Fedichev, Peter O",
  journal   = "Nat. Commun.",
  publisher = "nature.com",
  volume    =  13,
  number    =  1,
  pages     = "6529",
  month     =  nov,
  year      =  2022,
  url       = "http://dx.doi.org/10.1038/s41467-022-34051-9",
  issn      = "2041-1723",
  pmid      = "36319638",
  doi       = "10.1038/s41467-022-34051-9",
  pmc       = "PMC9626636"
}

@ARTICLE{Kalantar-Zadeh2015-ld,
  title    = "Dietary restrictions in dialysis patients: is there anything left
              to eat?",
  author   = "Kalantar-Zadeh, Kamyar and Tortorici, Amanda R and Chen, Joline L
              T and Kamgar, Mohammad and Lau, Wei-Ling and Moradi, Hamid and
              Rhee, Connie M and Streja, Elani and Kovesdy, Csaba P",
  journal  = "Semin. Dial.",
  volume   =  28,
  number   =  2,
  pages    = "159--168",
  month    =  feb,
  year     =  2015,
  url      = "http://dx.doi.org/10.1111/sdi.12348",
  issn     = "0894-0959, 1525-139X",
  pmid     = "25649719",
  doi      = "10.1111/sdi.12348",
  pmc      = "PMC4385746"
}

@ARTICLE{Liu2021-zu,
  title     = "Prediction of Mortality in Hemodialysis Patients Using Moving
               Multivariate Distance",
  author    = "Liu, Mingxin and Legault, V{\'e}ronique and F{\"u}l{\"o}p,
               Tam{\`a}s and C{\^o}t{\'e}, Anne-Marie and Gravel, Dominique and
               Blanchet, F Guillaume and Leung, Diana L and Lee, Sylvia Juhong
               and Nakazato, Yuichi and Cohen, Alan A",
  journal   = "Front. Physiol.",
  publisher = "ncbi.nlm.nih.gov",
  volume    =  12,
  pages     = "612494",
  month     =  mar,
  year      =  2021,
  url       = "http://dx.doi.org/10.3389/fphys.2021.612494",
  issn      = "1664-042X",
  pmid      = "33776784",
  doi       = "10.3389/fphys.2021.612494",
  pmc       = "PMC7993059"
}

@ARTICLE{Kalantar-Zadeh2001-mq,
  title     = "A malnutrition-inflammation score is correlated with morbidity
               and mortality in maintenance hemodialysis patients",
  author    = "Kalantar-Zadeh, K and Kopple, J D and Block, G and Humphreys, M
               H",
  journal   = "Am. J. Kidney Dis.",
  publisher = "Elsevier",
  volume    =  38,
  number    =  6,
  pages     = "1251--1263",
  month     =  dec,
  year      =  2001,
  url       = "http://dx.doi.org/10.1053/ajkd.2001.29222",
  issn      = "0272-6386, 1523-6838",
  pmid      = "11728958",
  doi       = "10.1053/ajkd.2001.29222"
}

@ARTICLE{Zoccali2011-ax,
  title    = "The complexity of the cardio-renal link: taxonomy, syndromes, and
              diseases",
  author   = "Zoccali, Carmine and Goldsmith, David and Agarwal, Rajiv and
              Blankestijn, Peter J and Fliser, Danilo and Wiecek, Andrzej and
              Suleymanlar, Gultekin and Ortiz, Alberto and Massy, Ziad and
              Covic, Adrian and Martinez-Castelao, Alberto and Jager, Kitty J
              and Dekker, Friedo W and Lindholm, Bengt and London, Gerard and
              {for EUropean REnal and CArdiovascular Medicine working group of
              the European Renal Association--European Dialysis and Transplant
              Association (ERA--EDTA)}",
  journal  = "Kidney Int. Suppl.",
  volume   =  1,
  number   =  1,
  pages    = "2--5",
  month    =  jun,
  year     =  2011,
  url      = "http://dx.doi.org/10.1038/kisup.2011.4",
  issn     = "0098-6577",
  pmid     = "25028622",
  doi      = "10.1038/kisup.2011.4",
  pmc      = "PMC4089616"
}

@ARTICLE{Rosner2014-jo,
  title     = "Electrolyte disorders associated with cancer",
  author    = "Rosner, Mitchell H and Dalkin, Alan C",
  journal   = "Adv. Chronic Kidney Dis.",
  publisher = "Elsevier",
  volume    =  21,
  number    =  1,
  pages     = "7--17",
  month     =  jan,
  year      =  2014,
  url       = "http://dx.doi.org/10.1053/j.ackd.2013.05.005",
  issn      = "1548-5595, 1548-5609",
  pmid      = "24359982",
  doi       = "10.1053/j.ackd.2013.05.005"
}

@ARTICLE{Lodebo2018-od,
  title     = "Is it Important to Prevent and Treat {Protein-Energy} Wasting in
               Chronic Kidney Disease and Chronic Dialysis Patients?",
  author    = "Lodebo, Bereket Tessema and Shah, Anuja and Kopple, Joel D",
  journal   = "J. Ren. Nutr.",
  publisher = "Elsevier",
  volume    =  28,
  number    =  6,
  pages     = "369--379",
  month     =  nov,
  year      =  2018,
  url       = "http://dx.doi.org/10.1053/j.jrn.2018.04.002",
  issn      = "1051-2276, 1532-8503",
  pmid      = "30057212",
  doi       = "10.1053/j.jrn.2018.04.002"
}

@ARTICLE{Harrell1982-qt,
  title    = "Evaluating the yield of medical tests",
  author   = "Harrell, Jr, F E and Califf, R M and Pryor, D B and Lee, K L and
              Rosati, R A",
  journal  = "JAMA",
  volume   =  247,
  number   =  18,
  pages    = "2543--2546",
  month    =  may,
  year     =  1982,
  url      = "https://www.ncbi.nlm.nih.gov/pubmed/7069920",
  issn     = "0098-7484",
  pmid     = "7069920",
  doi      = "10.1001/jama.1982.03320430047030"
}

@ARTICLE{Nie2022-db,
  title     = "Distinct biological ages of organs and systems identified from a
               multi-omics study",
  author    = "Nie, Chao and Li, Yan and Li, Rui and Yan, Yizhen and Zhang,
               Detao and Li, Tao and Li, Zhiming and Sun, Yuzhe and Zhen, Hefu
               and Ding, Jiahong and Wan, Ziyun and Gong, Jianping and Shi,
               Yanfang and Huang, Zhibo and Wu, Yiran and Cai, Kaiye and Zong,
               Yang and Wang, Zhen and Wang, Rong and Jian, Min and Jin, Xin
               and Wang, Jian and Yang, Huanming and Han, Jing-Dong J and
               Zhang, Xiuqing and Franceschi, Claudio and Kennedy, Brian K and
               Xu, Xun",
  journal   = "Cell Rep.",
  publisher = "Elsevier",
  volume    =  38,
  number    =  10,
  month     =  mar,
  year      =  2022,
  url       = "http://www.cell.com/article/S2211124722001863/abstract",
  issn      = "2211-1247",
  pmid      = "35263580",
  doi       = "10.1016/j.celrep.2022.110459"
}

@MISC{cookbook2012,
  title        = "The matrix cookbook",
  author       = "{Petersen, Kaare, Brandt and Pedersen, Michael, Syskind}",
  year         =  2012,
  url          = "https://www.math.uwaterloo.ca/~hwolkowi/matrixcookbook.pdf",
  howpublished = "Online"
}

@BOOK{Moore2016-rh,
  title     = "Applied Survival Analysis Using {R}",
  author    = "Moore, Dirk F",
  publisher = "Springer",
  year      =  2016,
  url       = "https://link.springer.com/10.1007/978-3-319-31245-3",
  isbn      = "9783319312453",
  doi       = "10.1007/978-3-319-31245-3"
}

@ARTICLE{Oye-Somefun2021-ud,
  title     = "Associations between elevated kidney and liver biomarker ratios,
               metabolic syndrome and all-cause and coronary heart disease
               ({CHD}) mortality: analysis of the {U.S}. National Health and
               Nutrition Examination Survey ({NHANES})",
  author    = "Oye-Somefun, Akinkunle and Kuk, Jennifer L and Ardern, Chris I",
  journal   = "BMC Cardiovasc. Disord.",
  publisher = "Springer Science and Business Media LLC",
  volume    =  21,
  number    =  1,
  pages     = "352",
  month     =  jul,
  year      =  2021,
  url       = "https://bmccardiovascdisord.biomedcentral.com/articles/10.1186/s12872-021-02160-w",
  copyright = "https://creativecommons.org/licenses/by/4.0",
  issn      = "1471-2261",
  pmid      = "34311708",
  doi       = "10.1186/s12872-021-02160-w",
  pmc       = "PMC8311936"
}

@ARTICLE{Bruera2014-oz,
  title    = "Variations in vital signs in the last days of life in patients
              with advanced cancer",
  author   = "Bruera, Sebastian and Chisholm, Gary and Dos Santos, Renata and
              Crovador, Camila and Bruera, Eduardo and Hui, David",
  journal  = "J. Pain Symptom Manage.",
  volume   =  48,
  number   =  4,
  pages    = "510--517",
  month    =  oct,
  year     =  2014,
  url      = "http://dx.doi.org/10.1016/j.jpainsymman.2013.10.019",
  issn     = "0885-3924, 1873-6513",
  pmid     = "24731412",
  doi      = "10.1016/j.jpainsymman.2013.10.019",
  pmc      = "PMC4197073"
}

@ARTICLE{Stolz2021-wd,
  title     = "Acceleration of health deficit accumulation in late-life:
               evidence of terminal decline in frailty index three years before
               death in the {US} Health and Retirement Study",
  author    = "Stolz, Erwin and Mayerl, Hannes and Hoogendijk, Emiel O and
               Armstrong, Joshua J and Roller-Wirnsberger, Regina and Freidl,
               Wolfgang",
  journal   = "Ann. Epidemiol.",
  publisher = "Elsevier",
  volume    =  58,
  pages     = "156--161",
  month     =  jun,
  year      =  2021,
  url       = "https://www.sciencedirect.com/science/article/pii/S1047279721000521",
  issn      = "1047-2797",
  doi       = "10.1016/j.annepidem.2021.03.008"
}

@ARTICLE{Kirkwood2015-zm,
  title     = "Deciphering death: a commentary on Gompertz (1825) 'On the
               nature of the function expressive of the law of human mortality,
               and on a new mode of determining the value of life
               contingencies'",
  author    = "Kirkwood, Thomas B L",
  journal   = "Philos. Trans. R. Soc. Lond. B Biol. Sci.",
  publisher = "The Royal Society",
  volume    =  370,
  number    =  1666,
  pages     = "20140379",
  month     =  apr,
  year      =  2015,
  url       = "https://royalsocietypublishing.org/doi/10.1098/rstb.2014.0379",
  issn      = "0962-8436, 1471-2970",
  pmid      = "25750242",
  doi       = "10.1098/rstb.2014.0379",
  pmc       = "PMC4360127"
}

\FloatBarrier
\newpage

\renewcommand{\theequation}{S\arabic{equation}}
\renewcommand{\thefigure}{S\arabic{figure}}
\renewcommand{\thetable}{S\arabic{table}}
\renewcommand{\thesection}{S\arabic{section}}
\setcounter{equation}{0}  
\setcounter{figure}{0}  
\setcounter{table}{0}  
\setcounter{section}{0}  

\section{Supplemental information}
By Glen Pridham$^{1,*}$, Karthik K.\ Tennankore, Kenneth Rockwood, George Worthen and Andrew~D.\ Rutenberg$^{1,\dagger}$.\\
$^1$Department of Physics and Atmospheric Science, Dalhousie University, Halifax, B3H 4R2, Nova Scotia, Canada.\\
$^2$Dalhousie University and Nova Scotia Health, 5820 University Avenue, Halifax, B3H 1V8, Nova Scotia, Canada. \\
$^3$Division of Geriatric Medicine, Dalhousie University, Halifax, B3H 2E1, Nova Scotia, Canada. \\
$^*$glen.pridham@dal.ca \\
$^\dagger$adr@dal.ca \\



The supplemental is structured as follows. We begin with additional results of interest which were excluded from the main text for want of space. In particular, we first compare the ability of the natural variables versus principal component analysis (PCA) to compress and prioritize survival information in Section~\ref{sec:si:cumC}. Next we perform a full simulation analysis of the terminal decline plots, together with a simplified analytical description in Section~\ref{sec:si:td}. Then we provide additional associations, stratified by sex and diabetes status in Section~\ref{sec:si:ass}. 

Next we move on to describing the data and data handling methods, starting with a description of the datasets (Section~\ref{sec:si:demo}) and ending with missing data handling (Section~\ref{sec:si:miss}). In the final portion of the supplemental we provide model diagnostics, starting with the parameterized model in Section~\ref{sec:si:fit}, and ending with a sensitivity analysis in Section~\ref{sec:si:sa}. In the sensitivity analysis section we verify our results are insensitive to: sub-population of interest (sex, diabetes and frailty status), specific choice of biomarkers to use in the network, and use of study window. The study window sets the time interval over which data are used for the analysis; in the main text it is 3~months to 5~years. The purpose is to prevent possible biases, but ultimately we show that it has no effect on the parameter estimates (Section~\ref{sec:window}).

\subsection{Cumulative survival prediction} \label{sec:si:cumC}
In the main text, we propose that the natural variables are useful, systems-level health biomarkers. To summarize, the natural variables have a knack for compressing survival information and are able to automatically identify which variables are most important for health based on their eigenvalue, $\lambda$ --- slower $\lambda$ are more important for predicting adverse outcomes. In contrast, fast $\lambda$ are incompatible with either stochastic accumulation or mallostasis. Fast-recovering stochastic accumulation is unpredictable since there isn't enough time for substantial accumulation and hence the effect is indistinguishable from noise. Fast-recovering mallostasis would lead to very-tightly regulated, deterministic decline which would prevent the abnormal values we associate with mortality --- it would look instead like programmed mortality (which would be absorbed into the time-dependence of the hazard). Fast (large $|\lambda|$) are therefore unimportant, at least for prediction of adverse outcomes. 

In this section we test this understanding by comparing the natural variables to other sets of biomarkers. A salient set of system-level health biomarkers should identify and compress health information. In particular, they should automatically prioritize biomarkers with high relevance to survival. We tested this prioritization and compression ability by comparing cumulative survival prediction between different sets of predictors, and compared to a random ordering. How quickly a set of variables reaches its maximum predictive power reflects how well the transformation prioritizes salient information of survival risk and, by implication, overall health and disease severity. 

We used a time-dependent Cox proportional hazard model for survival, using start-stop formatting \cite{Moore2016-rh}. We compare 4 sets of predictors: (i) the natural variables, $z$, (ii) principal component analysis (PCA), (iii) the (sorted) raw biomarkers, and (iv) a randomly-reordered set of principal components (PCs). The raw biomarkers are sorted by their univariate predictive power (C-index \cite{Harrell1982-qt}; described below). Our test is to build and compare cumulative survival models using these four sets of predictors. Each cumulative survival model is constructed by adding predictors one-by-one starting from the lowest-ranked (highest priority) predictor. The randomly-reordered PCA provides a null hypothesis that the ranking is no better than chance. The best performing model should achieve the maximum possible C-index with as few predictors as possible. 

The C-index is a measure of prognostic ability, defined as the proportion of individuals which will be correctly ranked as dying sooner between all pairs of individuals \cite{Harrell1982-qt}. We used the 632 estimator which is 63.2\% in-sample  plus 36.8\% out-of-sample  (via 100-sample bootstrap)  (i.e. $C_{632}\equiv 0.632 \cdot C_{train} + 0.368 \cdot C_{test}$) \cite{Hastie2017-wl}.

\begin{figure}[ht] 
     \centering
        \includegraphics[width=0.7\textwidth]{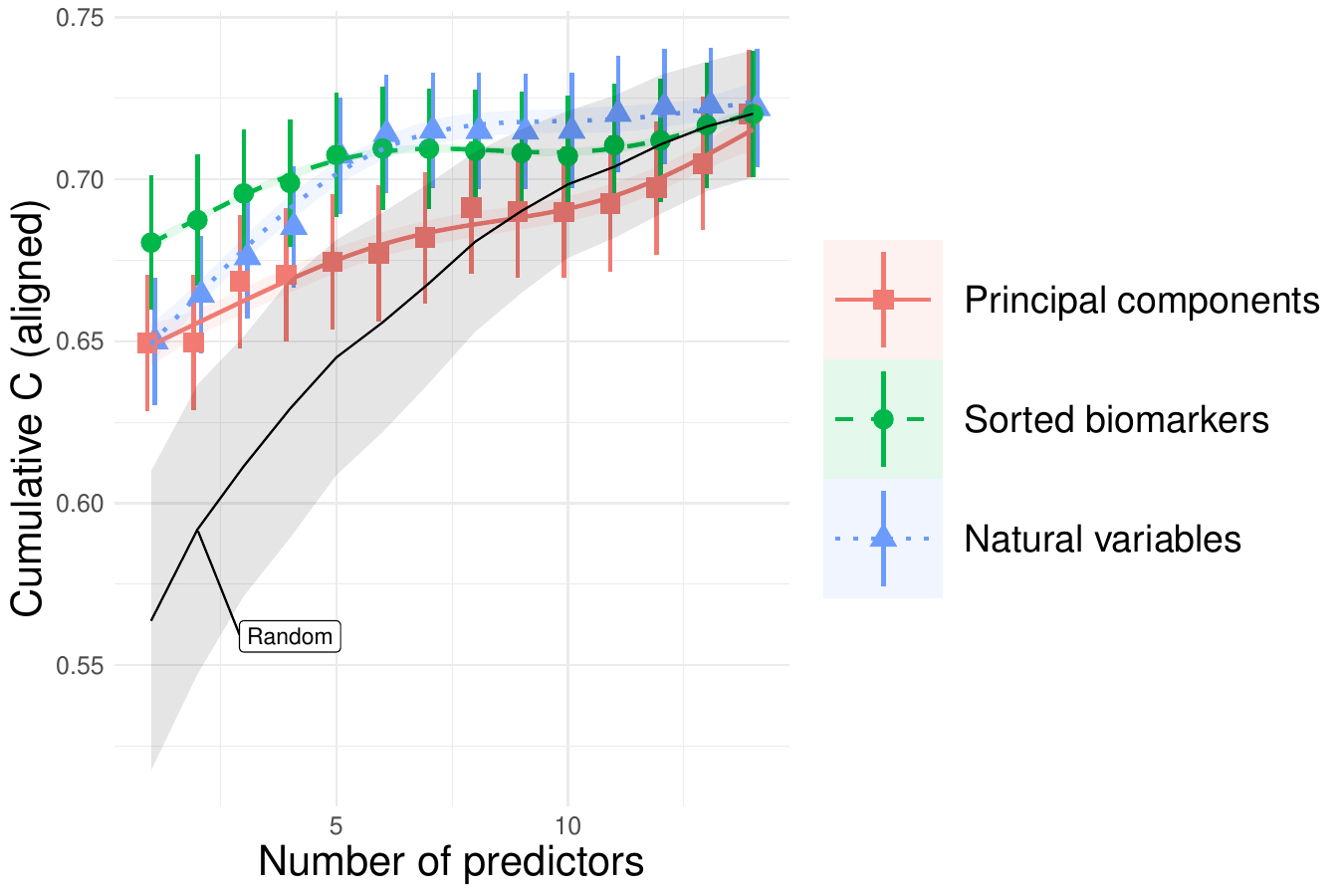}   
    \caption{Cumulative survival prediction by variable set. x-axis is cumulative number of predictors, y-axis is performance of associated survival model for each set of predictors (colours).  There is a clear tendency to concentrate survival information into the lowest natural variables (blue triangles), whereas survival information is spread across high and low ranks for PCA (red squares). The optimally-sorted raw biomarkers (green dots) performed the best but with substantial overlap with the natural variables. The grey band is PCA with random ordering (the black line is the bootstrap average). PCA only marginally out-performs a random ordering. All results have been bootstrapped 100 times. The red, blue, and green lines are the local polynomial regression fits \cite{Wickham2016-kw}.}
    \label{fig:si:cumc}
\end{figure}

For the main dataset, the natural variables demonstrated a strong ability to compress health information into the lowest orders, as illustrate by the cumulative survival C-index, Figure~\ref{fig:si:cumc}. We find that the health information has been efficiently compressed into the first 5 natural variables. In contrast, the principal components struggled to perform significantly better than a random ordering. To perform PCA we trained using the first time point. The $z$ variables performed as well as picking the best biomarkers, but did so automatically without knowledge of survival. This indicates a deep connection between the dynamical behaviour --- which determines $z$  --- and disease severity. Specifically, the $z$ are ranked by their resilience parameter, $\lambda$, indicating that the variables most relevant to health are those which demonstrate the worst resilience (smallest $|\lambda|$).

\begin{figure}[b]
     \centering
        \includegraphics[width=0.55\textwidth]{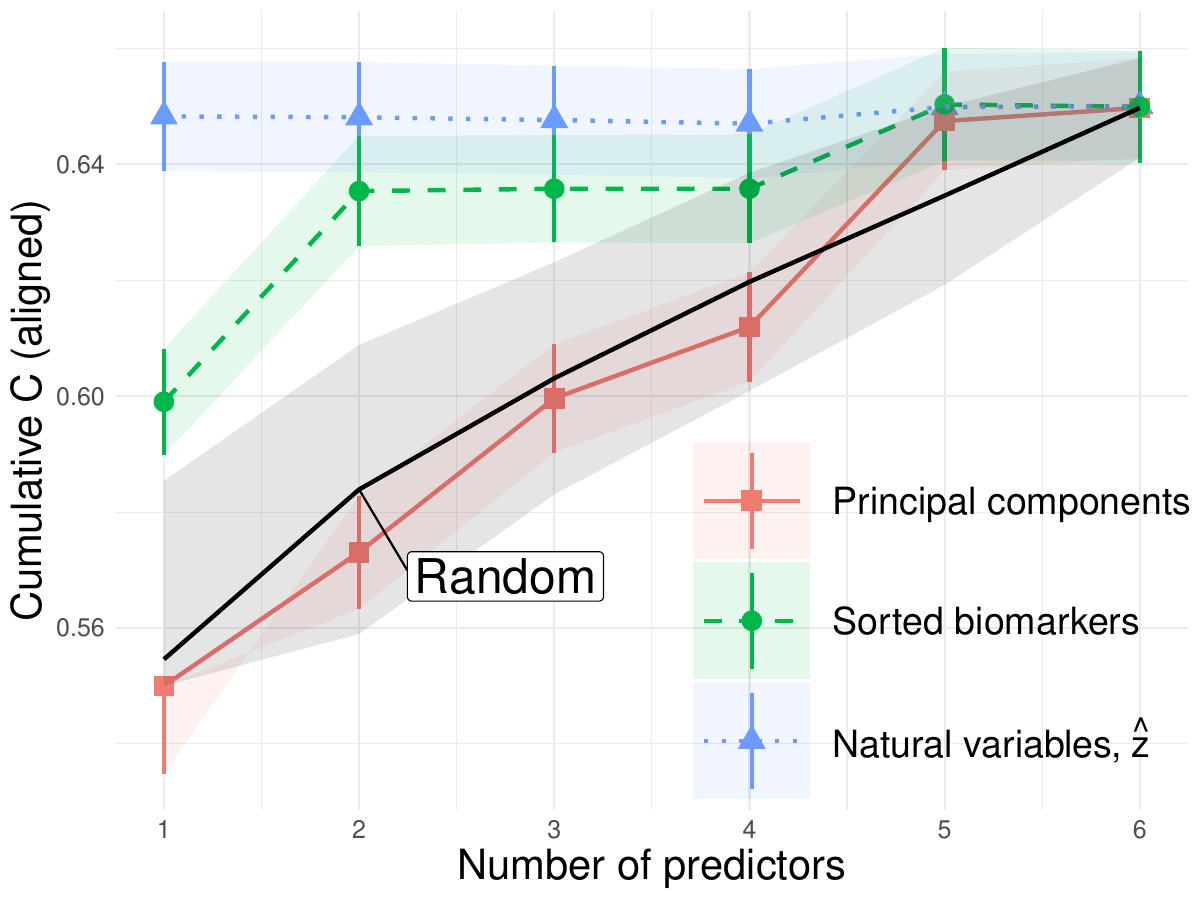}  
    \caption{Cumulative survival prediction by aggregate variable set --- validation data. x-axis is cumulative number of predictors, y-axis is performance of associated survival model for each set of predictors (colours). The survival information is concentrated into the lowest natural variable (blue triangles), versus PCA (red squares), and the raw biomarkers sorted by independent survival predictive power (green dots). PCA sorts by variance, but this ordering was no better than a random ordering (black band). Biomarkers are sorted by their marginal survival performance (C-index), yielding the ordering: creatinine, albumin, HCO3, phosphate, urea, then calcium.  The natural variables are the emulated variables, $\hat{z}$. }
    \label{fig:si:cumcvalidation}
\end{figure}

We performed the same analysis using the (cross-sectional) validation data with emulated natural variables, although that was limited to a maximum of 6 variables. Only a few of the natural variables were accurately emulated (see below, Figure~\ref{fig:si:validation_r2}). In Figure~\ref{fig:si:cumcvalidation} we see that $\hat{z}_1$ is able to independently achieve maximum performance, and significantly better than the first biomarker (creatinine). In contrast, PCA performs no better than a random ordering. In our model $z_1$ has the longest memory (auto-correlation time) and therefore we expect that we see $z_1$ as the dominant survival predictor.  Note that the C-index score for $\hat{z}_1$ and $z_1$ are approximately equal ($0.65$).


Note that PCA is a special case of our model wherein the noise is diagonal and the model has reached a steady-state \cite{mallostasis}. However, the noise cannot be diagonal if the network is asymmetrical since $\boldsymbol{\hat{\Sigma}}=\boldsymbol{\hat{\Sigma}}^T$ must be diagonalized by an orthogonal matrix but $\boldsymbol{W}\neq \boldsymbol{W}^T$ cannot be because the eigenvectors of a symmetric matrix form an orthogonal matrix (see e.g. Eq.~293 of the 2012 Matrix Cookbook \cite{cookbook2012})). Nevertheless, under those conditions the principal components are natural variables, possibly re-ordered (although this re-ordering appears to be uncommon in practice \cite{mallostasis}). In practice, as we showed in the main text, our estimated noise matrix was block-diagonal and hence the natural variables will generally not coincide with the principal components. 

\FloatBarrier

\subsection{Terminal decline} \label{sec:si:td}
We define a terminal decline plot as the mean of the conditional distribution $p(z|t)$ versus time-to-event, where $z$ is a survival predictor and $t$ is the time-to-event (death, censorship or transplant). The purpose of this plot is to use our knowledge of when individuals died to infer what changes preceded their death. This is necessary to understand stochastic accumulation since it is a random process. It is difficult to intuit the behaviour of such a distribution since knowledge of when a person will die can provide a great deal of information. Our primary goal is to understand stochastic accumulation and whether our combined dynamical and survival models describe it. 

Throughout this section we will include both a `case' group with $\beta \neq 0$ which emulates the observed data, and a hypothetical `control' group with $\beta\equiv0$. The latter represents a hypothetical group that are immune to the effects of $z$. 

In Section~\ref{sec:simttd} we present a stripped-down simulation study of model parameters and their effects on the terminal decline plots (e.g.\ Figure~4). This simulation starts in the steady-state, and analyzes only one natural variable, $z$. To summarize, we see that $\mu_t$ drives population-level changes whereas $\lambda$ and $\sigma$ drive individual-level changes, consistent with Eq.~6. We see log-linear behaviour followed by saturation as the time-to-event approaches $0$. When the case and control diverge depends only on $\lambda$ and the baseline hazard $h_0\equiv \text{scale}^{-\text{shape}} \equiv \zeta^{-\nu}$. The time between when control crosses case and when they die we call the horizon time since it should represent the time between when the signs of abnormally first begin to when the abnormality becomes fatal (i.e.\ from when $z$ is indistinguishable between the groups to when it is lethal). Together with our results from Section~\ref{sec:sttd} we can infer that the terminal decline curve shares the same features as the non-stochastic version but is truncated by the limited memory of the system, as determined by the auto-correlation time $|\lambda|^{-1}$: which is much faster than the typical survival horizon (5~years, half-life). This can be clearly seen in Figure~\ref{fig:si:survival_sim} for $|\lambda| > 4$ where there is no apparent survival advantage to having low $z$ after $0.25~$years. This is because the memory of the system lasts for only $|\lambda|^{-1} = 0.25~$years. 

To complement the survival simulation of terminal decline, we include an analytical model of the non-stochastic version of our model (in which individuals don't evolve over time). In Section~\ref{sec:sttd} we use the saddle point approximation of the mean to derive and explain the characteristic phenomenon for terminal decline in the simplified case of non-stochastic Gaussian statistics. This means that individuals enter the study with Gaussian-distributed predictor values that do not change in time (similar to what was observed for $z_1$). We assume Weibull survival statistics. If we compare a non-stochastic survival predictor ($\beta\neq 0$) to a hypothetical control ($\beta\equiv0$) we see some characteristic phenomena for the mode of the distribution. We see that both log-linear behaviour and a large gap versus control as $t\to0$ are a consequence of a $z$ being a strong survival predictor. The gap between the normal (case) group and control saturates at $\beta\text{Var}(z)$ for $t\to 0$ (i.e.\ no deaths). This gap is caused by abnormal individuals dying faster. The gap does not persist however, because abnormal individuals don't survive and therefore the case mode has to drop over time due to attrition. The normal group crosses the hypothetical control at the unique (``horizon'') time $\ln{(t_{h})} = -\ln{(h_0)}/\nu-\beta\langle z \rangle/\nu$, which corresponds to the typical individual's survival probability reaching its characteristic value $e^{-1}$ (i.e.\ $h_0 t^\nu e^{\beta \langle z \rangle}=1$). For strong survival predictors, log-linear behaviour is dominant with saturation just before death leading to a large gap between those who will die imminently and the remaining population. For weak predictors, the behaviour is sub-linear, the saturation is early and the gap is small. These features are also seen in the stochastic case, Section~\ref{sec:simttd}, although since the memory of the system is fixed by $|\lambda|^{-1}$ the effects are truncated at $t\approx |\lambda|^{-1}$, at which point the case-mean rapidly converges towards the control-mean.
\subsubsection{Simulated terminal decline} \label{sec:simttd}
\begin{figure}[!ht] 
     \centering
        \includegraphics[width=\textwidth]{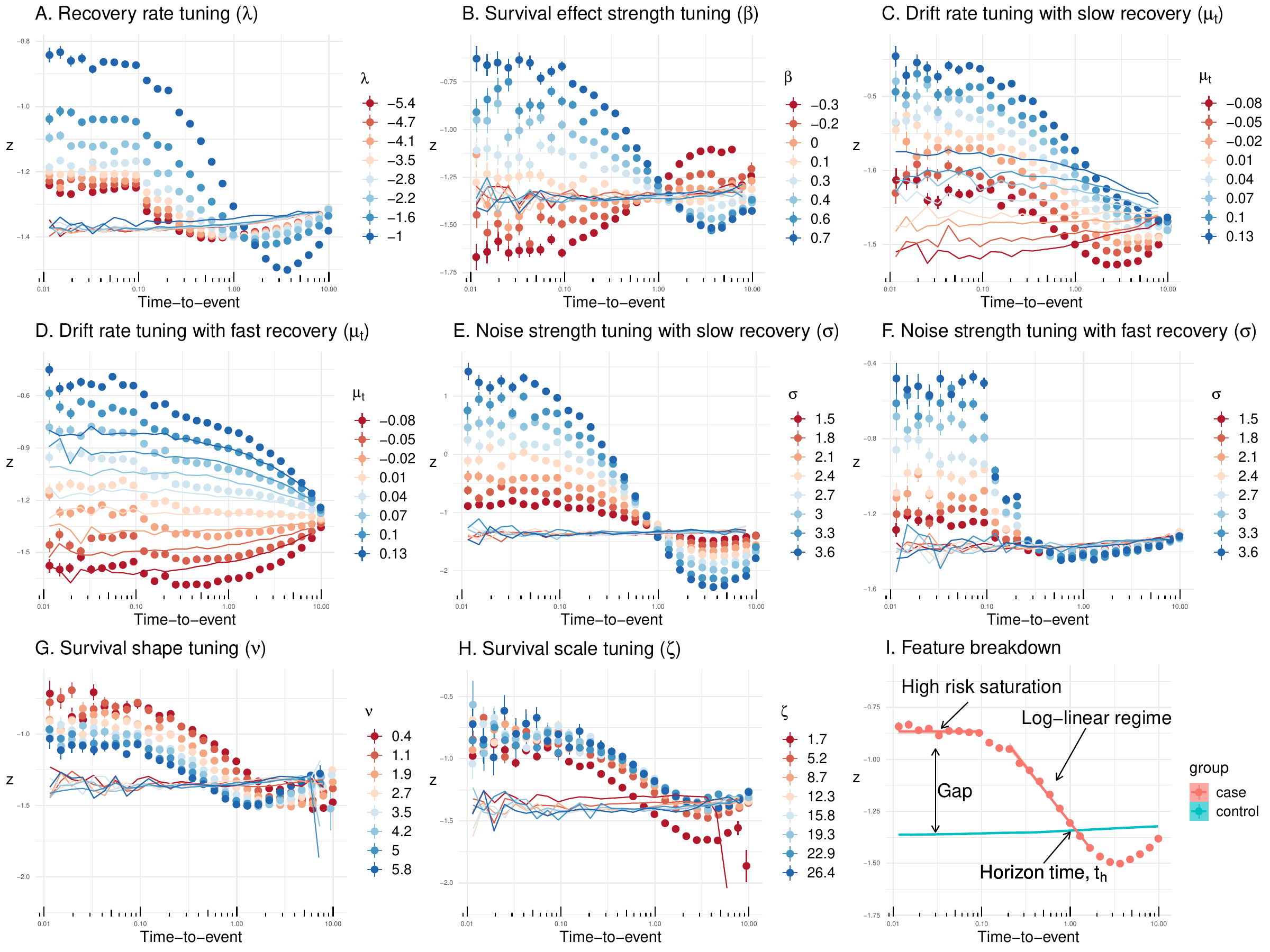}
    \caption{Stochastic simulation of univariate terminal decline. Points: simulated ``case'' data ($\beta=0.53$, unless specified); lines: simulated ``control'' data ($\beta=0$). Time-to-event plots depend on $\beta$, $\mu_t$ and $\lambda$. We varied over the observed ranges for all values. Default values were fits to $z_1$: $\lambda=-0.96~\text{years}^{-1}$, $\beta =0.53$, $\mu_t=-0.02~\text{years}^{-1}$ and $\sigma^2 = 4.9~\text{years}^{-1}$ ($\nu = 1.7$, $\zeta=4.6$, $h_0 = \zeta^{-\nu} = 0.080$; $\mu_0=-1.30$). \textbf{A.} varying $|\lambda|$ demonstrate that both the gap and horizon time depend strongly on $|\lambda|$. Larger $|\lambda|$ creates smaller variance (Eq.~\ref{eq:var}), which decreases the hazard (Eq.~\ref{eq:ht}), and smaller auto-correlation time (Eq.~\ref{eq:acf}). \textbf{B.} varying the survival strength, $\beta$, changes the slope and gap. \textbf{C.} varying $\mu_t$ with slow recovery rate (default $\lambda$). \textbf{D.} varying $\mu_t$ with fast recovery (large $|\lambda| = 5.4~\text{years}^{-1}$). $\mu_t$ controls the entire population. \textbf{E.} varying $\sigma$ with slow recovery (default). Individuals have a long time to accumulate stochastic dysfunction, deviating from normal at $t=1~\text{year}$. Increasing the noise strength increases this effect. \textbf{F}. varying $\sigma$ with fast recovery, large $|\lambda|=5.4~\text{years}^{-1}$. The shorter auto-correlation times means a much shorter horizon between worsening and eventual death. \textbf{G.} Varying of survival hazard shape, $\nu$. All shapes coincide at $S=e^{-1}$. Small shapes drop quickly then saturate at $S=e^{-1}$; large shapes drop suddenly to $0$ just before $S=e^{-1}$. The shape clearly affects the shape of the case curves, likely because $S < e^{-1}$ in this study (the half-life as 5~years). This means more deaths will be observed for small shapes, and quicker. \textbf{H.} Varying of survival hazard scale, $\zeta$. The scale controls the baseline hazard: large scale means long lifespans. There is only an effect when $\zeta$ is small enough to make the survival times dip into the time scale set by $|\lambda|^{-1}\approx 1~$year. \textbf{I}. breakdown of the key features of the time-to-death plot. As described in the text these features are a generic consequence of a Gaussian variable combined with our survival model.}
    \label{fig:si:survival_sim}
\end{figure}

While the hazard can be computed exactly (Eq.~6), the survival and terminal decline distributions considering all possible stochastic paths between measurements are non-trivial to calculate since they depend on a number of subtle effects, such as having multiple entries for a single individual. These effects are automatically accounted for using simulated data. We simulated a single variable with default parameters taken from our fit for $z_1$, with initial values starting in the steady-state at $t=0$. We then varied the model parameters ($\lambda$, $\mu_t$, $\sigma$, $\beta$, $\text{hazard shape}\equiv\nu$, and $\text{hazard scale}\equiv \zeta$) to characterize their influence on decline trajectories, Figure~\ref{fig:si:survival_sim}. $\mu_0$ was defined by the average value over all individuals (incorporates all static covariates: baseline age, sex and DM status). Since $\mu_0$ is just a global shift of the steady-state it was not varied (for this reason, in the main text we simply used $\mu_0\equiv 0$).

Higher $z$ values are at exponentially higher risk of death via $h\propto e^{\beta z}$. We compare to a hypothetical control which does not feel the effects of $z$ (i.e.\ $\beta=0$; who still die via the baseline hazard $\nu h_0 t^\nu$). We observe that the shape of the case curve depends heavily on $\lambda$, $\sigma$ and $\beta$ whereas both the case and control depend on $\mu_t$. This is because $\mu_t$ determines the homeostatic set point and so it will move $\langle z \rangle$ thus translating everybody in the population up or down (control lines). Conversely, decreasing $\lambda$ or increasing $\sigma$ increases the individual differences (variance) and therefore produces more extremely unhealthy individuals (thus moving up the upper saturation point) and extremely healthy individuals (thus moving down the lower saturation point: but only if $|\lambda|^{-1}$ is large enough to permit a long memory). When $|\lambda|$ is small (slow recovery), as is the default, the noise has time to accumulate individual differences through fluctuations which do not recover (E.) but when $\lambda$ is fast the fluctuations cannot accumulate and the individual either dies quickly, in this case within 16~weeks (0.3~years), or recovers completely with no increased risk of death (F.). The characteristic phenomena are summarized in Figure~\ref{fig:si:survival_sim}G, for comparison to experiment. 

The non-stochastic case, Section~\ref{sec:sttd}, is equivalent to the stochastic case in the limits $|\lambda|\to 0$ and $\sigma\to 0$. This leads to individuals that differ according to a normal distribution but do not evolve over time. Hence slow $z$ (e.g.\ $z_1$) with small noise, $\sigma$, should behave similarly to the non-stochastic case.

\subsubsection{Non-stochastic terminal decline} \label{sec:sttd}
Here we provide an analytical description of the terminal decline plot as well as the key features observed in Figure~\ref{fig:si:survival_sim}. For simplicity, we will assume variables are not changing over time. First, in order to compute the terminal decline mean we need to know the distribution of the conditional distribution $p(z|t)$, where $z$ is the predictor of interest and $t$ is the time to death. From Bayes' theorem we have
\begin{align}
    p(z|t) &= \frac{p(t|z)p(z)}{p(t)} = \frac{p(t,z)}{\int_{-\infty}^{\infty} p(t,z) dt}.
\end{align}
We can infer $p(z)$ and $p(t|z)$ from the data. Our dynamical model is Gaussian at each time point and hence we assume normal statistics
\begin{align}
    p(z) &= \frac{1}{\sqrt{2\pi\phi^2}} \exp{\bigg( -\frac{1}{2}\frac{(z-\bar{z})^2}{\phi^2} \bigg)}
\end{align}
where $\bar{z}$ is the mean and $\phi^2$ is the variance.

We know empirically that the $z$ satisfy the proportional hazards assumption and furthermore that the survival distribution is Weibull hence we have,
\begin{align}
    p(t|z) &= h_0 \nu t^{\nu-1} \exp{\big( \beta z - h_0t^\nu e^{\beta z} \big)}
\end{align}
where $h_0$, $\nu$ and $\beta$ are fit parameters.

Combining the ``prior'' ($p(z)$) with the ``likelihood'' ($p(t|z)$) we have
\begin{align}
    p(z|t) &= \frac{1}{Z} \exp{\big( -\frac{1}{2}\frac{(z-\bar{z})^2}{\phi^2} + \beta z - h_0t^\nu e^{\beta z} \big)}
\end{align}
where the denominator $Z$ is defined as the integral of the numerator. We can safely drop all terms that don't depend on $z$ since they will cancel out with $Z$. The `partition function', $Z$, is not analytically solvable since it involves a double exponential $\exp{(e^{\beta z})}$. A simple solution is to look at the mode of $p(z|t)$ as an approximation for the mean. This is a `saddle point' approximation.

Neglecting irrelevant terms which do not depend on $z$ we have
\begin{align}
    p(z|t) &\propto \exp{\big( -\frac{z^2}{2\phi^2} + \frac{z\bar{z}}{\phi^2} + \beta z -h_0t^\nu e^{\beta z}\big)} \equiv \exp{(H_0)} 
\end{align}
where $H_0$ is defined by the above equation. The saddle point of $H_0$ is given by the mode of $p(z,t)$. The derivative is
\begin{align}
    \frac{dH_0}{dz} &= -\frac{z}{\phi^2} + \frac{\bar{z}}{\phi^2} + \beta - \beta h_0t^\nu e^{\beta z}.
\end{align}
The derivative is $0$ at the saddle point, $z^*$, hence
\begin{align}
   z^* + \beta\phi^2 h_0t^\nu e^{\beta z^*} &= \bar{z} + \beta\phi^2 \label{eq:zsp}
\end{align}
which is a transcendental equation in $z^*$. For small $t$ the solution is
\begin{align}
    \lim_{t\to 0} z^* &=  \bar{z} + \beta\phi^2.
\end{align}
If $\beta z^*$ becomes large we have instead
\begin{align}
   \lim_{\beta z\to \infty}   z^* &= \frac{1}{\beta}\ln{\bigg(\frac{\bar{z} + \beta\phi^2}{\beta\phi^2 h_0}\bigg)} - \frac{\nu}{\beta}\ln{(t)}.
\end{align}
These two limits explain the saturation and log-linear behaviours observed in Figure~\ref{fig:si:survival_sim}. Note that $\beta$ will determine the direction that $z^*$ becomes large in as $t$ approaches $0$, hence the sign will always be appropriate, $\text{sign}(\beta (z(t=0)-\bar{z}))=1$.

For the control group we have $\beta \equiv 0$ and hence $p(z|t)=p(z)$. Hence we can solve for the horizon time from when the case first starts looking different from the control. This occurs for
\begin{align}
   \bar{z} + \beta\phi^2 h_0t_h^\nu e^{\beta \bar{z}} &= \bar{z} + \beta\phi^2
\end{align}
which is re-arranged to yield the horizon time
\begin{align}
   t_{h} &= \exp{\bigg(-\frac{1}{\nu}\ln{(h_0)} - \frac{\beta\bar{z}}{\nu} \bigg)}.
\end{align}

In general, $\bar{z}$ evolves smoothly according to an underlying ordinary differential equation. This is easily derived by differentiating Eq.~\ref{eq:zsp} with respect to $u\equiv \ln{(t)}$ which yields
\begin{align}
    \frac{dz^*}{du} &= -\frac{\nu}{\beta} s(w)
\end{align}
where $s(x)=1/(1+e^{-x})$ is the sigmoid function and
\begin{align}
    w &\equiv \ln{(h_0)}+2\ln{|\phi\beta|}+\nu\ln{(t)}+\beta z^*.
\end{align}

The sigmoid has three interesting regimes: $w \ll 0$, $w\approx 0$, and $w \gg 0$. These correspond to just before death, intermediate, and long before death. The first regime gives an approximate ODE
\begin{align}
    \frac{dz^*}{du}\bigg|_{w \ll 0} &\approx -\frac{\nu}{\beta} e^w,
\end{align}
whose solution is
\begin{align}
    z^* |_{w \ll 0} &\approx -\frac{1}{\beta}\ln{(h_0\phi^2\beta^2t^\nu+const)}.
\end{align}
This regime must occur for sufficiently small $t\to0$.

The second regime gives an approximate ODE
\begin{align}
    \frac{dz^*}{du}\bigg|_{w \approx 0} &\approx -\frac{\nu}{4\beta} \bigg(2+w\bigg) =  -\frac{\nu}{4\beta} \bigg(2+ \ln{(h_0\phi^2\beta^2)}+\nu\ln{(t)}+\beta z^* \bigg) ,
\end{align}
whose solution is
\begin{align}
    z^* |_{w \approx 0} &\approx -\frac{1}{\beta}\ln{(h_0\phi^2\beta^2)}-\frac{\nu}{\beta}\ln{(t)} + \frac{2}{\beta} + const\cdot t^{-\nu/4}.
\end{align}

Finally, the third regime gives an approximate ODE
\begin{align}
    \frac{dz^*}{du}\bigg|_{w \gg 0} &\approx -\frac{\nu}{\beta},
\end{align}
whose solution is
\begin{align}
    z^* |_{w \gg 0} &\approx -\frac{\nu}{\beta}\ln{(t)} + const.
\end{align}

We see that $z^\ast$ always includes $-\frac{\nu}{\beta}\ln{(t)}$, though with a constant offset that may change between the regimes. Accordingly, apart from saturation for small $t$, we
expect $z^*$ to be approximately piece-wise log-linear in time.

\subsection{Stratified associations} \label{sec:si:ass}
An important consequence of our conceptualization is that as natural variables become abnormal they should \textit{each} drive changes to multiple biomarkers, leading to a spectrum of signs associated with the dysfunction of any single $z$. Using regression models, we can infer what those signs should look like in terms of observables: serum biomarker values, clinical conditions, and causes of death. In the main text we performed this association analysis using all individuals pooled together. Here we consider stratifying the individuals and then testing for associations. Note that we are still fitting to the pooled set of individuals, and it is only after we have fit our model and transformed into the natural variables that we split up into groups and test for associations separately.

There are four main (overlapping) groups of interest: diabetics (DM=1), non-diabetics (DM=0), males (sex=0) and females (sex=1). We had many more males than females (64\% vs 36\%) but a nearly equal number of diabetics versus non-diabetics (57\% vs 43\%).

\begin{figure*}[!ht] 
     \centering
        \includegraphics[width=\textwidth]{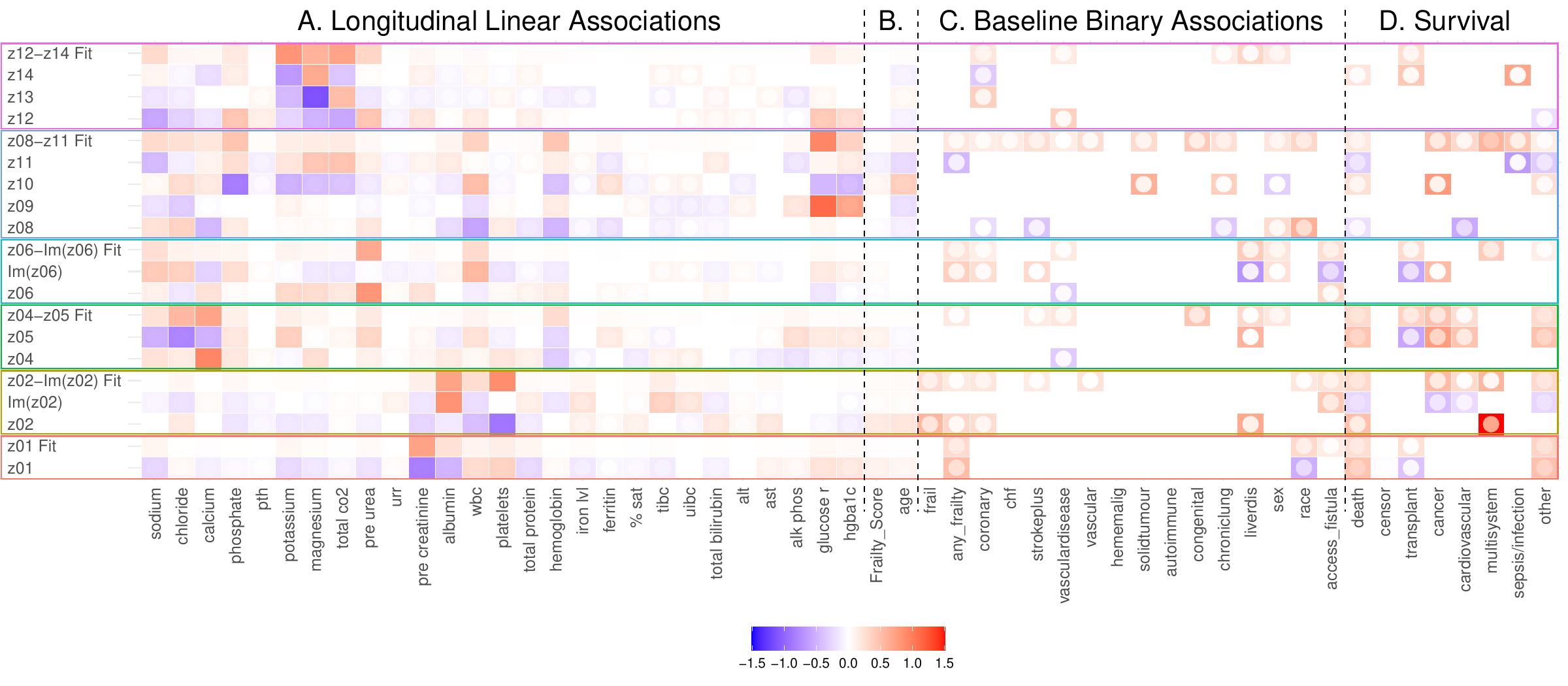}
    \caption{Associations for the natural variables, $z$, by module --- \textbf{diabetics only}. Diabetics and non-diabetics were both included in the initial fit.  \textbf{A.} linear regression against (continuous) longitudinally-measured biomarkers ($z$ within module are predictors). \textbf{B.} linear regression against ordinal baseline variables. \textbf{C.} logistic regression against binary baseline variables. \textbf{D.} Competing time-to-event regression \cite{De_Wreede2011-tj}. The variables are grouped by modules (outlines). Each module has a row of regression coefficients for each $z$ within the module and an overall fit quality row (e.g.\ z04-z05 Fit). To read, pick a module e.g.\ $z_1$ and read first the coefficients from left to right (what information is present), then read the next row in that module, until you reach the overall fit quality (how much information is present). Inner point is 95\% confidence interval closest to 0; non significant points have been whited out (no multiple-comparison corrections). Readers should look for biologically-consistent trends. Score ("Fit") depends on variable type: $R^2$ for linear, $2\times\text{AUC}-1$ for binary, and $2\times\text{C}-1$ for survival (all range from 0 worst to 1 best). C-index is marginal making it only a rough measure fit quality due to competing risks. Note the drop in survival information as we move up in rank. Colour scale is truncated at 1.5 for visualization. All continuous variables (\textbf{A}) were scaled to zero mean, unit variance.}
    \label{fig:zcor_dm}
\end{figure*}

\begin{figure*}[!ht] 
     \centering
        \includegraphics[width=\textwidth]{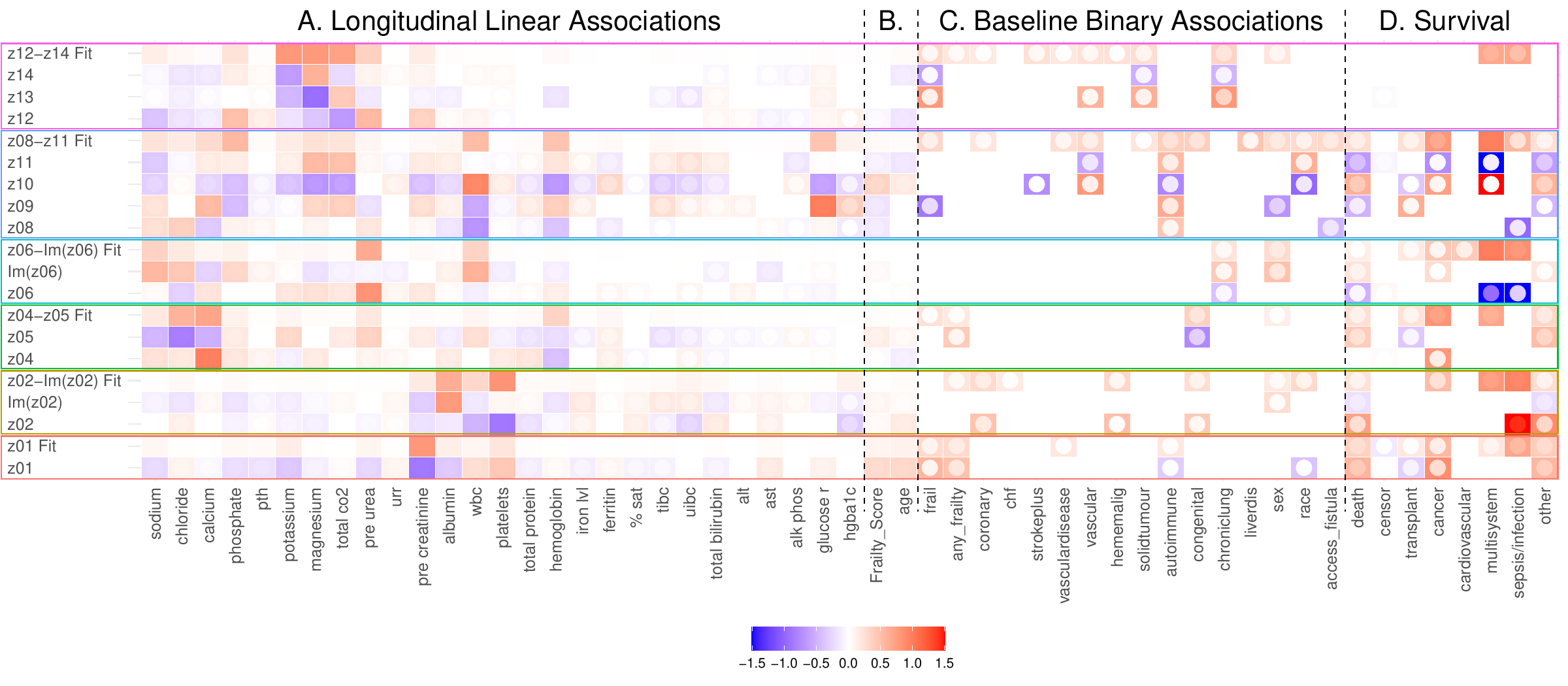}
    \caption{Associations for the natural variables, $z$, by module --- \textbf{non-diabetics only}. Diabetics and non-diabetics were both included in the initial fit.  \textbf{A.} linear regression against (continuous) longitudinally-measured biomarkers ($z$ within module are predictors). \textbf{B.} linear regression against ordinal baseline variables. \textbf{C.} logistic regression against binary baseline variables. \textbf{D.} Competing time-to-event regression \cite{De_Wreede2011-tj}. The variables are grouped by modules (outlines). Each module has a row of regression coefficients for each $z$ within the module and an overall fit quality row (e.g.\ z04-z05 Fit). To read, pick a module e.g.\ $z_1$ and read first the coefficients from left to right (what information is present), then read the next row in that module, until you reach the overall fit quality (how much information is present). Inner point is 95\% confidence interval closest to 0; non significant points have been whited out (no multiple-comparison corrections). Readers should look for biologically-consistent trends. Score ("Fit") depends on variable type: $R^2$ for linear, $2\times\text{AUC}-1$ for binary, and $2\times\text{C}-1$ for survival (all range from 0 worst to 1 best). C-index is marginal making it only a rough measure fit quality due to competing risks. Note the drop in survival information as we move up in rank. Colour scale is truncated at 1.5 for visualization. All continuous variables (\textbf{A}) were scaled to zero mean, unit variance.}
    \label{fig:zcor_nondm}
\end{figure*}

The diabetic associations are reported in Figure~\ref{fig:zcor_dm}. The non-diabetics are reported in Figure~\ref{fig:zcor_nondm}. The primary variables we are concerned with are random glucose (glucose r) and hgab1c (hemoglobin A1C) since they are the primary biomarkers related to glucose metabolism. Comparing the two figures we can see that $z_1$ for diabetics appears to have stronger associations with both glucose and hgab1c: both are clearly weaker for non-diabetics and hgab1c isn't even significant (last two columns of \textbf{A.}). Both show glucose is positively associated with high $z_1$ and thus with worsening wasting. This could indicate that $z_1$ is stressing the metabolic system, e.g.\ inflammatory cells increasing caloric demands. Perhaps more interesting, we see that $z_2$ is associated with multisystem failure in diabetics versus sepsis in non-diabetics. Clinically, there is substantial overlap between these two causes of death since they both culminate in multi-organ failure. 

\begin{figure*}[!ht] 
     \centering
        \includegraphics[width=\textwidth]{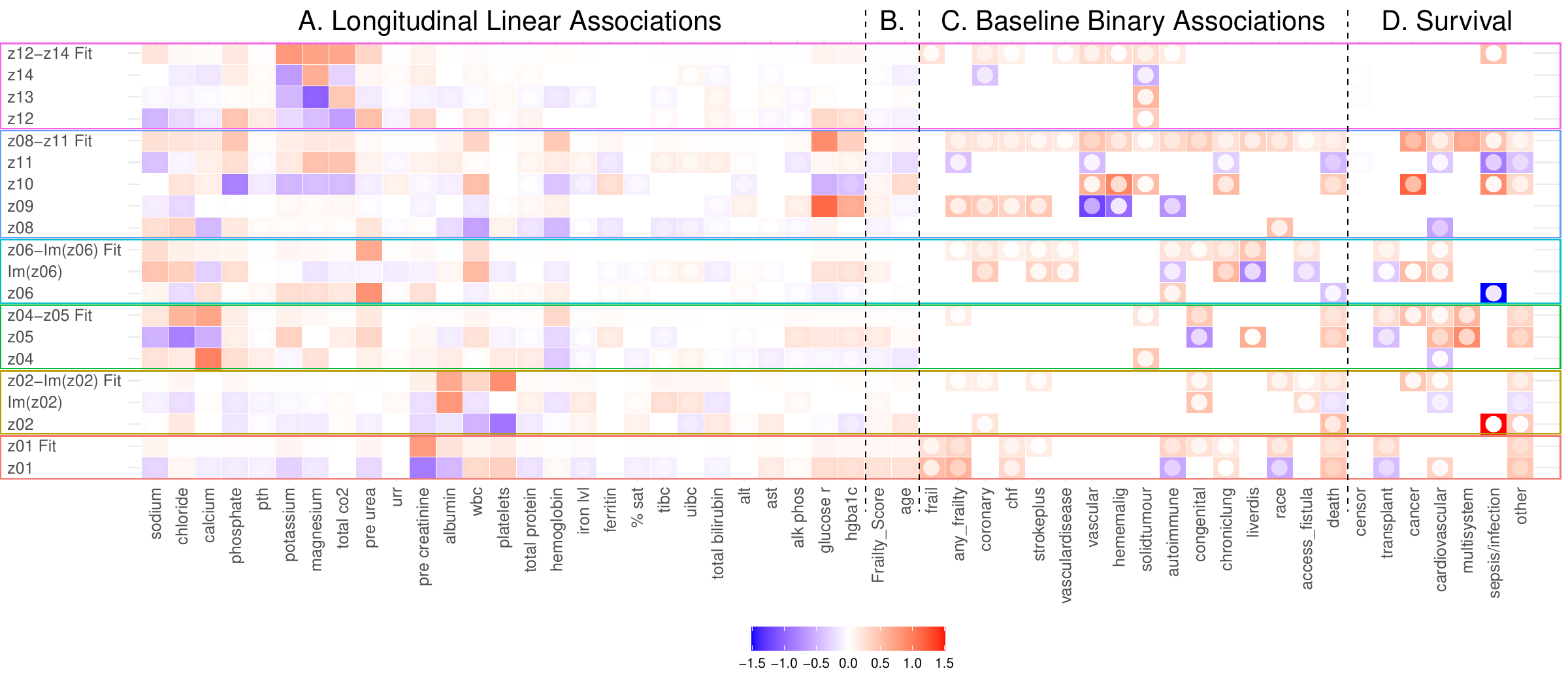}
    \caption{Associations for the natural variables, $z$, by module --- \textbf{males only}. Both sexes were included in the initial fit.  \textbf{A.} linear regression against (continuous) longitudinally-measured biomarkers ($z$ within module are predictors). \textbf{B.} linear regression against ordinal baseline variables. \textbf{C.} logistic regression against binary baseline variables. \textbf{D.} Competing time-to-event regression \cite{De_Wreede2011-tj}. The variables are grouped by modules (outlines). Each module has a row of regression coefficients for each $z$ within the module and an overall fit quality row (e.g.\ z04-z05 Fit). To read, pick a module e.g.\ $z_1$ and read first the coefficients from left to right (what information is present), then read the next row in that module, until you reach the overall fit quality (how much information is present). Inner point is 95\% confidence interval closest to 0; non significant points have been whited out (no multiple-comparison corrections). Readers should look for biologically-consistent trends. Score ("Fit") depends on variable type: $R^2$ for linear, $2\times\text{AUC}-1$ for binary, and $2\times\text{C}-1$ for survival (all range from 0 worst to 1 best). C-index is marginal making it only a rough measure fit quality due to competing risks. Note the drop in survival information as we move up in rank. Colour scale is truncated at 1.5 for visualization. All continuous variables (\textbf{A}) were scaled to zero mean, unit variance.}
    \label{fig:zcor_m}
\end{figure*}

\begin{figure*}[!ht] 
     \centering
        \includegraphics[width=\textwidth]{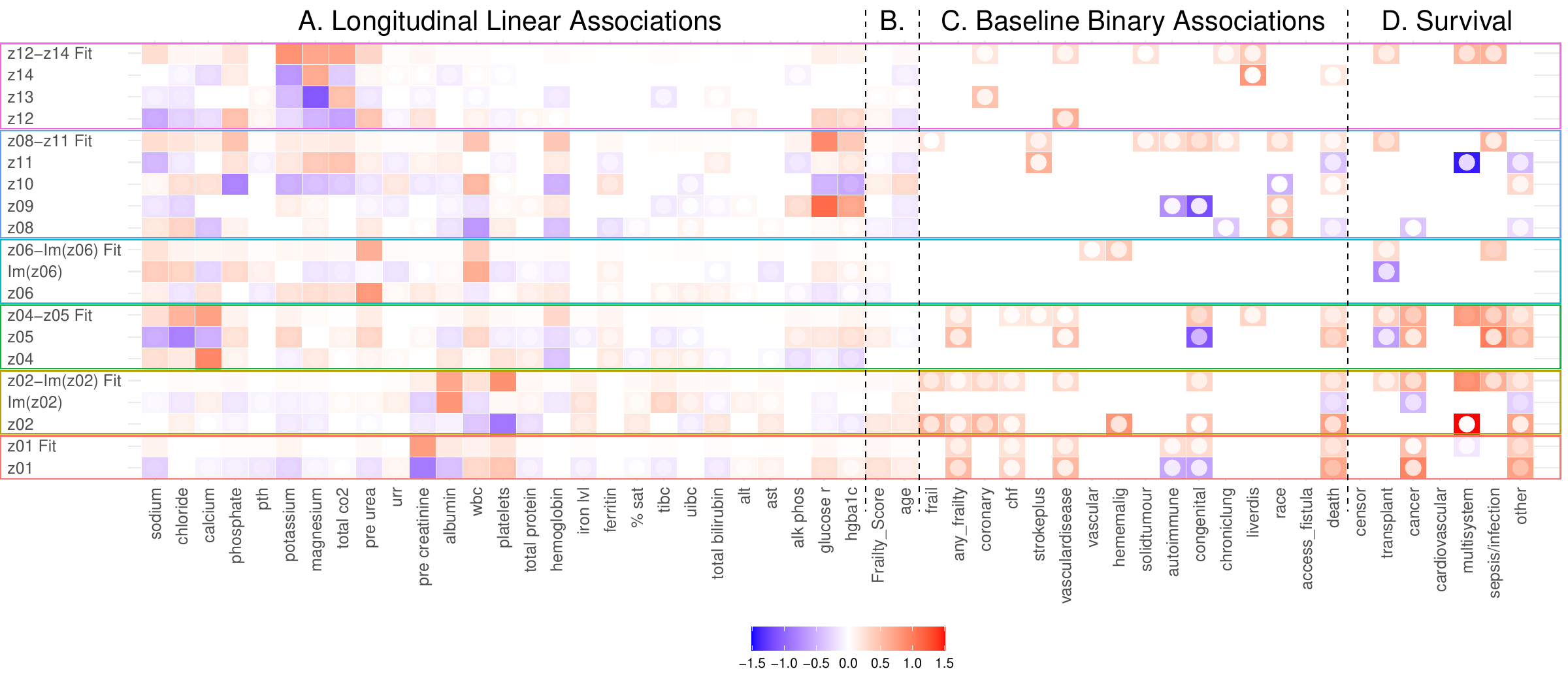}
    \caption{Associations for the natural variables, $z$, by module --- \textbf{females only}. Both sexes were included in the initial fit.  \textbf{A.} linear regression against (continuous) longitudinally-measured biomarkers ($z$ within module are predictors). \textbf{B.} linear regression against ordinal baseline variables. \textbf{C.} logistic regression against binary baseline variables. \textbf{D.} Competing time-to-event regression \cite{De_Wreede2011-tj}. The variables are grouped by modules (outlines). Each module has a row of regression coefficients for each $z$ within the module and an overall fit quality row (e.g.\ z04-z05 Fit). To read, pick a module e.g.\ $z_1$ and read first the coefficients from left to right (what information is present), then read the next row in that module, until you reach the overall fit quality (how much information is present). Inner point is 95\% confidence interval closest to 0; non significant points have been whited out (no multiple-comparison corrections). Readers should look for biologically-consistent trends. Score ("Fit") depends on variable type: $R^2$ for linear, $2\times\text{AUC}-1$ for binary, and $2\times\text{C}-1$ for survival (all range from 0 worst to 1 best). C-index is marginal making it only a rough measure fit quality due to competing risks. Note the drop in survival information as we move up in rank. Colour scale is truncated at 1.5 for visualization. All continuous variables (\textbf{A}) were scaled to zero mean, unit variance.}
    \label{fig:zcor_f}
\end{figure*}

The sex-stratified associations are reported in Figure~\ref{fig:zcor_m} for males and Figure~\ref{fig:zcor_f} for females. While the associations with biomarkers (\textbf{A}) looks similar between the sexes, the binary associations (\textbf{C}) and causes of death (\textbf{B}) are notably different. The robustness of the associations reflects the underlying robustness of our network estimate, since it controls the transformation into the natural variables. The network doesn't change much whether we fit to only males, only females or to the combined group, Section~\ref{sec:si:sa} (nor is the network sensitive to any other grouping).


\FloatBarrier

\subsection{Summary of datasets} \label{sec:si:demo}
Table~\ref{tab:demo} summarizes the study and validation populations, which were Canadian haemodialysis patients. We modelled longitudinal blood-based biomarkers routinely measured approximately every 6~weeks. We chose 3 months as the start of the observation window to avoid including individuals with acute kidney injury. We chose the half-life as the end of the observation window, 5~years. An alternative window is considered in Section~\ref{sec:si:sa}. Serum biomarkers were routinely measured, typically every 6~weeks, and included individual laboratory values across the domains of kidney function, dialysis clearance,  electrolytes, immune function, anemia and metabolic function. We selected the 14 which were regularly measured for modelling, including the remaining when testing for associations. Biomarkers with the prefix "pre" are measured before dialysis session and those measured afterwards have prefix "post".

\begin{table}[H]
    \centering
\begin{threeparttable}
\caption{Dataset Summary\tnote{*}} \label{tab:demo}
    \begin{tabular}{l l l}
    \hline
        Variable & Study (main)  & Validation   \\ \hline
        Observation window & 3 months--5 years & Baseline only\tnote{1} \\
        N & 713 & 61036\\
        Male sex & 64\% (454) & 61\% (37339)\tnote{2}\\ 
        White race & 12.3\% (82) & -- \\ 
        Baseline age & 65.3 (18) & 67 (20) \\
        Diabetes & 57\% (404) & 14\% (8802)\tnote{3}\\ 
        Frailty\tnote{4} & 14\% (74/529) & -- \\
        Any frailty\tnote{5} & 64\% (337/529) & -- \\ 
        Deaths\tnote{6} & 50\% (270) & 52\% (31853) \\
        Transplants\tnote{6} & 15\% (80) & 22\% (9480)\\ 
        Fistula Access & 23.1 (165) & -- \\ \hline
    \end{tabular}
\begin{tablenotes}
\item[*] median (IQR) for continuous variables, frequency (N) for binary variables.
\item[1] We excluded 169 individuals whom died before 3~months for consistency with the main dataset.
\item[2] 22 reported their sex as ``other''.
\item[3] only 524 were type-1 diabetic.
\item[4] clinical frailty score of 6 or greater \cite{Rockwood2005-gl,Rockwood2020-oa}. Only 529 patients had this information recorded (all in main dataset).
\item[5] clinical frailty score of 4 or greater (very mild frailty or worse). Only 529 patients had this information recorded.
\item[6] during sample period (excludes end-of-study censorship).
\end{tablenotes}
\end{threeparttable}
\end{table}

Patients were screened upon entry into the study and their pre-existing conditions were recorded. The frequencies of these pre-existing conditions are summarized in Table~\ref{tab:conditions} (main dataset only).

\begin{table}[H]
    \centering
\begin{threeparttable}
\caption{Conditions/Disease Summary (main dataset)} \label{tab:conditions}
    \begin{tabular}{ll}
    \hline
        Variable & Condition\tnote{*} \\ \hline
        Coronary & 35.6\% (254)  \\ 
        CHF\tnote{1} & 33.1\% (236)  \\ 
        Stroke & 15.4\% (110)  \\ 
        Vascular & 17.4\% (124)  \\ 
        Tumour & 10.0\% (71)  \\ 
        Hememalig\tnote{2} & 5.0\% (36) \\ 
        Chronic lung & 19.6\% (140) \\ 
        Liver disease & 3.6\% (26)  \\ 
        Autoimmune & 18.4\% (97)  \\ 
        Vascular & 13.3\% (70) \\ 
        Congenital & 11.6\% (61) \\  \hline
    \end{tabular}
\begin{tablenotes}
\item[*] frequency (N).
\item[1] Congestive heart failure.
\item[2] Hematologic malignancies.
\end{tablenotes}
\end{threeparttable}
\end{table}

\subsection{Data pre-processing} \label{sec:si:data}
The main dataset was from a population of patients receiving haemodialysis in Nova Scotia. The data needed cleaning, as described in this section. Our primary concern was excluding individuals with acute kidney injury, since our target population was individuals with chronic kidney disease receiving regular dialysis. Our secondary concern was ensuring that we avoided any possible sampling biases in the data.

We excluded individuals with acute kidney injury by (i) excluding all people who recovered, and (ii) by starting the study window at 3~months (Section~\ref{sec:si:sa} shows that the window doesn't affect our key parameter estimates). Some individuals had much more data than others due to either extra blood tests or being on dialysis for a very long time. This could cause them to have an inordinate influence on the model, so we restricted their effects by: (i) averaging together all multiple tests on the same day, (ii) including only 1 test within each 4~week interval thus setting the maximum sampling rate at 4~weeks, and (iii) ending the study window at the population half-life (5~years). Another major prospective issue is that individuals whom are suspected of having a medical condition could be tested more frequently, which is averted by our maximum sampling rate (there was no major change without this imposition, not shown).

In summary we performed the following exclusions on individuals:
\begin{itemize}
    \item excluded anybody who was recorded as having recovered, $N=54$ (indicates acute kidney injury),
    \item excluded anybody who didn't have both a baseline record and blood test records, $N=15$ (could indicate individuals not on dialysis),
    \item excluded any individuals who had no blood tests, $N=5$,
    \item excluded individuals with less than 2 time points, $N=33$,
    \item excluded individuals with a $\geq3$~month gap between initiation of dialysis and first measurement, $N=26$, and
    \item included but censored any individuals with a $\geq3$~month gap between the last measurement and their final recorded event. They were instead re-coded as being censored 1~day after their last measurement. This was to avoid individuals who stopped treatment or otherwise had unrealistic final values prior to death. $N=95$ individuals were affected, 37 of which were deaths converted to censorship.
\end{itemize}
After our exclusions our final population was 713 individuals.

For biomarker values / measurements:
\begin{itemize}
    \item we excluded all measurements outside the study window (which was 3~months--5~years),
    \item multiple measurements on the same day for an individual  were averaged together (total of 508 measurements, representing 0.08\% of total measurements),
    \item to further avoid individuals with excessive data we dropped all measurements taken within 4 weeks of the previous measurement (did not have a major effect on results, not shown),
    \item some individuals did not have a recorded exit date since they were ostensibly still receiving dialysis, they were assumed censored at their last bloodwork date + 1 day, $N=250$, and
    \item all biomarkers were standardized by the first time point mean and standard deviation (zero-mean, unit standard deviation).
\end{itemize}

After pre-processing, all individuals were regularly sampled with 96\% of measurements within 8~weeks of each other and the majority occurring 6~weeks apart (mean time between measurements: 6.12~weeks, standard deviation: 0.98~weeks).

\subsection{Biomarkers}
Our target population received regular haemodialysis, typically several times per week, and every $\sim$6~weeks had a blood test before and after their dialysis session. These blood tests were used for our study. Generally we used blood tests before dialysis, but where ambiguous we include the ``pre'' prefix for before and ``post'' for after (applies to metabolites). We exclusively used serum biomarkers to build our network. We picked the 14 most commonly measured biomarkers to avoid the confounding effect of measurement bias and minimize imputation bias, since the existence of a non-routine test could indicate an increased risk for that test being abnormal. 

In general, the biomarkers we used are non-specific and are each sensitive to at least two different important biological processes. From our perspective, this is because they are sensitive to a variety of disruptions to the underlying biological network that controls homeostasis. (The natural variables are more specific, and we see that they tend to coherently drive multiple biomarkers leading to a spectrum of signs within the observed biomarkers.) We provide a terse summary on probable associations between biomarker values and their biological meaning in Table~\ref{tab:si:biomarkers}, together with references.

Our choice of 14 main biomarkers was based on two primary criteria: (i) they should be regularly measured, as indicated by low missingness ($<$~25\%), and (ii) they should be non-redundant, as indicated by modest correlation with other biomarkers. Highly correlated biomarkers lead to collinearity issues which are an unwanted, and unnecessary headache to handle (PCA can be used to deal with collinearity within our model \cite{twins}). For example, calcium, calcium by phosphate (ca x p) and corrected calcium are all highly correlated and so we picked only calcium to include, excluding the others to avoid collinearity.

We also considered additional variables for both associations and for sensitivity analysis. For sensitivity analysis we included the 5 next-most commonly measured biomarkers with missingness $<$~70\% (Section~\ref{sec:si:morevar}). For associations we included all biomarkers with missingness $<$~90\%.

\begin{table}
    \centering
    \caption{Biomarker Summary} \label{tab:si:biomarkers}
\resizebox{\columnwidth}{!}{%
\begin{threeparttable}
    \begin{tabular}{lllllll} \hline
        Full name & Variable & Units & Missing\tnote{1} & Group\tnote{2} & Validation\tnote{3} & Biomarker of\tnote{4}\\ \hline
        Albumin & albumin & g/L & 14.9\% & main & y & inflammation; survival \cite{Karaboyas2020-nx}; liver function \\ 
        Calcium & calcium & mmol/L & 11.1\% & main & y & electrolyte balance \\ 
        Chloride & chloride & mmol/L & 10.5\% & main & ~ & electrolyte balance  \\ 
        Glucose (random) & glucose r & mmol/L  & 22.6\% & main & ~ & metabolism  \\ 
        Hemoglobin & hemoglobin & g/L & 13.3\% & main & ~ & anemia \cite{Wilhelm-Leen2009-xe}  \\ 
        Magnesium & magnesium & mmol/L & 16.8\% & main & ~ & electrolyte balance  \\ 
        Phosphate & phosphate & mmol/L & 12.3\% & main & y & electrolyte balance\\ 
        Platelets & platelets & billion/L & 14.7\% & main & ~ & inflammation \cite{Vardon-Bounes2019-os}; mortality and sepsis \cite{Vardon-Bounes2019-os}  \\ 
        Potassium & potassium & mmol/L & 4.8\% & main & ~ & electrolyte balance; dialysis clearance  \\ 
        Creatinine (pre dialysis) & pre creatinine &  $\mu$mol/L & 14.3\% & main & y & protein metabolism; dialysis clearance; survival \cite{Karaboyas2020-nx}  \\ 
        Urea (pre dialysis) & pre urea & mmol/L & 9.5\% & main & y & protein metabolism; dialysis clearance  \\ 
        Sodium & sodium & mmol/L & 4.8\% & main & ~ & electrolyte balance  \\ 
        Total CO$_2$ & total co2 & mmol/L & 12.1\% & main & y & blood PH \cite{Wilhelm-Leen2009-xe}  \\ 
        White blood cell count & wbc & billion/L & 14.1\% & main & ~ & inflammation and immune function  \\ 
        Transferrin saturation  & \% sat & & 61.0\% & extended & ~ & anemia \\ 
        Aspartate aminotransferase\tnote{5} & ast & & 62.3\% & extended & ~ & liver function \cite{Oye-Somefun2021-ud,Nie2022-db}  \\ 
        Parathyroid hormone & pth & & 48.2\% & extended & ~ & electrolyte balance \\ 
        Total bilirubin\tnote{5} & total bilirubin & & 60.2\% & extended & ~ & liver function \cite{Nie2022-db}\\ 
        Alkaline phosphatase\tnote{5} & alk phos & & 74.3\% & associations only & ~ & liver function \\ 
        Alt\tnote{5} & alt & & 61.9\% & associations only & ~ & liver function \\ 
        Ferritin\tnote{5} & ferritin & & 78.7\% & associations only & ~ & anemia \\ 
        Hemoglobin A1C & hgba1c & & 86.0\% & associations only & ~ & metabolism \\ 
        Iron lvl\tnote{5} & iron lvl & & 79.3\% & associations only & ~ & anemia \\ 
        Total iron binding capacity & tibc & & 79.1\% & associations only & ~ & anemia  \\ 
        Total protein & total protein & & 65.4\% & associations only & ~ & protein metabolism; nutrition  \\ 
        Unsaturated iron binding capacity & uibc & & 82.4\% & associations only & ~ & anemia\\ 
        Urea reduction ratio & urr & & 25.7\% & associations only & ~ & dialysis clearance \\ \hline
    \end{tabular}
\begin{tablenotes}
\item[1] Total fraction of data missing, considering all measured time points and before pre-processing exclusions.
\item[2] The ``main'' group were used for modelling, the ``extended'' group were used in sensitivity analysis (Section~\ref{sec:si:sa}), and the ``associations only'' group were used exclusively for the association matrices.
\item[3] Only biomarkers with ``y'' (yes) are present in the validation dataset.
\item[4] Most of the biomarkers used have multiple interpretations, these are the pertinent ones.
\item[5] Log-transformed for normality.

\end{tablenotes}
\end{threeparttable}
}
\end{table}

\FloatBarrier

\subsection{Missing data} \label{sec:si:miss}
We imputed 6.9\% of entries in the main dataset using expectation-maximization as described elsewhere \cite{mallostasis}. We used single imputation, meaning that we inserted an estimate for each unknown value (excluding dead/censored individuals). In brief, we start by labelling each unknown value out of the 14 biomarkers. We then initialize the imputation process by first imputing all previous values (carry forward) then imputing backwards any values still missing using future values (carry backwards). Then at each iteration the algorithm iterates between fitting the model parameters and imputing the model expectation value for any value labelled as missing \cite{mallostasis}. The model iterates 5~times then ends (default). We did not impute censored or dead individuals.

The specific number of imputations for each of the 14 biomarkers is reported in Table~\ref{tab:si:imp}.

\begin{table}
    \centering
    \caption{Imputation Summary} \label{tab:si:imp}
\begin{threeparttable}
    \begin{tabular}{ll} \hline
        Biomarker & Imputed \tnote{$\dagger$} \\ \hline
        sodium & 2.2\% ($N=347$) \\
        potassium & 2.4\% ($N=375$) \\
        pre urea & 3.9\% ($N=609$) \\
        chloride & 4.4\% ($N=675$) \\
        calcium & 4.7\% ($N=720$) \\
        phosphate & 5.2\% ($N=791$) \\
        total co2  & 5.6\% ($N=856$) \\
       albumin & 6.3\% ($N=957$) \\
       magnesium & 6.5\% ($N=986$) \\
       pre creatinine & 10.4\% ($N=1520$) \\
       glucose r & 10.7\% ($N=1562$) \\
       hemoglobin & 12.9\% ($N=1841$) \\
       wbc & 13.7\% ($N=1947$) \\
       platelets & 14.1\% ($N=1995$) \\ \hline
    \end{tabular}
\begin{tablenotes}
\item[$\dagger$] Percentage is fraction imputed divided by observed, $N$ is number of imputed values.
\end{tablenotes}
\end{threeparttable}
\end{table}

\subsection{Validation dataset} \label{sec:si:validation}
We used a large, cross-sectional dataset of Canadian patients to validate our results. The dataset had only 6 of the 14 blood tests used for our analysis, each measured at the initiation of dialysis. If the $z$ (natural variables) represent underlying biology then we should be able to estimate their effects given any set of biomarkers (although the accuracy will depend on the relationship between the biomarkers and the underlying biology captured by $z$). This is particularly important for a clinical setting where there may be limited data available. This motivates us to use an emulator to estimate each $z$ using the 6 available biomarkers (the emulation becomes exact if we could use all 14 biomarkers). The estimate from the emulator is denoted with a hat, $\hat{z}$.

\subsubsection{Emulator} \label{sec:si:emulator}
\begin{figure*}[!ht] 
     \centering
        \includegraphics[width=0.75\textwidth]{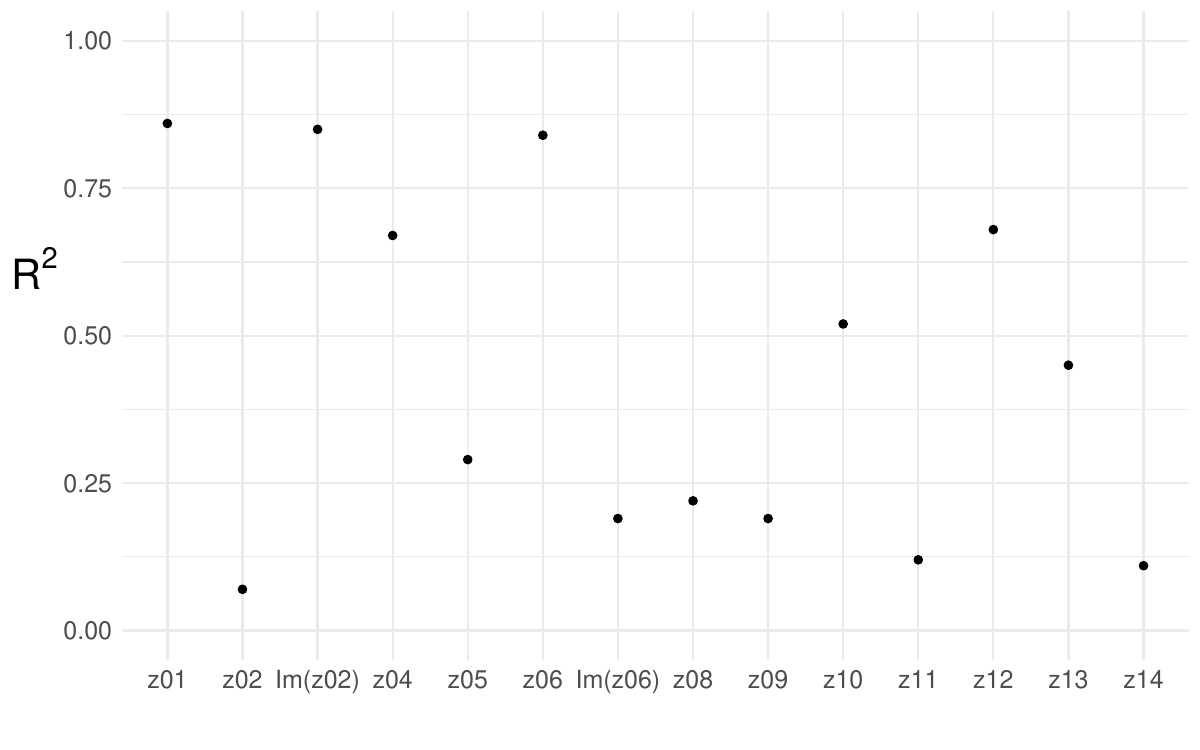}
    \caption{Emulator accuracy. Training $R^2$ for the emulated natural variables using the 6 biomarkers available for validation. We see good accuracy for some $z$, particularly $z_1$, whereas others are quite inaccurate (e.g.\ $z_2$). }
    \label{fig:si:validation_r2}
\end{figure*}

Each emulator is a linear model that predicts a particular $z$ using the 6 available biomarkers (via ordinary linear regression). We used the main dataset to train the emulator. For example, $\hat{z}_1 = 4.46 - 0.0046\cdot\text{creatinine} -0.080\cdot\text{albumin} + 0.031\cdot\text{urea} - 0.032\cdot\text{total co2} + 0.13\cdot\text{phosphate} + 0.23\cdot\text{calcium}$ is our emulator approximation of $z_1$. The specific transformations are available as CSV files on our GitHub page (\url{https:// github.com/GlenPr/stochastic_finite-difference_model}). This includes truncated emulators, such as $\hat{z}_1\approx 2.214-0.00431\times\text{creatinine} -0.0764\times\text{albumin}$ (which is a good approximation, $R^2=0.81$). The emulator accuracy varied considerably across the natural variables, $z$, as illustrated using the main dataset $R^2$ in Figure~\ref{fig:si:validation_r2}. In the main text we focused on $z_1$, which had the highest emulator accuracy.

\FloatBarrier

\subsubsection{Validation dataset pre-processing}
To avoid duplicates, we excluded all Nova Scotian individuals (potentially overlapping with our longitudinal analysis) from the validation set. To avoid coding errors in the biomarker values we also dropped the following values:
\begin{enumerate}
    \item Albumin $>$ 60~g/l
    \item Creatinine $> 2000~\mu$mol/l
    \item Urea $>$ 100~mmol/l
    \item Total CO$_2$ $>$ 40~mmol/l
    \item Calcium $>$ 3.5~mmol/l
    \item Phosphate $>$ 5~mmol/l.
\end{enumerate}

When computing survival we dropped 169 individuals whom died before 3~months for consistency with our main dataset. Note that the effects of the study window on the main dataset were minor and did not affect our network and hence is unlikely to affect our study conclusions, are shown in Section~\ref{sec:si:sa}.

\FloatBarrier

\subsection{Simulation} \label{sec:si:sim}
For the main simulation (Figure~5), we simulated 10000 synthetic individuals using the parameter estimates from the SF model (Eq.~1). Survival, censorship and transplant used a Weibull hazard with time-dependent proportional hazard (Eq.~5), for which we used an optimized model using $z$ which included linear terms for baseline age, $z_1$-$z_6$ and $z_{11}$ as described below (including linear and quadratic terms in the raw biomarkers gave similar results, not shown). Starting values and covariates (baseline age, sex, DM status and sampling dates) were sampled directly from the population (with replacement). For unknown sampling dates due to leaving the study we imputed 6~weeks. We simulated for 43 time steps, the same as was recorded in the data ($\sim$4.96~years); each time step was approximately 6~weeks. Individuals were censored using a time-dependent proportional hazard models for survival, censorship and transplant at each time step. The parameters needed for the simulation are available on the GitHub page \url{https:// github.com/GlenPr/stochastic_finite-difference_model}.

The simulation uses Euler's method with the step size approximately 6~weeks (exact for an individual if known), starting from known initial values and with known initial covariates. We used parameter estimates from the main text. Individual trajectories are generated for 5~years. The natural variables are then generated from the simulated biomarkers using $P^{-1}$ from the eigen-decomposition of the network. Afterwards we impose events using time-to-event statistics: death, censorship or transplant. This was performed by stepping through the simulated data and performing accept-reject sampling for events, whichever event happened first takes precedence (rejected if it happened after the end of the time step). Time-to-event statistics assumed a time-dependent Weibull distribution with proportional hazard term. Predictors were selected using the likelihood ratio test as follows. We fit using all linear powers of each $z$, baseline age, DM status and sex status, then rejected all non-significant terms at $p=0.05$ based on the likelihood ratio test (using \texttt{anova.coxph} in \texttt{R}). The final models are reported in Table~\ref{tab:si:simsurv} (and on the GitHub page).

\begin{table}[!ht]
    \centering
\begin{threeparttable}
\caption{Simulation Survival Parameters} \label{tab:si:simsurv}
    \begin{tabular}{llll} \hline
        Variable & Outcome & Type & Value \\ \hline
        z01 & death & PH\tnote{*} & 0.370 \\ 
        z02 & death & PH & 0.523 \\ 
        Im(z02) & death & PH & -0.214 \\ 
        z04 & death & PH & 0.0720 \\ 
        z05 & death & PH & 0.494 \\ 
        z06 & death & PH & -0.617 \\ 
        z11 & death & PH & -0.209 \\ 
        Baseline age & death & PH & 0.0269 \\ 
        Shape, $\nu$ & death & Weibull & 1.580 \\ 
        Base hazard, $h_0$ & death & Weibull & 0.01021 \\ 
        Baseline age & censor & PH & -0.0167 \\ 
        Shape, $\nu$ & censor & Weibull & 2.42 \\ 
        Base hazard, $h_0$ & censor & Weibull & 0.0768 \\ 
        z01 & transplant & PH & -0.248 \\ 
        z04 & transplant & PH & 0.0118 \\ 
        z05 & transplant & PH & -0.384 \\ 
        Im(z06) & transplant & PH & -0.332 \\ 
        Baseline age & transplant & PH & -0.0446 \\ 
        Shape, $\nu$ & transplant & Weibull & 1.088 \\ 
        Base hazard, $h_0$ & transplant & Weibull & 0.2735 \\ \hline
    \end{tabular}
\begin{tablenotes}
\item[*] PH: proportional hazard, defined as the $\vec{\beta}$ in $h= \nu h_0 t^{\nu-1} e^{\vec{\beta}^T\vec{x}}$.
\end{tablenotes}
\end{threeparttable}
\end{table}

\FloatBarrier

\subsection{Fit quality} \label{sec:si:fit}
Our goal is to capture the behaviour of the population \textit{in silico} such that we can analyze their health quantitatively. This means that the central diagnostic is ensuring that we have realistic population-level behaviour for our model. This is fundamentally a fit quality control.

\begin{figure}[H] 
     \centering
        \includegraphics[width=\textwidth]{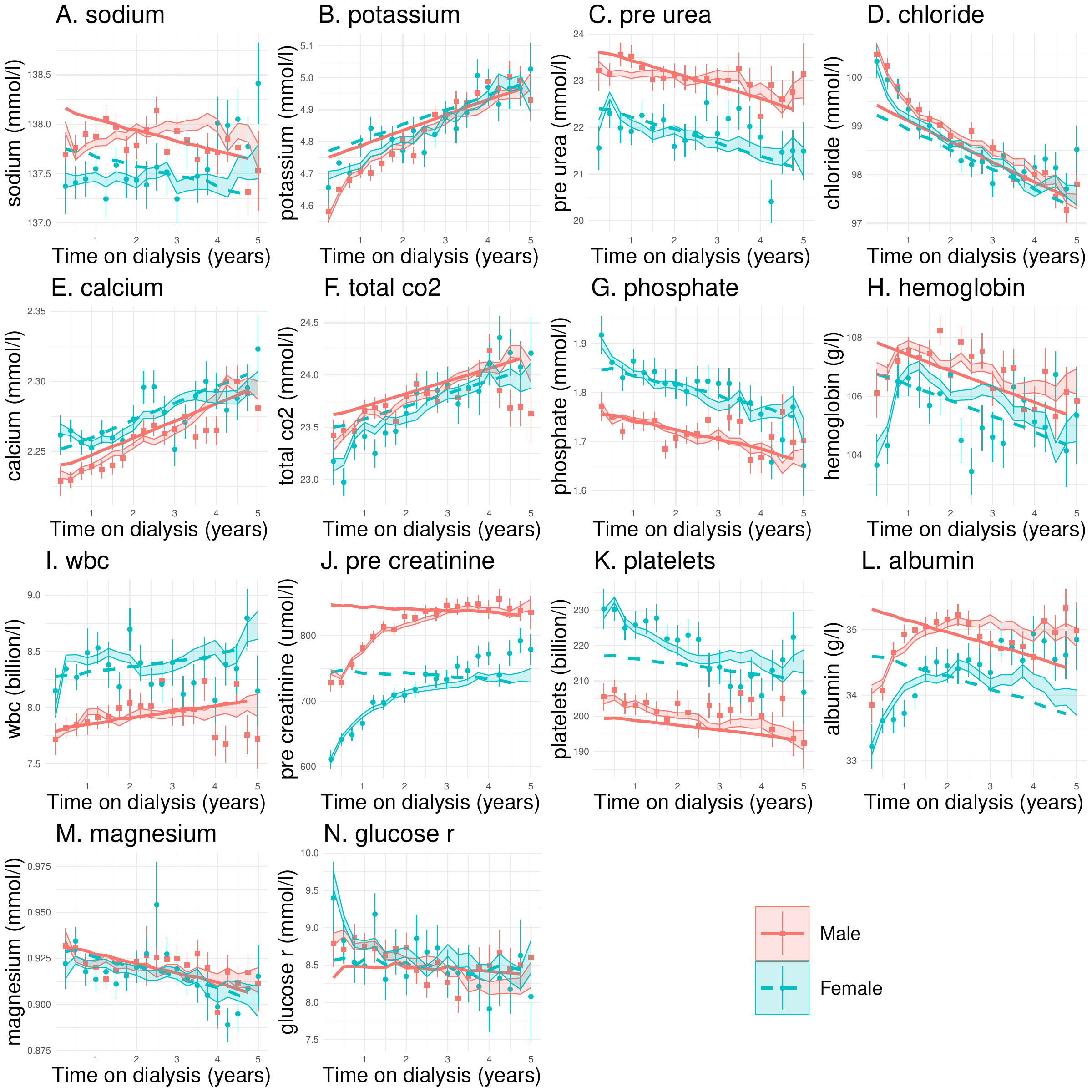}  
    \caption{`Fits' of simulated data to raw biomarkers (mean $\pm$ standard error). Points: data grouped by 3 month bins. Bands: simulated data. Lines: dynamical equilibrium ($\mu(t)$). We find good agreement between the simulated and real data.}
    \label{fig:yfits}
\end{figure}

In Figure~\ref{fig:yfits} we compare a simulated population to the observed data, stratified by sex (sim details are in Section~\ref{sec:si:sim}). The simulation includes dynamical behaviour according to the SF model (Eq.~1), and survival, censorship and transplant according to Weibull statistics (Eq.~5 and Table~\ref{tab:si:simsurv}). We see excellent agreement between the real data (points) and the simulation (bands). The lines are $\mu(t)$ and represent the homeostatic set point, which looks reasonable. Creatinine, albumin, and hemoglobin all showed an interesting transient period at the beginning of the study. Our model has no difficulty replicating this behaviour, since it permits a transient phase prior to the steady-state.

\begin{figure}[H] 
     \centering
        \includegraphics[width=.8\textwidth]{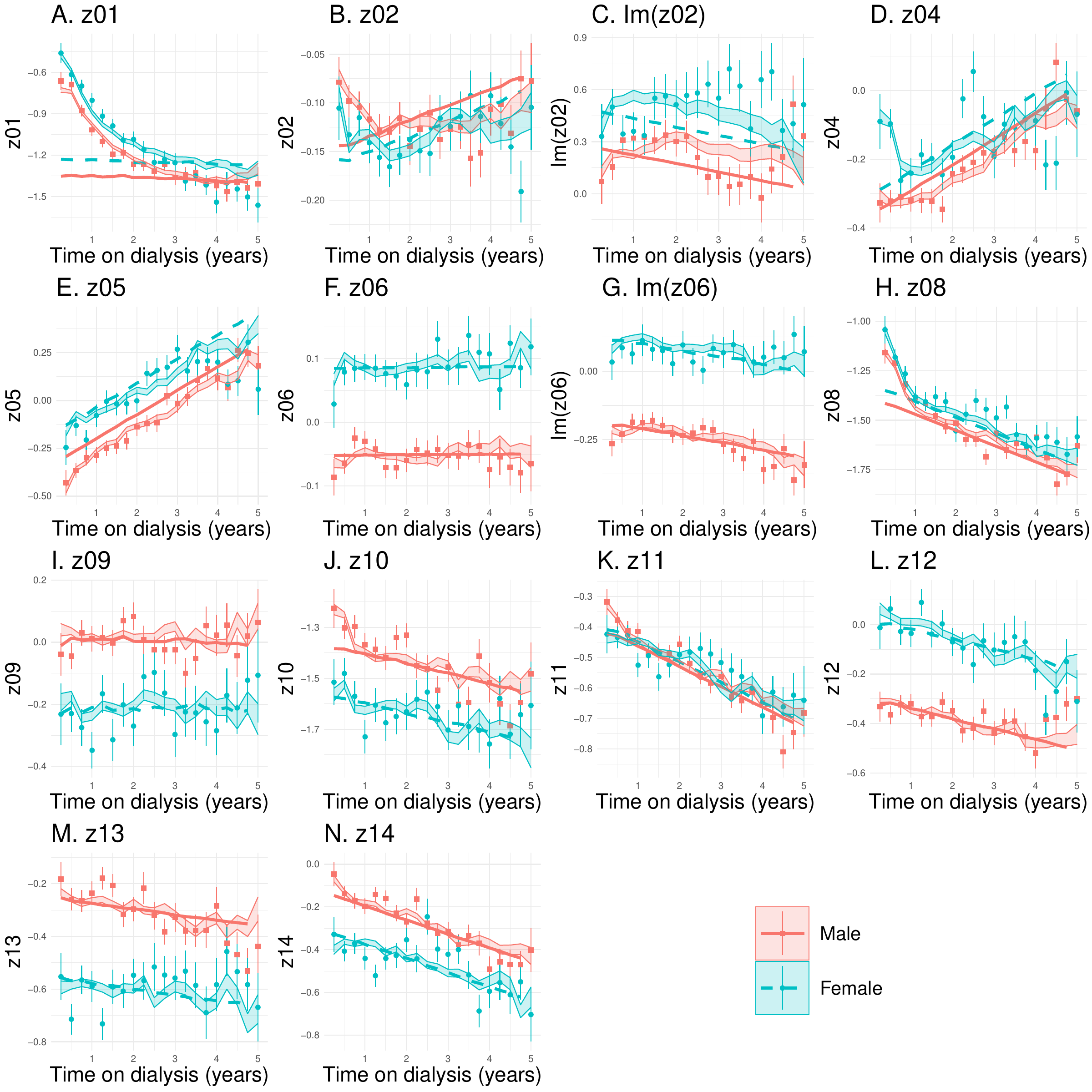}  
    \caption{`Fits' of simulated data to natural variables (mean $\pm$ standard error). Points: data grouped by 3 month bins. Bands: simulated data. Lines: dynamical equilibrium ($\mu(t)$). We find good agreement between the simulated and real data.}
    \label{fig:zfits}
\end{figure}

In Figure~\ref{fig:zfits} we again see good agreement between the real data (points) and the simulation (bands), this time for the natural variables. We also see reasonable steady-state behaviour. The correct steady-state behaviour is drift parallel to $\mu(t)$ with a small lag of size $-\mu_t/|\lambda|$ (Eq.~3 with $\lambda t\to -\infty$). 

\begin{figure}[!ht] 
     \centering
        \includegraphics[width=\textwidth]{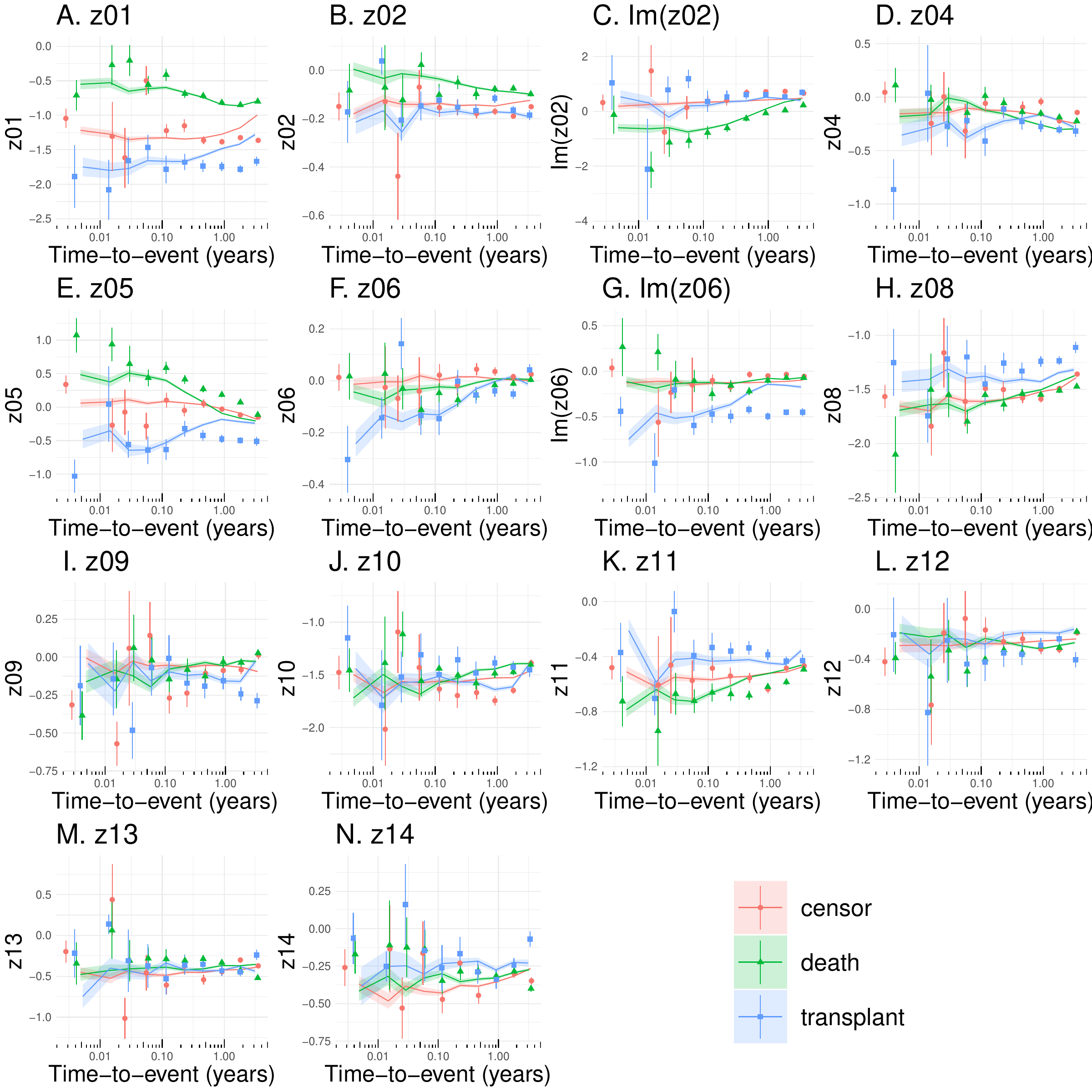}
    \caption{Full terminal decline plots with simulation (mean $\pm$ standard error). Points are real data, binned by cuts (0,2,4,...,28 = 256)~weeks. Bands are simulated data.}
    \label{fig:si:ttd}
\end{figure}

Finally, we include the terminal decline plots in full, Figure~\ref{fig:si:ttd}. These are qualitatively identical to the real data, and fit reasonably well considering that they aren't fit directly to any of the data.

\FloatBarrier

\subsection{Risk dynamics}
In Section~3.2 we present the results from a second model for the data based on risk strata. Here we provide additional details. We considered a second model for two reasons: (i) as a sensitivity analysis on our interpretation of the main model, and (ii) as an alternative perspective for those whom prefer to think in terms of risk groups. Our secondary model is to first discretize each $z$ value into risk strata using the baseline tertiles to convert into low, normal and high--risk groups. We then observe the transition behaviour between risk groups during the study period of 3~months to 5~years.

The transition time between states is estimated using start-stop formatting \cite{Moore2016-rh} to generate a survival curve. We observed exponential behaviour and hence employed a parametric, exponential estimator of the underlying hazard \cite{eha}. The transition time, $\tau$, is then defined as the inverse of the estimated transition hazard.

\FloatBarrier

\subsection{Survival model diagnostics}
For survival prediction, we considered three parameteric models: exponential, Weibull and Gompertz,  using the \texttt{eha} package \cite{eha}. Weibull fit best (others not shown). Variable predictors were permitted to vary over time using start-stop formatting \cite{Moore2016-rh}. Predictors included demographical variables (age, sex and DM status), raw biomarkers, natural variables, and principal components, depending on the specific model (as described in the respective sections). In general, we found that the Weibull distribution fit well with the proportional hazard assumption. The diagnostics for this fit are included in this section.

The Weibull distribution assumes a hazard of form
\begin{align}
    h(t|\vec{x}(t)) &= \nu h_0 t^{\nu-1}e^{\vec{\beta}^T\vec{x}(t)}, \label{eq:si:h}
\end{align}
where $\nu$, $h_0$, and $\vec{\beta}$ are model parameters to be estimated and $\vec{x}$ is the start-stop encoded set of predictors (biomarkers, natural variables, etc).

\begin{figure}[H] 
     \centering
        \includegraphics[width=.75\textwidth]{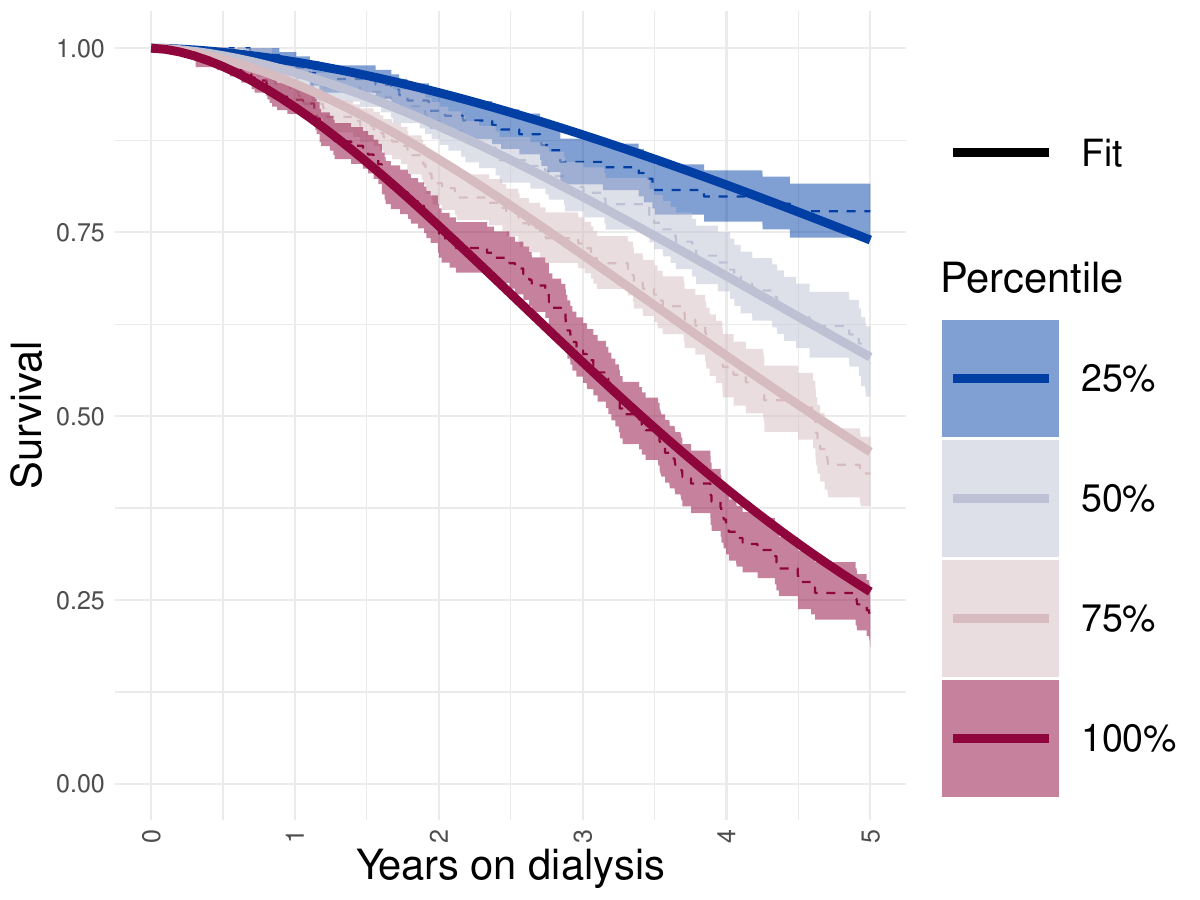}  
    \caption{The Weibull distribution with time-dependent proportional hazard for $z_1$ (solid lines) fits the data excellently (bands). Bands and dashed line are quartiles, solid lines are the fit \cite{eha} (band width is standard error). Hazard model: $h = \nu h_0 t^{\nu-1}e^{\beta z_1(t)}$ (Eq.~\ref{eq:si:h}).}
    \label{fig:si:simfit}
\end{figure}

The proportional hazard assumption states that the hazard has form
\begin{align}
    h &= h_0 e^{\vec{\beta}^T\vec{x}} \label{eq:si:ph}
\end{align}
where $h$ is the hazard, $h_0$ is the baseline hazard (excluding covariates), $\vec{x}$ is a vector of covariates (which each individual has), and $\vec{\beta}$ are a set of parameters to be estimated. Observe that if $\vec{x}$ is binary then the hazard is fully flexible whereas if $\vec{x}$ takes multiple values then the hazard is constrained by the form of Eq.~\ref{eq:si:ph}. In Figure~\ref{fig:si:simfit} we see that the time-dependent Weibull model fits our data very well, as demonstrated by $z_1$.

\begin{figure*}[!ht] 
     \centering
        \includegraphics[width=\textwidth]{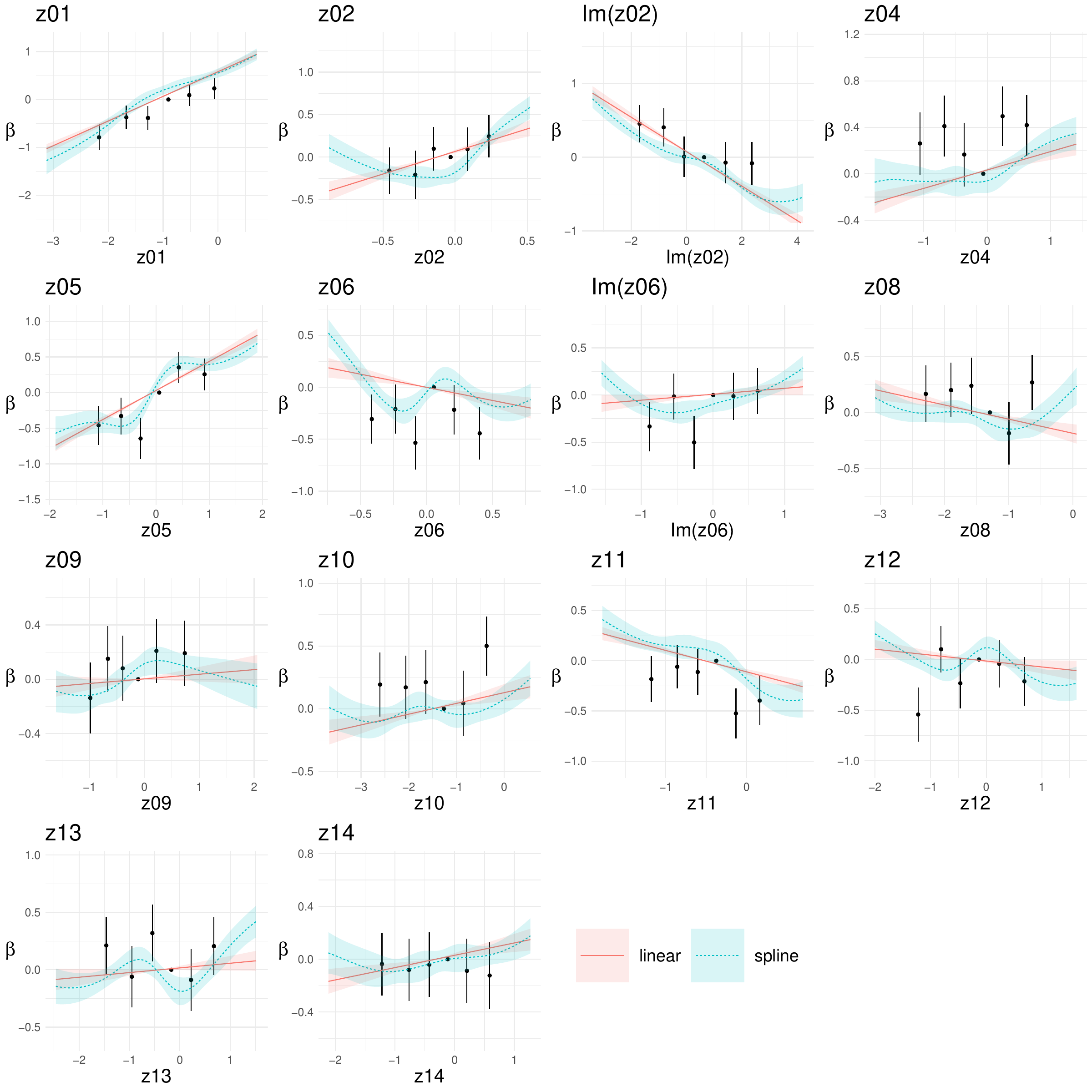}  
    \caption{The natural variables, $z$, satisfy the proportional hazard assumption. Points are one-hot encoded quantiles, solid red line is the linear proportional hazard model, and dashed blue line is a basis spline with 7 degrees of freedom. A linear relationship indicates that the proportional hazard assumption is valid --- the y-intercept is unimportant since it can always be absorbed into the baseline hazard ($h_0$). The central (4th) quantile is used as the reference ($\beta\equiv0$), uncertainty in which cannot be estimated but will shift the entire set up or down by a constant amount. The x-axis is the average value within each quantile (error bars are typically too small to see). Bands and error bars are standard errors.}
    \label{fig:phz}
\end{figure*}

\begin{figure*}[!ht] 
     \centering
        \includegraphics[width=\textwidth]{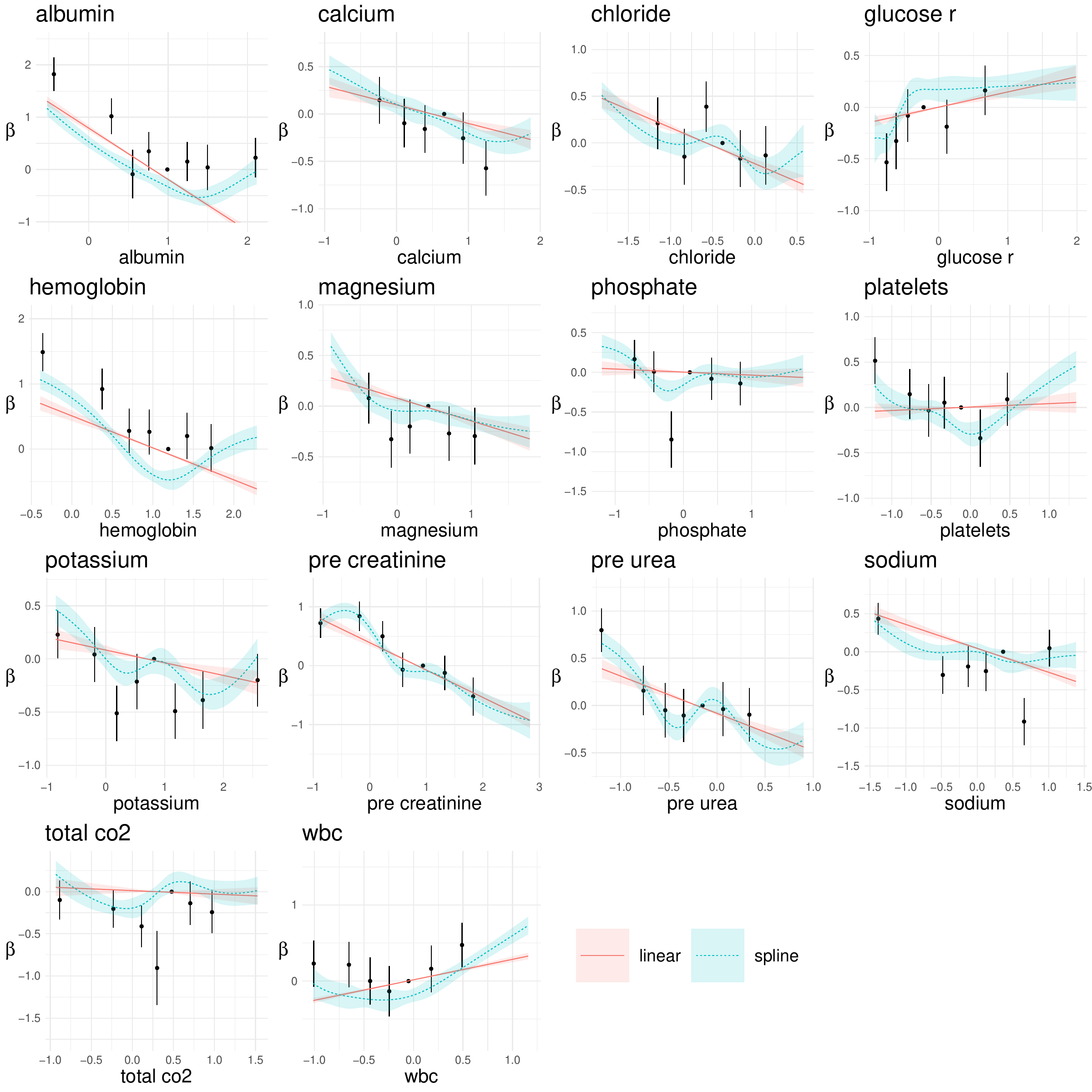}  
    \caption{The raw biomarkers variables do not satisfy the proportional hazard assumption. Points are one-hot encoded quantiles, solid red line is the linear proportional hazard model, and dashed blue line is a basis spline with 7 degrees of freedom. A well-fitted line would indicate that the proportional hazard assumption is valid. Several of the biomarkers have saturating, `J'-shaped curves: albumin, glucose, platelets, hemoglobin, wbc, and creatinine. Biomarkers have been standardized to the first measurement (unit variance, zero mean). The central (4th) quantile is used as the reference ($\beta\equiv0$), uncertainty in which cannot be estimated but will shift the entire set up or down by a constant amount. The x-axis is the average value within each quantile (error bars are typically too small to see). Bands and error bars are standard errors.} 
    \label{fig:phy}
\end{figure*}

We also tested the linear proportional hazard assumption, $\ln{(h)}\propto \beta x$ for each biomarker and natural variable. We compared two non-linear proportional hazard models to the linear model. The first is simply one-hot encoded quantiles (hextiles) with the central (4th) as reference (hence no error bar). The second is a 7~degree of freedom basis spline (we used the \texttt{splines2} package \cite{splines2}). The linear proportional hazard assumption is well-founded if both the quantiles (points) and splines (dashed blue lines) are monotonic and approximately linear. (Note that a universal shift of the y-axis doesn't matter since it can be absorbed into the baseline hazard e.g.\ the quantiles are occasionally shifted up or down relative to the splines, such as in hemoglobin.) In general, we observe that the natural variables, Figure~\ref{fig:phz}, are quite close to linear or are close to 0 and therefore not survival predictors ($z_1$-$z_5$ were the dominant survival predictors). In contrast, many of the raw biomarkers were clearly non-linear, Figure~\ref{fig:phy}.

The key difference is that several of the raw biomarkers have saturating, `J'-shaped curves (also called `hockey stick--shaped'). In particular, low albumin is highly predictive of death but high albumin has no discriminating power (high versus very high have the same risk). The same can be said for hemoglobin. White blood cell count (wbc) has the opposite curve where high is predictive but low has no predictive power. In contrast, the $z$ are all either monotonic and nearly linear; or are weak survival predictors ($z_6$ and higher, look at scale). This indicates that the linear, proportional hazard assumption is probably adequate for the $z$ but is unlikely to be sufficient for several of the raw biomarkers. 

It is remarkable that creatinine and albumin have opposing saturation effects, which perfectly cancel in $z_1$ ($z_1$ is primarily a weighted sum of negative creatinine minus albumin $R^2=0.81$; Section~\ref{sec:si:validation} explains). This supports the proposed connection between the natural variable dynamics and survival.

\subsubsection{Expected hazard}
The dynamical model modifies the empirical hazard since individuals are evolving over time (Eq.~6). Here we show the math behind Eq.~6 using the more general multivariate version, which ends up simplifying to Eq.~6 for our dataset. The key is the observation that at any given time our model is normally-distributed and the hazard function has no memory (in contrast to the survival function).

Let $\vec{y}$ be a multivariate normal random variable with mean $\langle \vec{y} \rangle$ and covariance $\boldsymbol{C}$. The expectation of the proportional hazard is
\begin{align}
    \langle e^{\vec{\beta}^T\vec{y}} \rangle &= \int_{y_1} \dots \int_{y_p} \bigg|2\pi \boldsymbol{C}\bigg|^{-1/2} \exp{\bigg( (\vec{y}-\langle \vec{y} \rangle)^T\boldsymbol{C}^{-1}(\vec{y}-\langle \vec{y} \rangle) + \vec{\beta}^T\vec{y} \bigg)} \prod_i dy_i.
\end{align}
Since $\boldsymbol{C}$ is symmetric and positive definite, it can be eigen-decomposed into $\boldsymbol{C} = \boldsymbol{U}\boldsymbol{D}\boldsymbol{U}^T$ where $\boldsymbol{U}\boldsymbol{U}^T=I$ and $D_{ij} = d_i^2\delta_{ij}$ is diagonal. Hence we have
\begin{align}
    \langle e^{\vec{\beta}^T\vec{y}} \rangle &= \bigg|2\pi \boldsymbol{C}\bigg|^{-1/2}\int_{y_1} \dots \int_{y_p}  \exp{\bigg( (\boldsymbol{U}^T\vec{y}-\langle \boldsymbol{U}^T\vec{y} \rangle)^T\boldsymbol{D}^{-1}(\boldsymbol{U}^T\vec{y}-\boldsymbol{U}^T\langle \vec{y} \rangle) + (\boldsymbol{U}^T\vec{\beta})^T(\boldsymbol{U}^T\vec{y}) \bigg)} \prod_i dy_i.
\end{align}
Define $\tilde{\vec{y}} \equiv \boldsymbol{U}^T\vec{y}$, and $\tilde{\vec{\beta}} \equiv \boldsymbol{U}^T\vec{\beta}$ then we have a set of decoupled normal random variables,
\begin{align}
    \langle e^{\vec{\beta}^T\vec{y}} \rangle &= \bigg|2\pi \boldsymbol{C}\bigg|^{-1/2}\int_{y_1} \dots \int_{y_p}  \exp{\bigg( \sum_i (\tilde{y}_i-\langle \tilde{y}_i \rangle)^TD^{-1}_{ii}(\tilde{y}_i-\langle \tilde{y}_i \rangle) + \sum_i \tilde{\beta}_i\tilde{y}_i \bigg)} \prod_i dy_i \nonumber \\
    &= \bigg|2\pi \boldsymbol{C}\bigg|^{-1/2}\int_{y_1} \dots \int_{y_p}  \exp{\bigg( \sum_i (\tilde{y}_i-\langle \tilde{y}_i \rangle - \tilde{\beta}_iD_{ii})^TD^{-1}_{ii}(\tilde{y}_i-\langle \tilde{y}_i \rangle  - \tilde{\beta}_iD_{ii}) + \sum_i \tilde{\beta}_i\langle \tilde{y}_i \rangle + \frac{1}{2} \sum_i \tilde{\beta}_i^2 D_{ii}  \bigg)} \prod_i dy_i \nonumber \\
    &= \bigg|2\pi \boldsymbol{C}\bigg|^{-1/2}\int_{y_1} \dots \int_{y_p}  \exp{\bigg( \sum_i (\tilde{y}_i-\langle \tilde{y}_i \rangle - \tilde{\beta}_iD_{ii})^TD^{-1}_{ii}(\tilde{y}_i-\langle \tilde{y}_i \rangle  - \tilde{\beta}_iD_{ii}) \bigg)} \prod_i dy_i \cdot \exp{\bigg( \sum_i \tilde{\beta}_i\langle \tilde{y}_i \rangle + \frac{1}{2} \sum_i \tilde{\beta}_i^2 D_{ii}  \bigg)}  \nonumber \\
\langle e^{\vec{\beta}^T\vec{y}} \rangle &= \exp{\bigg( \vec{\beta}^T\langle \vec{y} \rangle + \frac{1}{2} \vec{\beta}^T\boldsymbol{C}\vec{\beta}  \bigg)} \label{eq:mvexp}
\end{align}
where in the last line I've simply transformed back to $\vec{y}$ and used the fact that the first term was simply the expectation of a multivariate normal random variable which is $1$ due to the normalization constraint. The univariate (marginal \cite{cookbook2012}) case is simply Eq.~6.

The covariance, $\boldsymbol{C}$, is unlikely to matter for our dataset. We observed that the $z$ were correlated through the noise, forming modules. Within each module, most of the $|\beta_j|\approx 0$ were small with typically at most one being much larger from zero. Hence while Eq.~\ref{eq:mvexp} indicates that the covariance modifies the hazard, within the data we saw that $\boldsymbol{C}$ was block-diagonal and within each block there was only one large $|\beta_j|$. This means that $\vec{\beta}^T\boldsymbol{C}\vec{\beta}  \approx \beta_j^2C_{jj}$ within each module, where $j$ is the dominant survival predictor (the leading order correction would be to sum over the module indices $j,k \in \text{module}$ giving $\sum_{j,k} \beta_jC_{jk}\beta_k$).

\FloatBarrier

\subsection{Sensitivity analysis} \label{sec:si:sa}
We test how sensitive our results are to variations in the dataset used. In particular, which groups of individuals, which biomarkers and which study window. Since the key analysis step is estimating the network, our primary interest is in how much the network parameterization changes if we change the dataset used to estimate it. If the network changes little, then it follows that the eigenvectors will not change and thus the natural variables will also not change. No change would also be evidence that our results are generalizable to new datasets. 

We find that the networks ($\boldsymbol{W}$) change little as we vary the dataset used. Importantly, the network appears to be robust to adding or subtracting individuals or adding more variables. This likely reflects the robustness of the underlying estimator, which is linear regression with weights close to unity. 
However, the dynamical equilibrium ($\vec{\mu}$) does show non-trivial differences between groups, such as diabetics vs non-diabetics, and males vs females. This suggests that the interactions between variables are more universal across biological conditions than are the steady-state values. That said, it is difficult to estimate the dynamical equilibrium parameters since that requires forecasting based on the drift rate and interaction network (i.e.\ inverting $\boldsymbol{W}$) which leads to larger uncertainties in $\vec{\mu}$ (as compared to $\boldsymbol{W}$). 

\subsubsection{Fitting by sex and diabetes status}
We fit our model to different groups of individuals. We consider 3 primary conditions across which people vary: sex, diabetes status, and frailty status (using the clinical frailty scale of aging health, CFS \cite{Rockwood2020-oa}). After pre-processing the dataset, we grouped individuals and separately fit to each group. The result from the main text is referred to herein as the ``base'' result. In the present section, we simultaneously compare males vs females and diabetics vs non-diabetics.

\begin{figure}[H] 
     \centering
        \includegraphics[width=0.9\textwidth]{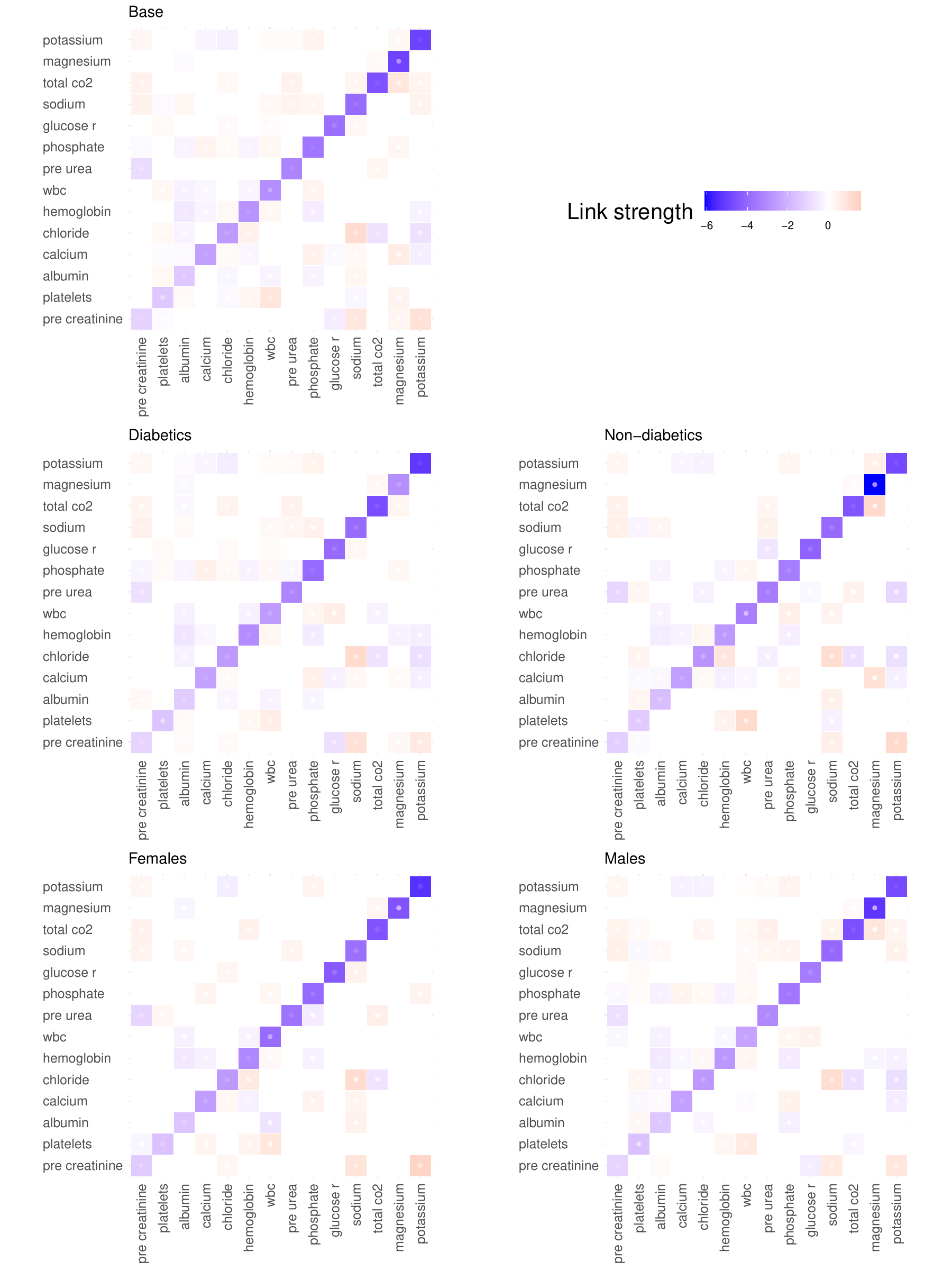}  
    \caption{Network estimates stratified by sex and diabetes status versus base network. The overall structures are similar to the main result (``Base''). See Figure~\ref{fig:si:Wpardmsex} for a direct comparison of links. Non-significant links are whited out at $p>0.05$.}
    \label{fig:si:Wdmsex}
\end{figure}

First we consider the network estimate, Figure~\ref{fig:si:Wdmsex}. There do not appear to be any major differences, although there are clearly some differences in terms of which links are statistically significant (non-significant links are whited out). This may simply reflect a loss of statistical significance due to the reduced number of individuals, which is roughly $1/2$ in all cases.

\begin{figure}[H] 
     \centering
        \includegraphics[width=\textwidth]{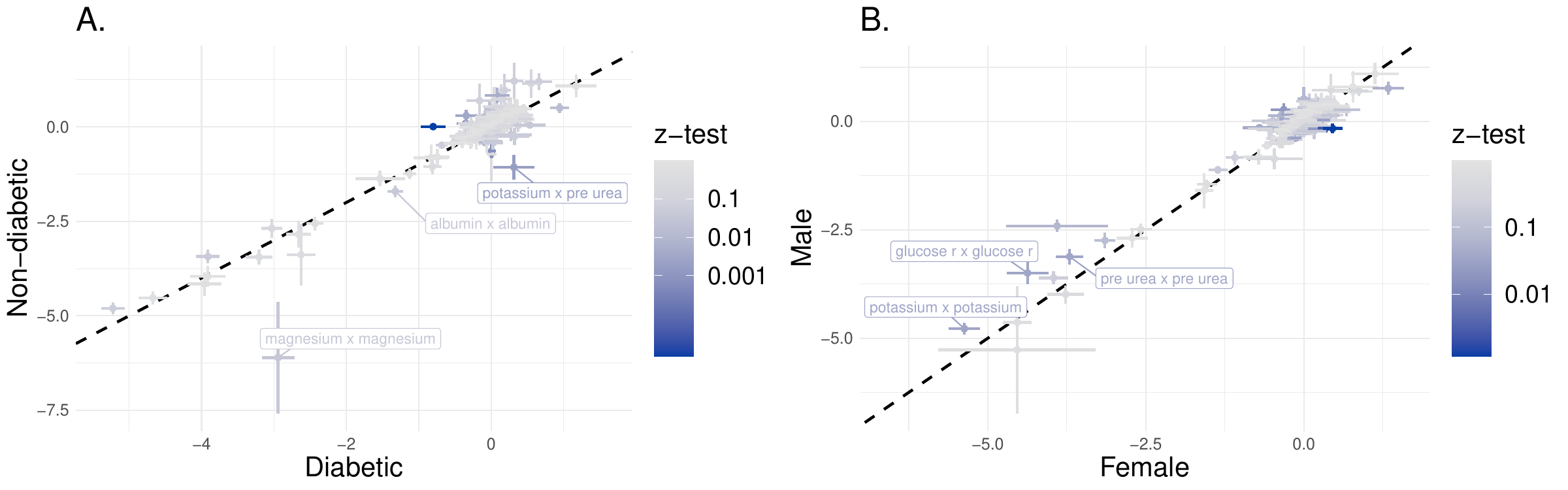}  
    \caption{Network link estimates by \textbf{A.} diabetes status, and \textbf{B.} sex. Significantly-different links with coefficients great in magnitude than $1$ have been labelled ($p < 0.05$). Dashed line indicates $x=y$.}
    \label{fig:si:Wpardmsex}
\end{figure}

A more direct comparison is to compare the network coefficients directly, Figure~\ref{fig:si:Wpardmsex}. The network coefficients quantify the strength of the (auto-regressive) relationships (i.e.\ links). Since each variable was normalized at the baseline, coefficients near $0$ can be considered unimportant (because all variables are on roughly the same scale). We see that for both group comparisons the links are very strongly correlated. Most links are near $0$ with a few diagonal links that are large and negative. The differences do not appear to be major overall, but are visually larger based on diabetes status as opposed to sex. The similarities between networks ensures that the natural variables will be similar as well.

\begin{figure}[H] 
     \centering
        \includegraphics[width=\textwidth]{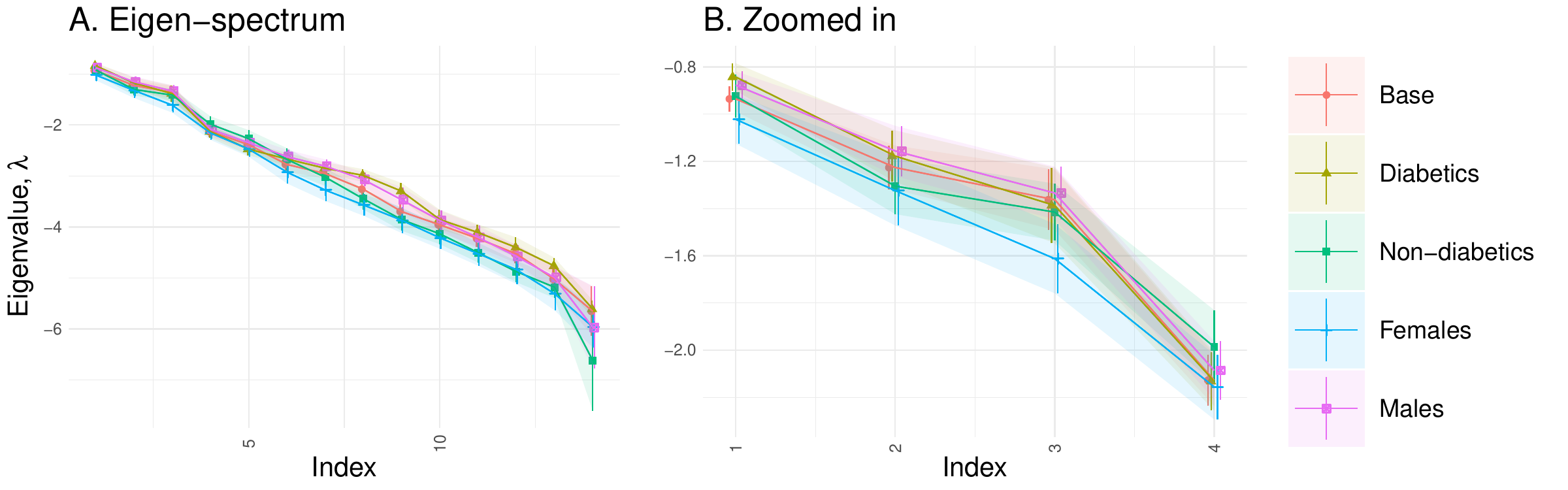}  
    \caption{Comparison of eigenvalue estimates by diabetes status and sex. We see no significant differences in the eigen-spectrum. \textbf{A.} spectrum. \textbf{B.} spectrum zoomed into the first 4 eigenvalues (lowest stability/resilience).}
    \label{fig:si:eigdmsex}
\end{figure}

We compared the eigenvalues in Figure~\ref{fig:si:eigdmsex}. In our model, the eigenvalues capture stability against short-term stressor events. Large magnitude, negative eigenvalues are the most stable (positive are unstable). The similarity between each of these groups implies that there is no loss of resilience in the diabetics nor difference between the sexes.

\begin{figure}[H] 
     \centering
        \includegraphics[width=\textwidth]{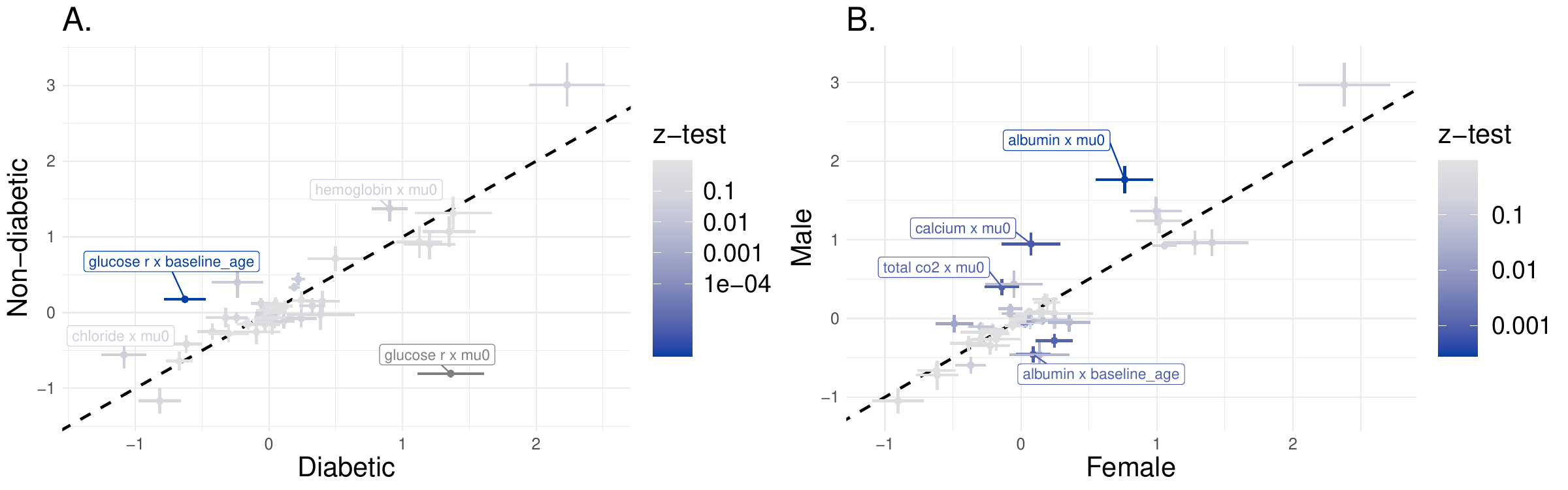}  
    \caption{Comparison of dynamical equilibrium estimates by diabetes status and sex. The differences appear more pronounced in the equilibrium estimates as opposed to the network links (Figure~\ref{fig:si:Wpardmsex}). Labelled points are either significantly different at $p<0.001$, or significant at $p<0.05$ and at least 1 in magnitude. Baseline age has been scaled to units of 40~years. ``mu0'' is the intercept parameter, $\mu_0$. Dashed line indicates $x=y$.}
    \label{fig:si:lambdadmsex}
\end{figure}

Finally, we compare the dynamical equilibrium estimates, $\vec{\mu}$ in Figure~\ref{fig:si:lambdadmsex}. The differences are visually larger than those of the networks (Figure~\ref{fig:si:Wpardmsex}). As expected, glucose is higher in diabetics ($\mu_0$). Diabetics on dialysis are purposefully maintained at higher glucose to avoid dangerously low blood sugar levels. This was the largest difference, although there were many smaller but still significant differences between sexes and diabetes status'. Our experience from prior work with this model has been that $\boldsymbol{W}$ is less sensitive to such status variables while $\vec{\mu}$ is much more sensitive. This general statement appears to be true here as well. By including the binarized status variables in the fit we should be able to mitigate this effect, by permitting status-specific equilibrium values (which we did). 

\subsubsection{Fitting by frailty status}
Many patients had a Clinical Frailty Scale (CFS) score recorded at baseline. CFS is a measure of worsening health associated with aging \cite{Rockwood2005-gl}. We used the modern grouping \cite{Rockwood2020-oa}, simplified into 5 groups: fit (CFS $\leq 2$), managing (CFS = 3), very mild frailty (CFS = 4), mild frailty (CFS = 5) or frail (CFS $\geq 6$). Only 529 individuals had baseline CFS recorded and hence data were limited to only about 100 individuals per group (50 individuals can be enough to get reasonably accurate parameter estimates \cite{mallostasis}.) We compared the fitted parameter values for these 5 groups.

\begin{figure}[H] 
     \centering
        \includegraphics[width=0.9\textwidth]{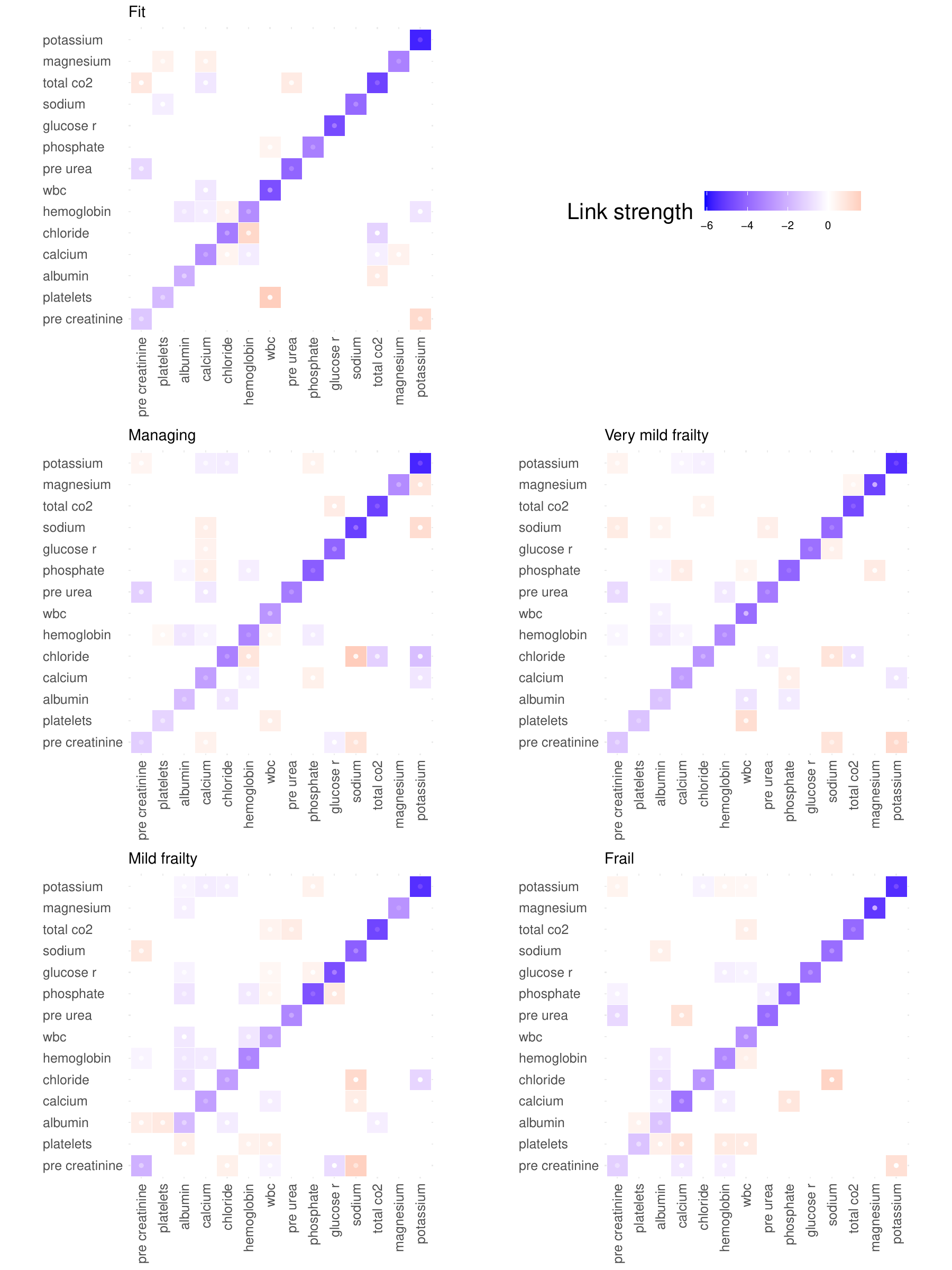}  
    \caption{Network estimates stratified by frailty status. The overall structures are similar to the main result (``Base''). See Figure~\ref{fig:si:Wparfrail} for a direct comparison of links. Non-significant links are whited out at $p>0.05$.}
    \label{fig:si:Wfrail}
\end{figure}

\begin{figure}[H] 
     \centering
        \includegraphics[width=\textwidth]{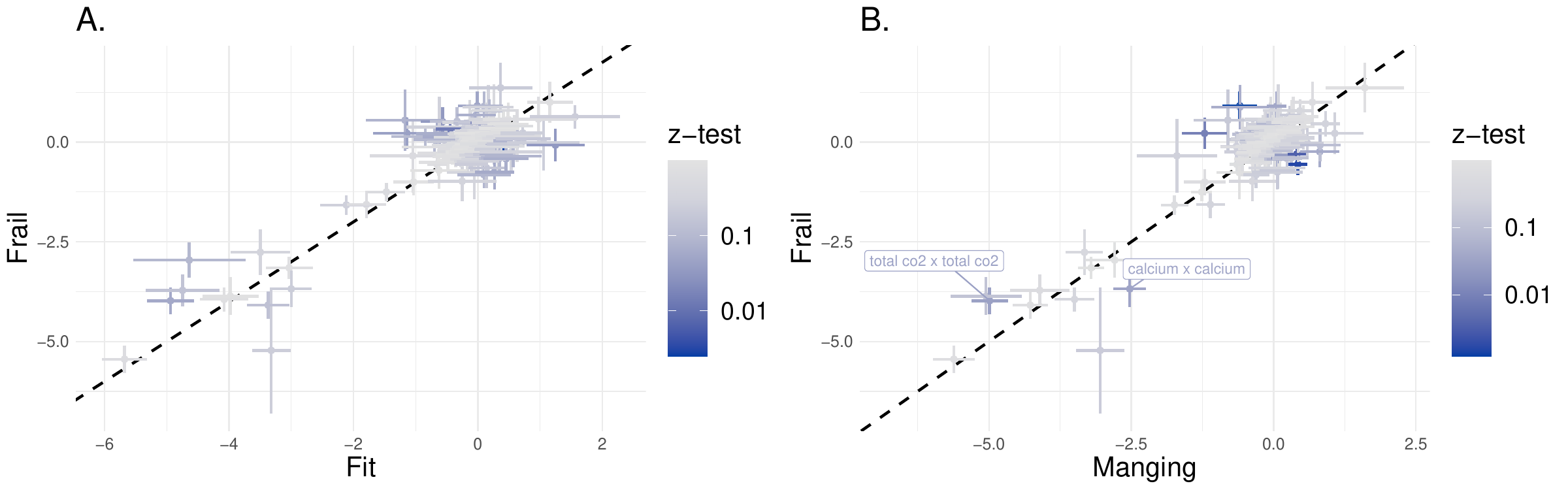}  
    \caption{Network link estimates by \textbf{A.} diabetes status, and \textbf{B.} sex. Significantly-different links with coefficients great in magnitude than $2$ have been labelled ($p < 0.05$). Dashed line indicates $x=y$.}
    \label{fig:si:Wparfrail}
\end{figure}

The estimated networks appear similar but there are clearly differences, Figure~\ref{fig:si:Wfrail}. These may reflect differences in statistical significance related to the relatively small groups. If we compare directly the parameters of the least and most frail, we see that the network parameters are strongly correlated, Figure~\ref{fig:si:Wparfrail}. The differences in the coefficients appear to be random, unlike the diabetics or males/females (Figure~\ref{fig:si:Wpardmsex}). This is demonstrated by the cloud of parameters with large error bars and no major outliers. This suggests that there aren't major differences in network parameters between the frailty groups.

\begin{figure}[H] 
     \centering
        \includegraphics[width=\textwidth]{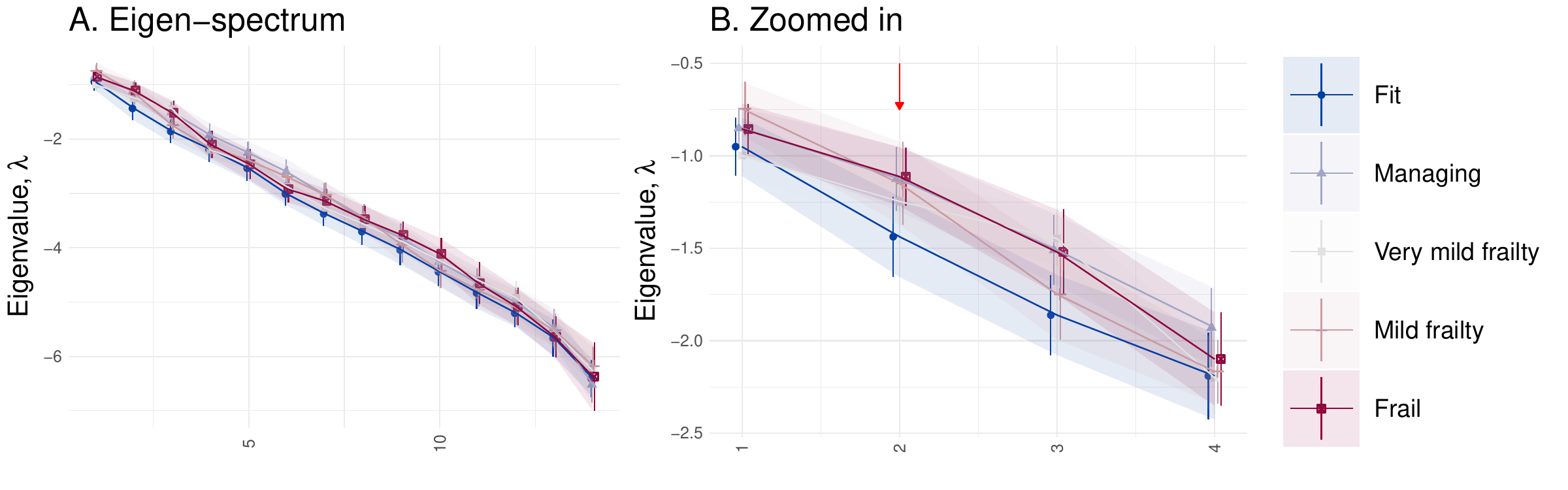}  
    \caption{Comparison of eigenvalue estimates by frailty. \textbf{A.} spectrum. \textbf{B.} spectrum zoomed into the first 4 eigenvalues (lowest stability/resilience). There is visually a trend towards less stable eigenvalues with increasing frailty (closer to 0), but the effect is smaller than error in most cases. Only in $\lambda_2$ does the error interval differ between the fit and frail groups (arrow). Eigenvalues closer to 0 are indications of smaller stability and hence weaker resilience.}
    \label{fig:si:eigfrail}
\end{figure}

Frailty is associated with a loss of resilience \cite{Fulop2010-ho}. In our model stability is captured by the eigenvalues. Large magnitude, negative eigenvalues are the most stable; positive are unstable. We compared the eigenvalues in Figure~\ref{fig:si:eigfrail}. Coloured strata are clearly visible, indicating a trend of lower resilience with increasing frailty ($\lambda$ closer to $0$). This trend was not significant using linear regression and the F-test, $p=0.42$ ($\lambda_1$) and $p=0.14$ ($\lambda_2$) --- but would be interesting to investigate with a larger dataset. It is interesting that the effect was strongest in $\lambda_2$. This could indicate that there is a minimum resilience for $\lambda_1$ that prevents it from going lower. Regardless, the differences are again small.

\begin{figure}[H] 
     \centering
        \includegraphics[width=\textwidth]{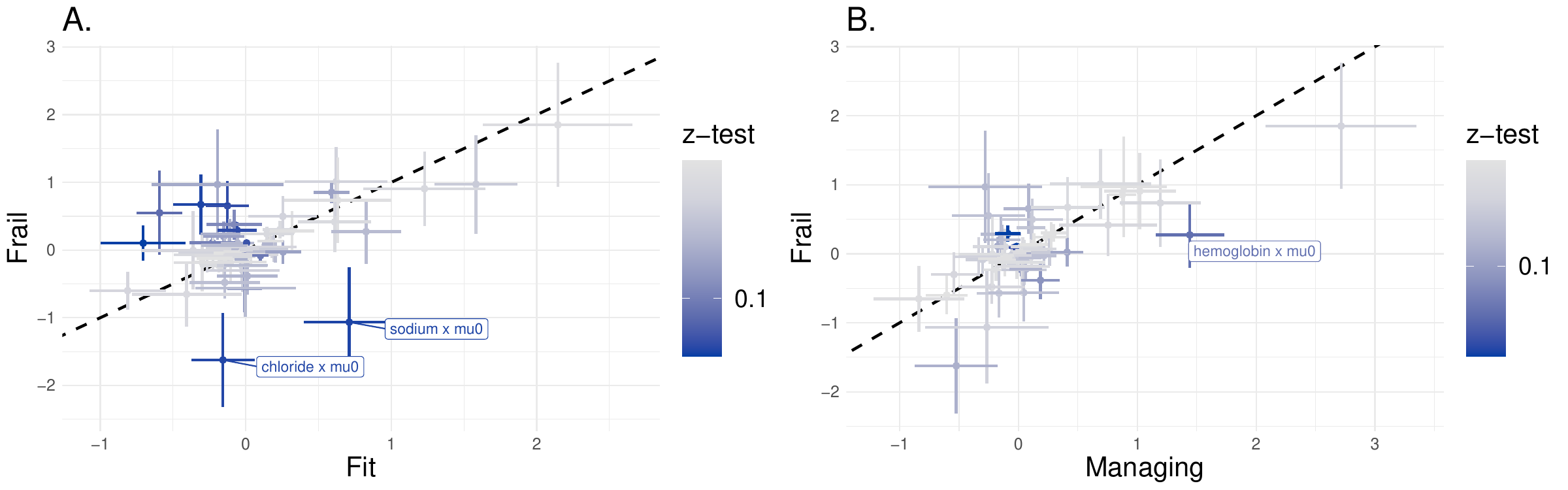}    \caption{Comparison of dynamical equilibrium estimates by frailty status. Labelled points are either significantly different at $p<0.01$, or significant at $p<0.05$ and at least 1 in magnitude. ``mu0'' is the intercept parameter, $\mu_0$. Dashed line indicates $x=y$.}
    \label{fig:si:lambdafrail}
\end{figure}

In Figure~\ref{fig:si:lambdafrail} we compare the most frail to the least frail for their dynamical equilibrium parameters, $\vec{\mu}$. As with the network links, the differences appear to be random. Hemoglobin and salt (sodium and chloride) stand out as differences between the groups. The lack of consistency between Fit vs Frail and Managing vs Frail suggests these differences are random.

\subsubsection{Fitting to a different set of biomarkers} \label{sec:si:morevar}
As mentioned in Section~\ref{sec:si:data}, we selected a subset of 14 blood tests which were regularly sampled. This permits us to build bigger networks by including more blood tests. This provides a sensitivity analysis for the estimated network: does adding new variables change the estimated network? We considered adding the 5 next-most commonly measured biomarkers: bilirubin, parathyroid hormone (pth), aspartate aminotransferase (ast), uric acid and transferrin saturation (\% sat).

\begin{figure}[H] 
     \centering
        \includegraphics[width=\textwidth]{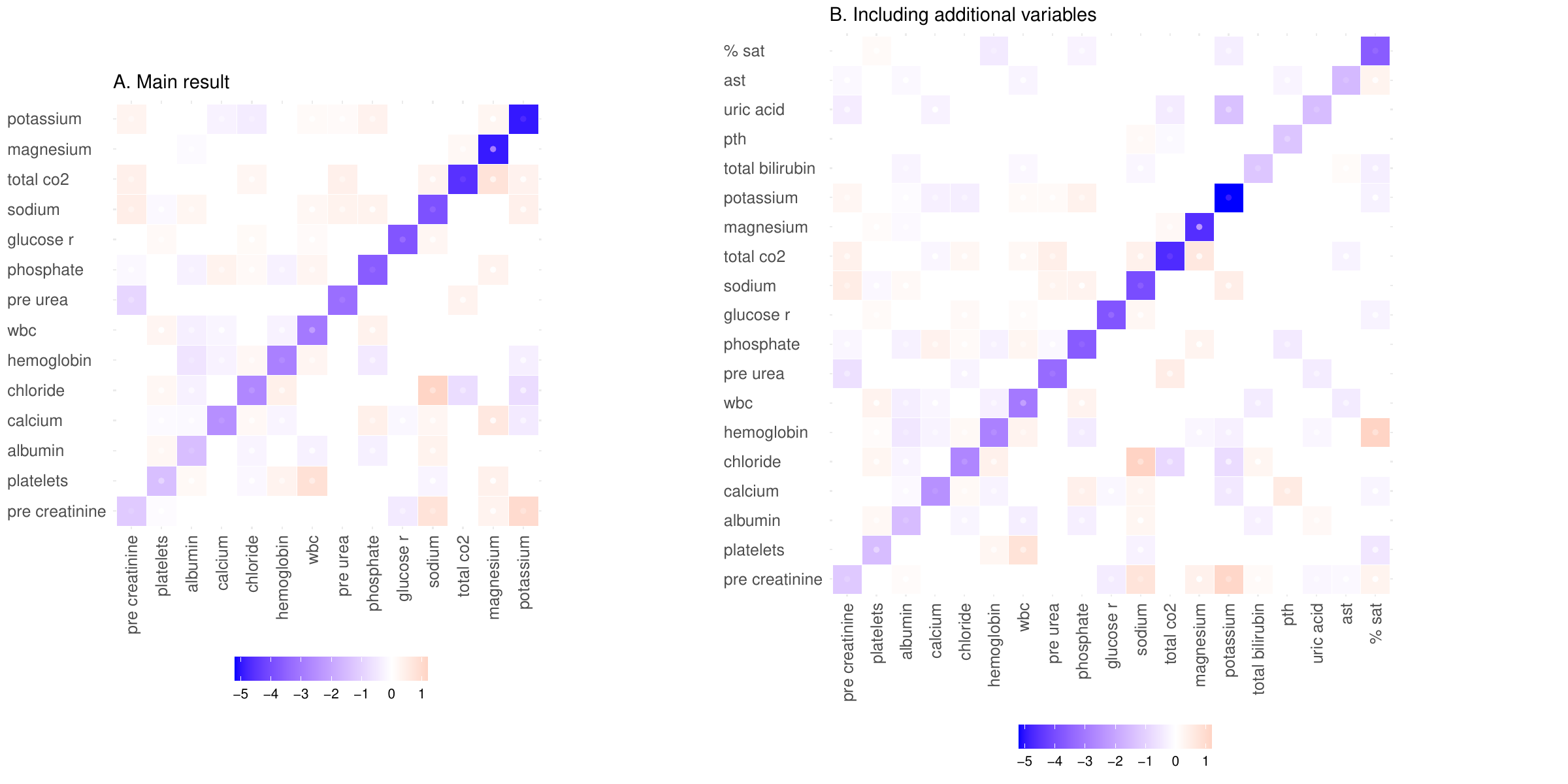}        \caption{Network estimates using the main set of variables and including an auxiliary set. The ordering of the shared variables is preserved in A. and B.. The (sub)networks look nearly identical. Non-significant links are whited out at $p>0.05$.}
    \label{fig:si:wmorevar}
\end{figure}

\begin{figure}[H] 
     \centering
        \includegraphics[width=.75\textwidth]{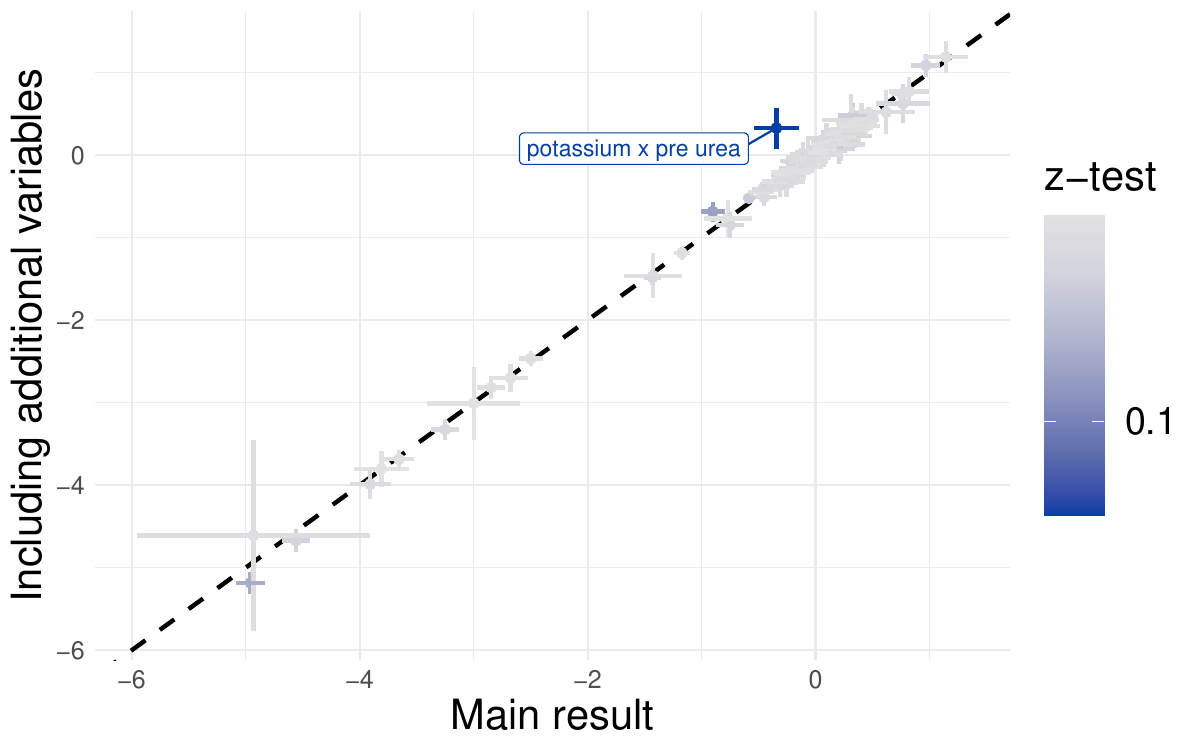}        
        \caption{Effect of adding more variables on network estimate. Only three of the coefficients were significantly different at $p<0.05$ when additional variables were included in the fit (labelled). Dashed line indicates $x=y$.}
    \label{fig:si:wmorevarpar}
\end{figure}

Looking at the specific network coefficients, we observed that they were Pearson correlated at $\rho=1.00$ ($p<2\cdot10^{-16}$), Figure~\ref{fig:si:wmorevarpar}. Hence the networks are nearly identical.

\begin{figure}[H] 
     \centering
        \includegraphics[width=.75\textwidth]{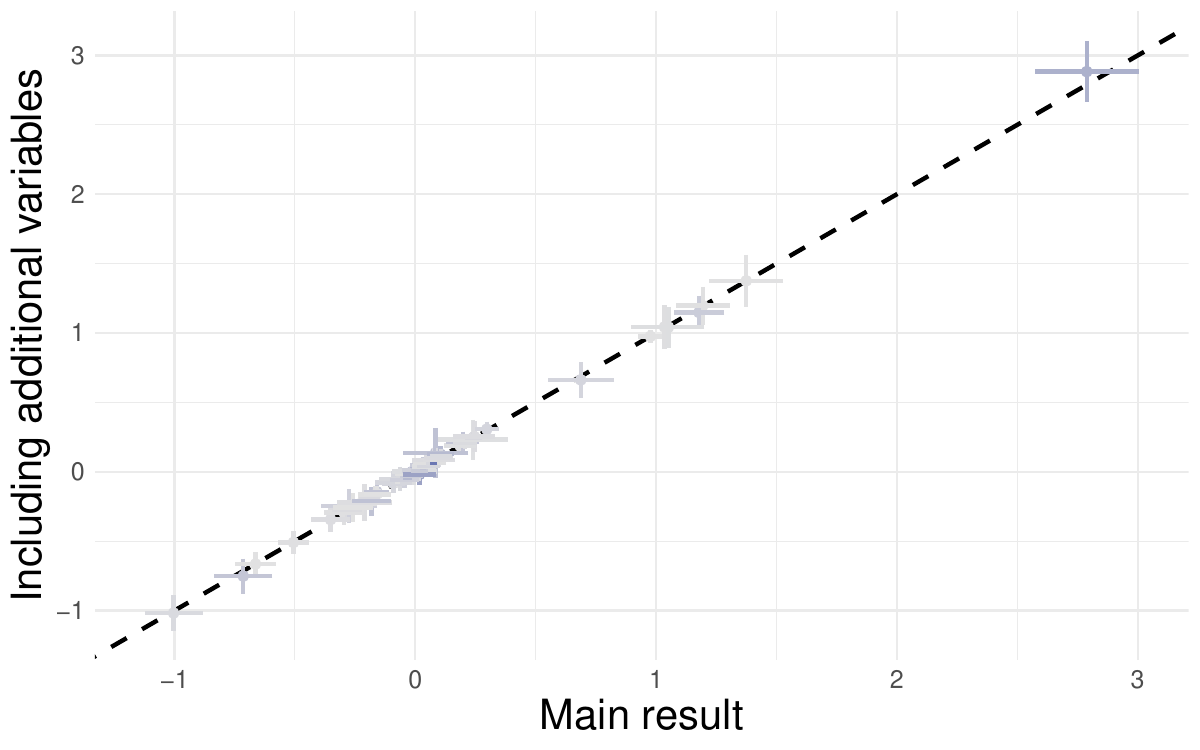}        
        \caption{Effect of adding more variables on dynamical equilibrium estimate $\vec{\mu}$. We see no difference when including additional variables. Dashed line indicates $x=y$.}
        \label{fig:si:lambdamorevarpar}
\end{figure}

The dynamical equilibrium parameters were almost identical, Figure~\ref{fig:si:lambdamorevarpar}. The two were Pearson correlated at $\rho=1.00$ ($p<2\cdot10^{-16}$).

Adding more variables did not change our parameter estimates. This is important since, by necessity, there are countless other possible variables that we have not measured and included. We infer that we have some robustness against excluding these variables.

\subsubsection{Fitting without a study window} \label{sec:window}
Here we relax our use of a study window. In the main text we restricted our attention to the interval from 3~months to 5~years. Here we consider simply including all time points. We find that the network estimate is nearly identical and therefore the natural variable transformation will remain the same. Similarly, the estimated dynamical equilibrium parameters are also nearly identical.

\begin{figure}[H] 
     \centering
        \includegraphics[width=\textwidth]{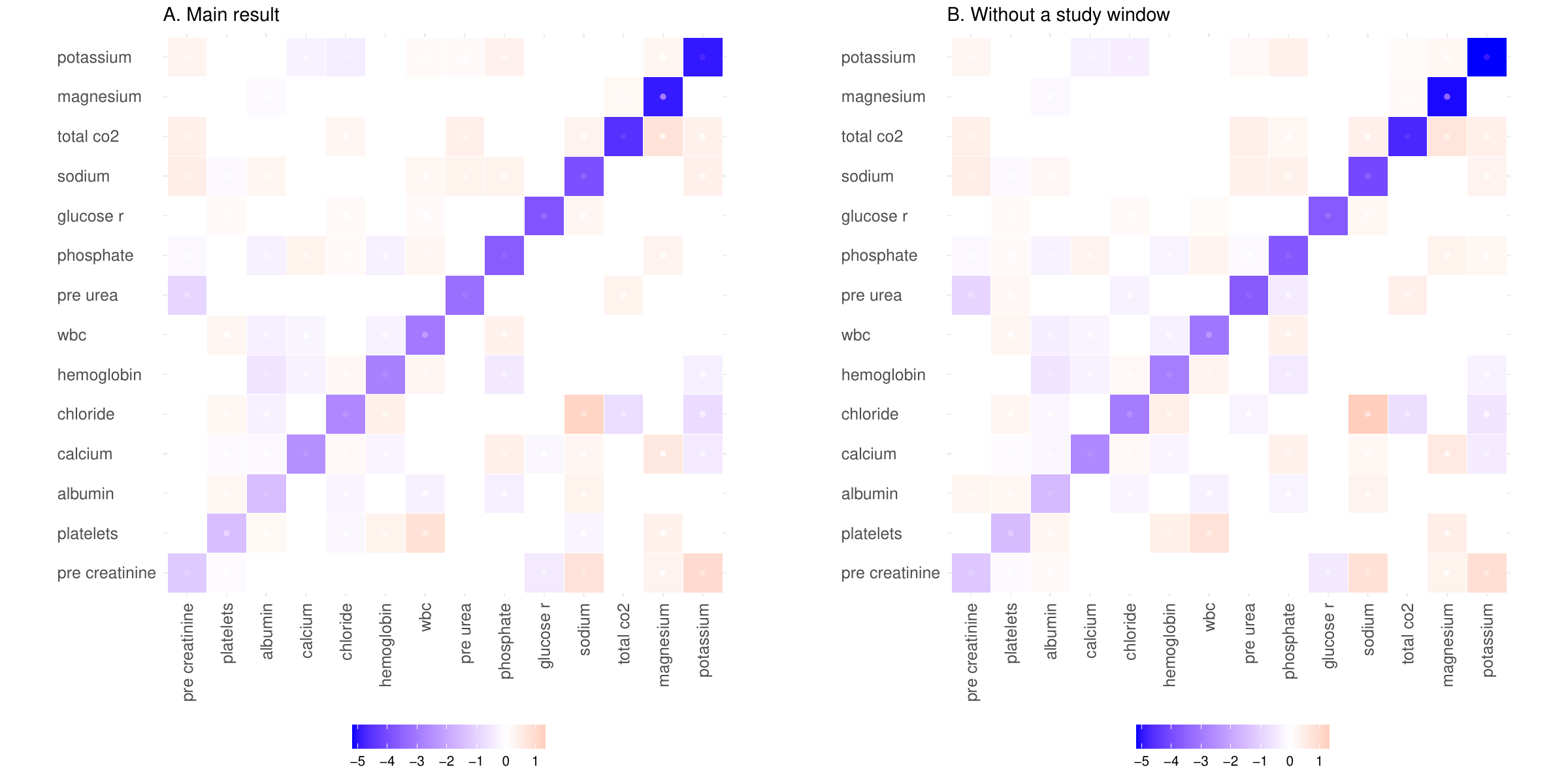}        \caption{Network estimates using the study window 3~months to 5~years (\textbf{A.}) versus no window (\textbf{B.}). The (sub)networks look very similar. Non-significant links are whited out at $p>0.05$.}
    \label{fig:si:wnowindow}
\end{figure}

\begin{figure}[H] 
     \centering
        \includegraphics[width=.75\textwidth]{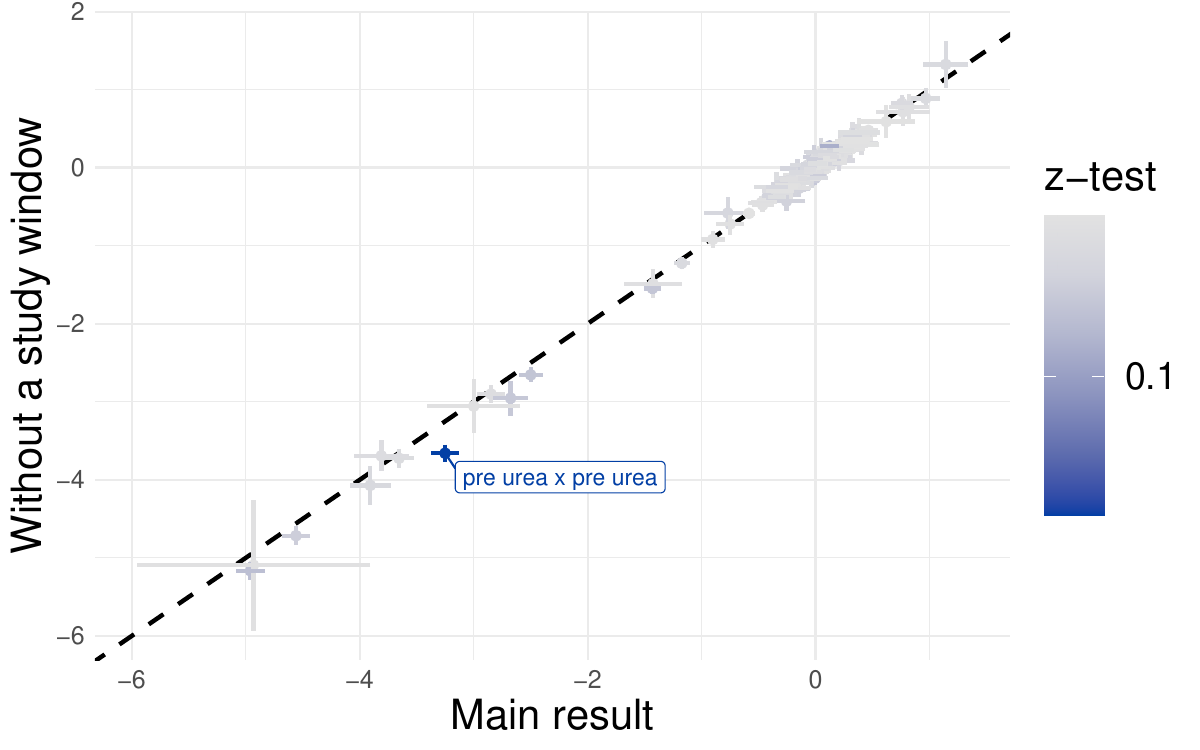}        
        \caption{Effect of imposing the study window on network estimate. None of the coefficients were significantly different at $p<0.05$ when additional variables were included in the fit. Dashed line indicates $x=y$. Labelled are parameters were significantly different at $p < 0.05$.}
    \label{fig:si:wnowindowpar}
\end{figure}

Looking at the specific network coefficients, we observed that the networks are nearly identical. They were Pearson correlated at $\rho=1.00$ ($p<2\cdot10^{-16}$), Figure~\ref{fig:si:wnowindowpar}.

\begin{figure}[H] 
     \centering
        \includegraphics[width=.75\textwidth]{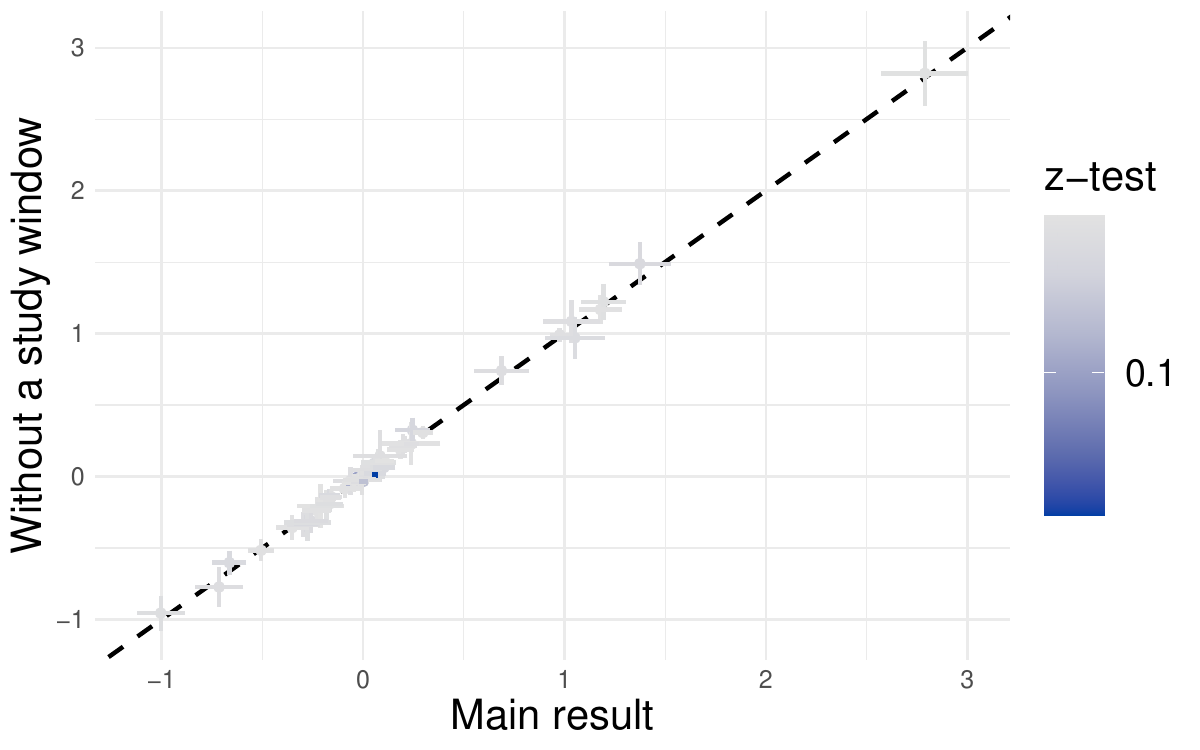}        
        \caption{Effect of imposing the study window on dynamical equilibrium estimate $\vec{\mu}$. The estimates are nearly identical. Dashed line indicates $x=y$.}
    \label{fig:si:lambdanowindowpar}
\end{figure}

The equilibrium parameters for $\mu$ were also almost identical, as shown in Figure~\ref{fig:si:lambdanowindowpar}. They were still Pearson correlated at $\rho=1.00$ ($p<2\cdot10^{-16}$). Hence the choice of study window should not affect our results.

\FloatBarrier

\end{document}